\documentclass[twocolumn]{aastex631}

\shorttitle{100 Pulsar Census below 100 MHz, and variations in pulsar DM and RM}
\shortauthors{Kumar et al.}
\graphicspath{{./}{figures/}}
\DeclareUnicodeCharacter{2212}{-}
\usepackage{amsmath}
\begin{document}

\title{A Multifrequency Census of 100 Pulsars below 100 MHz with LWA: A Systematic Study of Flux Density, Spectra, Timing, Dispersion, Polarization, and Its Variation from a Decade of Observations}

\correspondingauthor{Pratik Kumar}
\email{pratik.kumar@curtin.edu.au}

\author[0000-0003-0101-1986]{P. Kumar}
\affiliation{International Centre for Radio Astronomy Research (ICRAR), Curtin University, Bentley, WA, Australia}
\affiliation{Department of Physics and Astronomy, University of New Mexico, 210 Yale Blvd NE, Albuquerque, NM 87106, USA}

\author[0000-0001-6495-7731]{G. B. Taylor}
\affiliation{Department of Physics and Astronomy, University of New Mexico, 210 Yale Blvd NE, Albuquerque, NM 87106, USA}

\author{K. Stovall}
\affiliation{Department of Physics and Astronomy, University of New Mexico, 210 Yale Blvd NE, Albuquerque, NM 87106, USA}

\author[0000-0003-1407-0141]{J. Dowell}
\affiliation{Department of Physics and Astronomy, University of New Mexico, 210 Yale Blvd NE, Albuquerque, NM 87106, USA}

\author{S. M. White}
\affiliation{Space Vehicles Directorate, Air Force Research Lab.,  Kirtland AFB, NM 87117}
\affiliation{Department of Physics and Astronomy, University of New Mexico, 210 Yale Blvd NE, Albuquerque, NM 87106, USA}



\begin{abstract}
We present a census of 100 pulsars, the largest below 100 MHz, including 94 normal pulsars and six millisecond pulsars, with the Long Wavelength Array (LWA). Pulse profiles are detected across a range of frequencies from 26 to 88 MHz, including new narrow-band profiles facilitating profile evolution studies, and breaks in pulsar spectra at low frequencies. We report mean flux density, spectral index, curvature, and low-frequency turnover frequency measurements for 97 pulsars, including new measurements for 61 sources. Multi-frequency profile widths are presented for all pulsars, including component spacing for 27 pulsars with two components. Polarized emission is detected from 27 of the sources (the largest sample at these frequencies) in multiple frequency bands, with one new detection. We also provide new timing solutions for five recently-discovered pulsars. Low-frequency observations with the LWA are especially sensitive to propagation effects arising in the interstellar medium. We have made the most sensitive measurements of pulsar dispersion measures (DMs) and rotation measures (RMs), with median uncertainties of 2.9$\times$10$^{-4}$ pc cm$^{-3}$ and 0.01 rad m$^{-2}$, respectively, and can track their variations over almost a decade, along with other frequency-dependent effects. This allows stringent limits on average magnetic fields, with no variations detected above $\sim$20 nG. Finally, the census yields some interesting phenomena in individual sources, including the detection of frequency- and time-dependent DM variations in B2217+47, and the detection of highly circularly polarized emission from J0051+0423.

\end{abstract}



\section{Introduction} \label{sec:intro}
Even after the discovery of pulsars below 100 MHz, later associated with rapidly-rotating highly-magnetized neutron stars, low-frequency radio observations have remained limited, particularly due to challenges in instrumentation and the associated dispersive delays due to propagation. These effects are often characterized as dispersion and scattering due to the interstellar medium (ISM), which have a strong frequency dependence of $\nu^{-2}$ and $\nu^{-3}$ or higher, respectively, leading to smearing of the observed signal. Additionally, the spectral turnover in pulsars at low frequencies along with an increased Galactic background emission which follows a power law of index of $-$2.6 \citep[][]{Haslam} adds to the complexity, making it difficult to observe and discover pulsars at low frequencies. Although disruptive to pulsar observations, some of these frequency-dependent effects constitute a cache of interesting and complex phenomena and can be studied with more sensitivity at low frequencies. The inception of new low-frequency instruments with improved sensitivity in the last decade, including the Low Frequency Array \citep[LOFAR;][]{Vanhaarlem}, NenuFAR \citep[][]{zarka}; the Murchison Widefield Array \citep[MWA;][]{Tingay}); the upgraded Ukrainian T-shaped Radio telescope \citep[UTR-2;][]{Ryabov}; and the Long Wavelength Array \citep[LWA;][]{gtaylor}, has improved observing capabilities. Alongside development in computing which allows the implementation of techniques such as coherent de-dispersion, low-frequency pulsar observation has become more relevant in the last decade.

\subsection{Dispersion and Scattering by the ISM}\label{sec:introdm}
Dispersion and scattering introduce a frequency-dependent time delay due to the intervening plasma and lead to asymmetric pulse broadening due to multipath propagation. High sensitivity to these effects also makes low-frequency ideal to measure deviation in the expected behavior \citep[see, e.g.,][]{Lewandowski, Bansal} and test cold-plasma model for dispersion \citep[][]{Phillips}. An associated application is to provide high precision dispersion measure (DM) and understand the temporal variation in noise due to propagation in the ISM. This could provide useful information to improve high-precision pulsar timing array (PTA) experiments, geared towards the detection of low-frequency gravitational waves from supermassive binary black holes \citep[see, e.g.,][]{Phillips2,Cordshan}. Low frequency observations can also help improve solar wind modeling \citep[see, e.g.,][]{Tiburzi,Kumar} and provide direct detection of scattering screens along the Line-of-sight (LOS) to some pulsars \citep[see, e.g.,][]{Donner}.

\subsection{Polarization, Rotation Measure and Emission}\label{sec:intorrm}
Some pulsars are also substantially linearly polarized and can be used to investigate the magnetic field of the intervening ISM \citep[][]{Lyne} and heliosphere, and explore the density structure of Earth's ionosphere \citep[see, e.g.,][]{Malins}, via Faraday Rotation (FR), the integrated sum of the product of LOS electron density and magnetic field, and is proportional to $\lambda^{2}$. Hence, allowing high-precision RM measurements at low frequencies, and detecting weakly polarized sources. However, Earth's ionosphere is very dynamic, and accurate determination of pulsar RMs is limited by the ability to correct for the FR induced by the ionosphere on different time scales. 

Assuming no correlation between spatial electron density and the parallel component of the average ISM magnetic field ($B_\parallel$) we can obtain the relation,

\begin{equation} \label{eq:5}
    B_{\parallel} = 1.232\, \frac{RM}{DM}\, \mu G
\end{equation}

\noindent Independent measurements of pulsar RM using a large sample can be used to investigate and study the structure of Galactic magnetic fields \citep[GMF, see, e.g.,][]{Sobey}, while long-term continuous monitoring can allow us to detect magnetic field variations due to intervening structures such as HII regions, magnetized filaments of supernova remnants, and the motion of pulsar through a region of changing GMF \citep[see, e.g,][and references therein]{Wahl}. The associated time scales can be further used to infer their origin such as annual or solar cycle-related timescales, or more stochastic in nature. Pulsar polarization can also be used the understand the properties of pulsar emission itself \citep[see, e.g.,][]{Radhakrishnan,vonhoensbroech,Noutsos}.


\subsection{Profile evolution, Mode changes, and Spectral Turnover}\label{sec:introprof}
Radio pulsars emit coherent emission from relativistic plasmas, which can help us understand plasma physics in extreme conditions. But the exact mechanism remains unclear despite decades of work \citep[e.g.,][]{Melrose21}. Pulsars also undergo profile evolution \citep[][]{Phillips3} and spectral turnover at low frequencies \citep[][]{Sieber}, whose exact quantification is challenging due to propagation effects as well the strong Galactic background which adds to the systematics. Also, intrinsic profile evolution models like radius-to-frequency mapping (RFM) postulate that low frequencies arise higher in the magnetosphere, thus resulting in a wider emission profile \citep[][]{cordes}, and is shown to be prominent below 1 GHz \citep[][]{Thorsett}. Morphological investigation of emission has resulted in modeling in terms of ``core" and ``conal" components, originating near the magnetic poles and from a number of concentric cones \citep[][]{Rankin}, where ``core" components produce a steeper-index emission \citep[][]{basumitrageorge} and hence is important to understand pulsar spectra. Mode-changing phenomena, interpreted as chemical or structural change, that result in pulsar alternating between a ``normal" and ``abnormal" pulse profile has also been observed \citep[][]{Bartel2}. These changes are found to be simultaneous across frequencies but non-uniform. Thus, to improve our understanding of pulsar emission, and to study the dependence on observing frequency, measurements of pulse profiles are required across a wide range of frequencies for a large population of pulsars.

Quantifying pulsar radio spectra is important to understand the underlying emission mechanism. At moderate frequencies above a few 100 MHz, the spectrum is typically modeled as a power law of the form S$_{\nu}$ $\propto$ $\nu^{\alpha}$, where typically $\alpha\,\sim\, -1.4$ \citep[][]{bates}. However, there are deviations, especially at very low frequencies below 200 MHz where pulsar spectra show a turnover \citep[see, e.g.,][and references therein]{Sieber, malofeev}. \citet{malofeev} found the following relationship between the frequency of the spectral peak and the period of normal pulsars: $\nu_{\mathrm{max}} = 120\,P^{-0.36}$, where the peak frequency $\nu_{\mathrm{max}}$ is in MHz and $P$ is the pulse period in seconds. Spectral turnover at higher frequencies (around 1 GHz) has also been reported \citep[see, e.g.,][]{2021kijak}, while other pulsars do not have any turnover at detected frequencies. MSPs are a special class of pulsars that do not show a low-frequency turnover in their spectra \citep[][]{Kuzmin}, however more recent studies like \citet{rahulsharan} indicate a low-frequency turnover in MSPs. This necessitates the spectral characterization for a large sample of pulsars.

\begin{figure}[htbp!]
\centering
\includegraphics[width=0.45\textwidth]{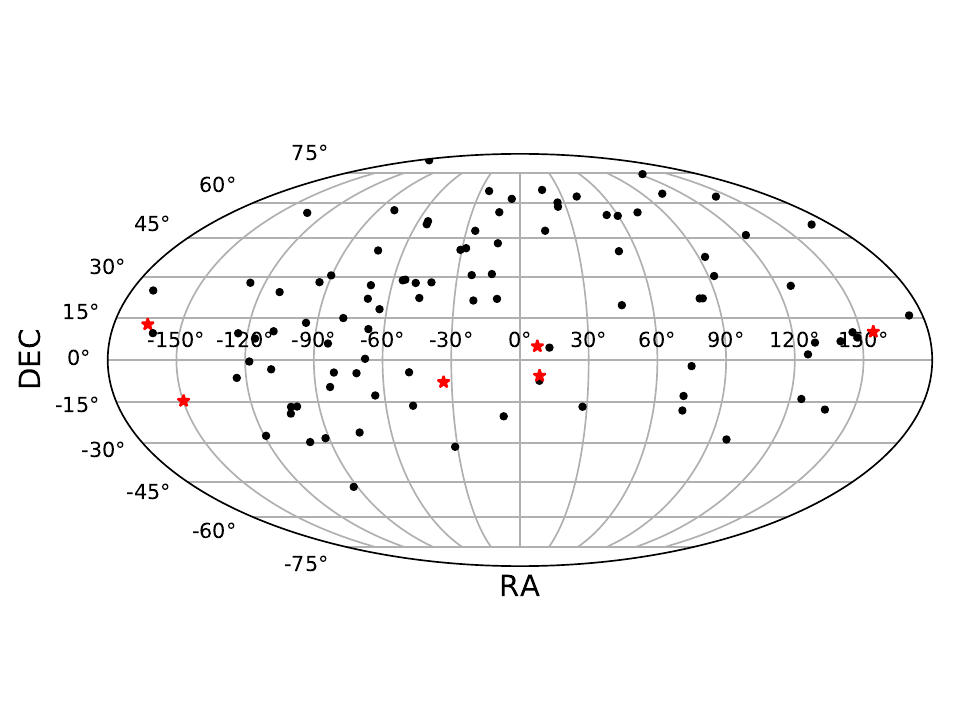}
\caption{The distribution of LWA pulsars on the sky, where black and red colors denote normal and millisecond pulsars, respectively.}
\label{fig:coverage}
\end{figure}

To study these effects in detail we have carried out continuous monitoring of pulsars at the lowest radio frequencies using the LWA. Here, we present a decade-long study of the different pulsar properties described above using a sample of 100 pulsars. Figure \ref{fig:coverage} shows the distribution of these sources on the sky. In section \ref{sec:observe} we provide the details of the observational setup and the data collected. Section \ref{sec:dataredux} describes the details of data reduction and analysis methods used to measure pulsar properties. In section \ref{sec:resultsanddiscussions} we present the relevant results and discussions. We finally conclude with section \ref{sec:summary}.

\section{Observations and LWA Pulsar Archive} \label{sec:observe}
The LWA \citep[][]{gtaylor} is a low-frequency interferometric dipole array radio telescope with three operational stations, and several more under construction or in the planning phase. Among these LWA1 was the first and is co-located with the Karl G. Jansky Very Large Array, and is used for our continuous pulsar monitoring program.
For complete details of data reduction, observing strategy, and data storage please refer to \citet{Stovall}. Briefly, observations are conducted with two simultaneous beams, each with two tunings of the center frequency with a bandwidth of 19.6 MHz, varying between 0.5-2 hours based on source brightness with 3-6 week cadence and additional solar campaigns, and are constrained to occur at transit to minimize elevation-dependent sensitivity of beams. Continuous monitoring of pulsars with LWA1 started in 2013, with an increasing number of sources over the years.

Completed reduced data products are stored in a public archive\footnote{LWA Pulsar Archive: \\ \url{https://lda10g.alliance.unm.edu/PulsarArchive/}}, where they are accessible (to anyone) for analysis, which currently amounts to more than 13000 hours of observations and spans up to a decade for some sources. The data reduction pipeline makes use of the LWA Software Library\footnote{\url{https://github.com/lwa-project/lsl}} \citep[LSL,][]{Dowell}, and the standard pulsar data reduction tools \textit{TEMPO}\footnote{\url{https://tempo.sourceforge.net/}}, \textit{PSRCHIVE}\footnote{\url{http://psrchive.sourceforge.net/}} (\citealt{vanstraten}), \textit{DSPSR}\footnote{\url{http://dspsr.sourceforge.net/}}, and \textit{PRESTO}\footnote{\url{https://github.com/scottransom/presto}}. For the purpose of this study, we mainly use the reduced data products from each individual tuning in the standard \textit{PSRCHIVE} archive format, which stores the data in a multi-dimensional format of time/sub-integration, frequency, phase bins, and polarization. The data is stored in 30-second averaged sub-integrations, where the number of frequency and phase bins varies between 128-4096 and 128-1024 respectively, for different sources. Other reduced data products are used for diagnostic purposes.

\section{Data reduction and Analysis}\label{sec:dataredux}
\subsection{Pulsar Timing, DM and scattering}\label{sec:timing}
We use the long-term timing data for all the pulsars in our sample to refine timing solutions and investigate DM variations and frequency-dependent DM behavior, as well as other properties. As discussed above, the pulse profile evolves with frequency due to intrinsic changes as well as propagation effects \citep[][]{Craft}, which can significantly affect the measurement of quantities such as DM \citep[see e.g.,][]{Ahuja}. Ignoring profile evolution can lead to bias in the Time of Arrival (TOA) of pulses, since the timing is determined by template matching and referencing to terrestrial clock and abstract time standards, leading to loss of precision and sensitivity. We use the pulsar data at one or more frequencies, depending on pulsar brightness and the scatter in the pulses obtained. For example, in some cases, the pulsar is not sufficiently bright at the highest center frequency of 79.2 MHz so we only use data corresponding to other frequencies. In other cases, the pulsar has poor S/N at the lowest frequency band centered at 35.1 MHz, and hence that tuning is left out, while in others the scatter in the pulse profile is too great to obtain a good template and so lower frequencies are not used. Since we have four independent tunings, we initially treat them separately creating a template profile for each, and combine the obtained TOAs at a later stage to improve our timing solution as well as our ability to measure the DM over a wide range of frequencies. The preparation stage of the date involves improving the S/N and averaging by using the tools in \textit{PSRCHIVE}. We obtain the typical pulse profile at each center frequency by choosing the best available data for that pulsar. We then apply a median RFI mask to the data using the routine \textit{zap}, removing data points higher than six times the median value for a smoothing window size of 13 channels in frequency. Additionally, we visually inspect the data after this process to remove any low-level RFI and mask it using the routine \textit{psrzap}. Once completed, the average pulse profiles are created by averaging this masked data over time and frequency, to one channel and sub-integration. This is followed by modeling these averaged profiles as Gaussian components and aligning them in phase to the highest available center frequency for that pulsar. These phase-aligned Gaussian profiles are used as the final templates for timing.

The timing analysis starts by applying some averaging to the entire coherently-dispersed raw data available for a given pulsar across multiple epochs. We begin by removing RFI from the data, following the same approach as before, except for visual inspection at the later stage. This is followed by averaging the data to one sub-integration and eight channels for any given tuning. Then TOAs are obtained independently for each tuning by cross-correlation with the appropriate templates using the routine \textit{pat} in \textit{PSRCHIVE}. The resulting TOAs are then combined into a single file for final timing. We then use \textit{TEMPO} and the available pre-existing pulsar ephemerides for the LWA derived using a small fraction of the data, as described in \citet{Stovall}. Initially, we only fit for the pulsar's spin rate and astrometric parameters, and then visually inspect the quality of TOAs based on the fitting residuals. If there is too much scatter in the residuals due to noise in the data, we go back to the data-averaging step and further average down to four, two, or one channel per tuning to improve the TOA quality. After improving the TOAs we fit for the spin, astrometric parameter, and the DM to see if there is any significant change from the pre-existing values, followed by fitting for the first derivative of pulsar spin, until the reduced chi-square ($\chi^{2}$) cannot be improved further. In the final step, we freeze the DM and fit for all other parameters including DMx, which is a variation of DM from the fiducial value, for each epoch. TOA fitting is done in an iterative process where any extreme outliers are removed at each step until we obtain a $\chi^{2}$ value of $\sim$ 1. In a few pulsars, a factor is used to scale the measurement uncertainty such that $\chi^{2}\,=\,1$ can be achieved: this factor, known as ``EFAC", depends on instrument characteristics and can be in the range $\sim$2-3. The resulting DMx values are stored for further analysis. Additionally, for pulsars where we have data from two or more frequency tunings, we independently evaluate DMx from each band and look for frequency-dependent effects in the DM that may be due to scattering. 

\subsection{Average Profile and Components}\label{sec:avgprof}
To obtain the average pulse profile of a pulsar we add data from multiple epochs, such that only epochs with comparable S/N are combined in the final result. We start by selecting the epoch with the highest S/N in the available data and other epochs with comparable S/N are selected automatically. The epochs are added in phase using the updated timing solutions, as described in section \ref{sec:timing}. Before adding the data, it is cleaned for RFI as described in section \ref{sec:timing} except for the visual inspection followed by averaging to one channel and one sub-integration. Then these individual epochs are added in phase using the routine \textit{psradd} from \textit{PSRCHIVE}, scaling the different epochs by their respective S/N. After the addition, the average pulse profile so obtained is inspected along with the individual epochs to identify any poor data, and the process is repeated by removing one or more epochs as appropriate. This produces the final representative pulse profiles. The process is repeated for each frequency tuning, to obtain four separate frequency profiles wherever possible.
These average pulse profiles are then used to measure the component width and separation in the case of multi-component profiles. The individual components are modeled as Gaussian components using the script \textit{pygaussfit.py} available in \textit{PRESTO}, to obtain the phase, full width at half maximum (FWHM), and amplitude of each component. Based on these measurements we create a Gaussian profile to measure the profile S/N, pulse width at half maximum (w$_\mathrm{50}$) as well as ten percent of the maximum (w$_\mathrm{10}$). The S/N is measured over the entire pulse, rather than a peak-to-peak measurement. 
\subsection{Flux Density and Spectra}\label{sec:fluxspec}
For a low-frequency dipole array like the LWA, which uses electronic beam steering, the system equivalent flux density (SEFD), varies with zenith angle, frequency, and local sidereal time (LST). To study the dependence of LWA1 SEFD on zenith angle and observing frequency, the authors in \citet{schinzel} performed drift scan observations of bright known radio sources (Cyg A, Cas A, Tau A, and Virgo A) over a range of zenith angle and frequency. The results from these observations were used to determine LWA1's SEFD at the zenith for Stokes I, using the drift scan method described in \citet{Ellingson}. The derived SEFD was then converted to a value at the zenith by applying an empirically derived power-ratio correction for Stokes I, given by
\begin{equation}\label{eq:6}
    SEFD(zenith) = \frac{SEFD(obs)}{(166.625*E^{-1.251}+0.173)}
\end{equation}

\noindent where $E$ is the target elevation in degrees. The value of zenith SEFD was found to be fairly constant over frequency bins at 8.119\,kJy, but below 40 MHz a value of 10.146\,kJy was found to fit better. Using the empirical relation and the pulsar transit elevation, we then calculated the SEFD for each of our pulsar observations. The error in LWA1's zenith SEFD is about 25$\%$, but to account for any unaccounted variations due to changes in LST and other factors, we assume a total error of 50$\%$ in the LWA1 SEFD measurements.

These SEFD measurements are then applied to the average pulse profile to calculate the average flux density of each pulsar for each tuning. Using the pulsar radiometer equation from \citet{handbook} below, 

\begin{equation} \label{eq:7}
    S_{\nu} = \gamma \frac{ \text{SNR}  T_\mathrm{sys}}{G \sqrt{n_\mathrm{p} \delta \nu T}}\sqrt{\frac{W_\mathrm{eq}}{P-W_\mathrm{eq}}}
\end{equation}

\noindent where $\gamma$=1 is a correction factor, SNR is the signal-to-noise ratio of the pulsar, $\frac{G}{T_\mathrm{sys}}$=SEFD, $W_\mathrm{eq}$ is the equivalent width of the pulse, P is the pulse period, T is the total integration time, $\delta \nu$ is the effective bandwidth of the data, and $n_\mathrm{p}$ is the number of polarizations, we obtain the pulsar flux density, S$_{\nu}$. We use the value of w$_{50}$ for the equivalent width of the pulse in these calculations. 

These flux density measurements along with some other available flux density measurements for these pulsars, 
are then used to fit for the spectral index ($\alpha$), and the ``curvature" ($\beta$) which accounts for the turnover at low frequencies. We perform a non-linear least-squares fit on this data using the fitting algorithm  \textit{spxfit} from the CASA package, widely used for spectral index and higher-order spectrum fitting in radio astronomical routines\footnote{https://casa.nrao.edu/docs/taskref/spxfit-task.html}. We fit the spectrum using the polynomial,

\begin{equation} \label{eq:8}
    S_{\nu} = S_0 \times e^{\alpha*\tilde{\nu}+\beta*\tilde{\nu}^{2}}
\end{equation}
 \noindent where $S_0$ is the reference frequency, $\tilde{\nu}$ = $\log\frac{\nu}{\nu_0}$, and $\nu_0$
 is a reference frequency. The choice of $\nu_0$ can affect the spectral fit to the data. To minimize residuals, $\nu_0$ is chosen as the nearest integer power of 10, from the geometric mean of all available frequencies, and is given by,
 \begin{equation}\label{eq:9}
     \nu_0= 10^{integer((\log_{10}\nu_1 *\log_{10}\nu_2...*\log_{10}\nu_n)^{\frac{1}{n}}))}
 \end{equation}

Using equation \ref{eq:8} and \ref{eq:9}, we apply a spectral fit to our data and measurements from other pulsar flux density catalogs, weighted by the fractional error in flux density measurements, 
to obtain the value of $S_0$, $\alpha$, and $\beta$, and use these to calculate the turnover frequency, $\nu_{turnover}$, as the low-frequency inflection point of the curve. Additionally, we also perform a linear fit to the data using the same equation as above, while freezing $\beta=0$. The AIC criteria from \citet{jankowski441psr} is used to determine which curve is the best representation of the collected data. 
 
\subsection{Polarization and RM}
\subsubsection{Detection of polarization and pulsar RM}\label{sec:rm}
Although the level of linear polarization in pulsars can often be high \citep[][]{kerr}, due to the effect of RFI and depolarization due to low resolution at low frequencies, it is generally more difficult to fit an RM compared to a DM. We start by cleaning RFI in the data using the method described above in section \ref{sec:timing}. The next step involves averaging the ``zapped" file to 256 channel and phase bins and one sub-integration. This cleaned average data is then converted to the Stokes basis (I,Q,U,V) for further processing. The RM is determined using the \textit{RMFIT}\footnote{https://psrchive.sourceforge.net/manuals/rmfit/} routine in \textit{PSRCHIVE}. The instrumental polarization of LWA1 is quite low, at less than 10$\%$ and 5$\%$ for linear and circular polarization, respectively \citep[][]{Obenberger}, for elevation angles within 45$^{\circ}$ of zenith. Since most of the pulsars where we detect polarization have small zenith angles (below 45$^{\circ}$), except for B0149-16, B0628-28, and B2327-20, we do not apply any instrumental polarization calibration. For pulsars with available low-frequency RM measurements, the initial values from these catalogs were used as starting points \citep[see, e.g.,][]{Dike}. For other cases, we used the ATNF pulsar catalog\footnote{\url{https://www.atnf.csiro.au/research/pulsar/psrcat/}} value \citep[][]{atnf}. The pulsar RM is obtained by a two-step process. First, we use the brute-force search around the chosen catalog value in the range $\pm$ 10 rad m$^{-2}$ to get a refined measurement by fitting a Gaussian to the linear polarized flux against trial RM values. In the second step, further refinement is done around this value using the iterative differential position angle refinement method\footnote{https://psrchive.sourceforge.net/manuals/rmfit/}, which computes the weighted differential PA, after splitting the data into two half bands, until this value is smaller than its threshold, for a 3 sigma threshold. For pulsars without any available RM priors, we first perform a brute-force search in the range of $\pm$100 rad m$^{-2}$ for an initial guess, then carry out the two steps above using that value as a starting point.
The same method is used to measure the pulsar RM on all available epochs and for each frequency tuning.

\subsubsection{Ionospheric Correction}\label{sec:ionrm}
In order to accurately measure the ISM contribution to the RM, we need to correct for the contribution due to Earth's ionosphere. These corrections require the electron density profile and the magnetic field with height along the LOS to the pulsar. For Earth's magnetic field, we use the International Geomagnetic Reference field 13 \citep[IGRF-13;][]{Alken2021}, which gives a set of spherical harmonic coefficients to describe the temporal portion of Earth's magnetic field ($B_\mathrm{E}$) and is continually revised typically every five years. The electron density of the ionosphere can be readily obtained using one of the available global total electron content (TEC) maps, which uses a set of GPS receivers located in remote locations to obtain TEC values and then interpolates between these stations to obtain values at a desired location. Additionally, we can make use of the LWA GPS receiver, co-located with LWA1 to get these measurements, which uses the dual frequency of 1574.4 and 1227.6 MHz and measures TEC based on the difference in signal travel time between the two frequencies. Given a single receiver, this only produced measurements along the vertical direction and needs to be projected along the LOS to the pulsar to get the effective contribution. Since the ionosphere is composed of a set of layers, the densest and largest among which is the ``\textit{F}" layer, between 150 and 500\,km, following the recommendations by \citet{Malins} we model the ionosphere as a thin layer located at 350\,km to calculate these corrections. The value of the TEC and (B$_\mathrm{E}$) are then computed at the ionospheric pierce point (IPP), which is the location where the LOS to the pulsar intersects this thin shell of the ionosphere. Based on these measurements, the ionospheric correction to the RM is given by, 

\begin{equation}\label{eq:10}
    RM_\mathrm{ion} = 2.63\times10^{-13}(TEC*B_\mathrm{IPP})
\end{equation}
\noindent where $B_\mathrm{IPP}$, is the parallel component of $B_\mathrm{E}$ at IPP along the LOS. For more details on ionospheric correction and possible sources of biases, please refer to \citet{Malins}.

\section{Results and Discussions}\label{sec:resultsanddiscussions}
\subsection{Average Pulse Profiles}\label{sec:resultsprof}
We report the detection of pulsed emission from 100 previously known pulsars, presented in Table \ref{tab:dmperiod}. The average pulse profiles are only reported for 97 sources since B0136+57 and B1322+83 are too weak to obtain adequate S/N and B0531+21 is only detected via single pulses. The average pulse profiles are reported at the center frequency of each observed tuning and are shown in Figure \ref{fig:psrprof}. 
\figsetend
We obtained up to four narrow-band pulse profiles for each pulsar, depending on its brightness and the scatter in the pulse profile. Although decametric emission from radio pulsars below 100 MHz has been reported previously, the sample of radio pulsars detected at these frequencies has remained small, due to propagation effects and low-frequency turnover in pulsar spectra, as described above. \citet{zakharenko} and \citet{kravtsov} have reported the detection of emission from 40 and 20 pulsars, respectively, at the UTR radio telescope, but the detections were reported using narrow-band observations at a single central frequency. Broadband observations have been carried out with LOFAR-LBA \citet{bilous2020} and NenuFAR \citet{bondonneau}, with detections of pulsed emission from 43 and 64 pulsars, respectively. Although these were among the largest samples below 100 MHz, they were also limited in their scope, as the observations were conducted using a large bandwidth of $\sim$40-50 MHz at one central frequency. Prior to this study, the largest sample of low-frequency pulsar observations covering a broad range of frequencies was presented in \citet{Stovall}, who reported the detection of 44 pulsars using the LWA. Our follow-up study using the current sample of $\sim$100 pulsars with the LWA more than doubles the number of pulsars well studied below 100 MHz. This is the single largest collection of pulsar detections at decametric wavelengths and provides narrow-band pulse profiles down to 26 MHz at multiple frequencies for 53 pulsars for the first time.

In addition to the average pulse profile, we also use our multi-epoch data to look for any sudden changes in the pulse profile such as mode changes and other anomalous behavior. \citet{Bansalecho} reported the detection of pulse evolution using this data, which was attributed to reflection from a dense cloud in the ISM. Similarly, we also detect the known mode-changing behavior of pulsar B0943+10 \citep[][]{Suleimanova} as shown in Figure \ref{fig:psrprof}, where the pulsar switches between a bright (B) and quiet (Q) mode. We also looked for any mode-changing behavior in the B0823+26, which has been reported at low frequencies by \citet{Sobey2} with the quiet mode being 100 times fainter than the bright mode. No such behavior was seen in our data. The typical mode duration for this pulsar is several hours, with the B mode being more abundant (about 5 times longer) than the weaker Q mode. LWA observes this pulsar for 30 mins per epoch, so it is possible that we missed the Q mode. However, the data has been collected for many years and this seems unlikely, which suggests that low sensitivity can explain the non-detection of the weaker Q mode. Using the multi-frequency narrow-band profiles in Figure \ref{fig:psrprof}, we can also visually inspect the effects of broadening and evolution in pulse profiles due to scattering in the ISM. From the current sample, 17 pulsars, including B0329+54, B0450+55, B0818-13, B0823+26, B0919+06, B1508+55, B1541+09, B1600-27, B1642-03, B1749-28, B1821+05, B1822-09, B1831-04, B1842+14, B1857-26, B1911-04, and B2217+47, show noticeable effects of scattering on the pulse profile across our frequency range.

\subsection{Component Analysis and Pulse Width}\label{sec:resultscomp}
For the 97 pulsars for which we provide average pulse profiles, we also performed a decomposition of pulse components for pulsars which clearly indicated multiple components, ignoring the effect of scattering. For the current sample, 65, 27, and 5 pulsars show one, two, and three components respectively. Pulsars with 4 or more components were not detected. The Gaussian fitting process for component fitting for all pulsars has already been described in section \ref{sec:avgprof}. The five pulsars exhibiting three components are B0834+06, B0950+08, B1133+16, B1257+12, and B1919+21, whose component properties have already been described in \citet{Stovall}. However, B1919+21 was reported as having two components in \citet{Stovall}. For the 28 pulsars with two components (B0943+10 in both modes), we provide the properties of individual components in Table \ref{tab:components}. Using the Gaussian fitting technique described in section \ref{sec:avgprof}, we performed the fit as a function of frequency for each pulsar, with the profiles as shown in Figure \ref{fig:psrprof}. The pulse components for these pulsars are labeled as 1 and 2, where 1 is the leading component of the pulse. Using a profile of all components, the average w$_{50}$ and w$_{10}$ parameters for all pulsars were calculated and listed in Table \ref{tab:fluxdensity}.
We see a general trend of decrease in pulse width with frequency, consistent with the predictions of RFM \citet{2003Kijakgil}, but there are some exceptions. In some cases, the pulse width is the same at 49.8 MHz and 64.5 MHz, where we are most sensitive, while in some others it remains almost unchanged or increases slightly at the lowest frequencies. These could be attributed to the collective effect of LWA sensitivity, changes in pulsar brightness with frequency due to spectral turnover, and scattering due to the ISM. Previous studies like \citet{Mitrarankin} have also shown that RFM is associated with outer-cone pairs, which may lead to the deviations we see in the case of some LWA pulsars. A detailed RFM analysis of individual pulsars is left to a future article.

It is also worth noting that, in a few pulsars, close visual inspection of the pulse profile shows the presence of more components than are obtained via the Gaussian fitting procedure, again likely due to limited sensitivity. Profile for B0329+54 indicated the presence of three components of which one is faint and the other is difficult to distinguish due to scattering. Similarly, B0823+26 and B1612+07 also indicate the presence of another faint component merged in the scattering tail. Pulsar B2306+55 has a triple peak structure with each peak having a narrow width and J0051+0451 shows four components, of which three are faint and consistent over the four observing bands. These results are consistent with the observation of these pulsars at higher frequencies, as presented in the EPN \footnote{\url{https://psrweb.jb.man.ac.uk/epndb/}} pulsar database. 

\begin{figure}[htbp!]
\includegraphics[width=\linewidth]{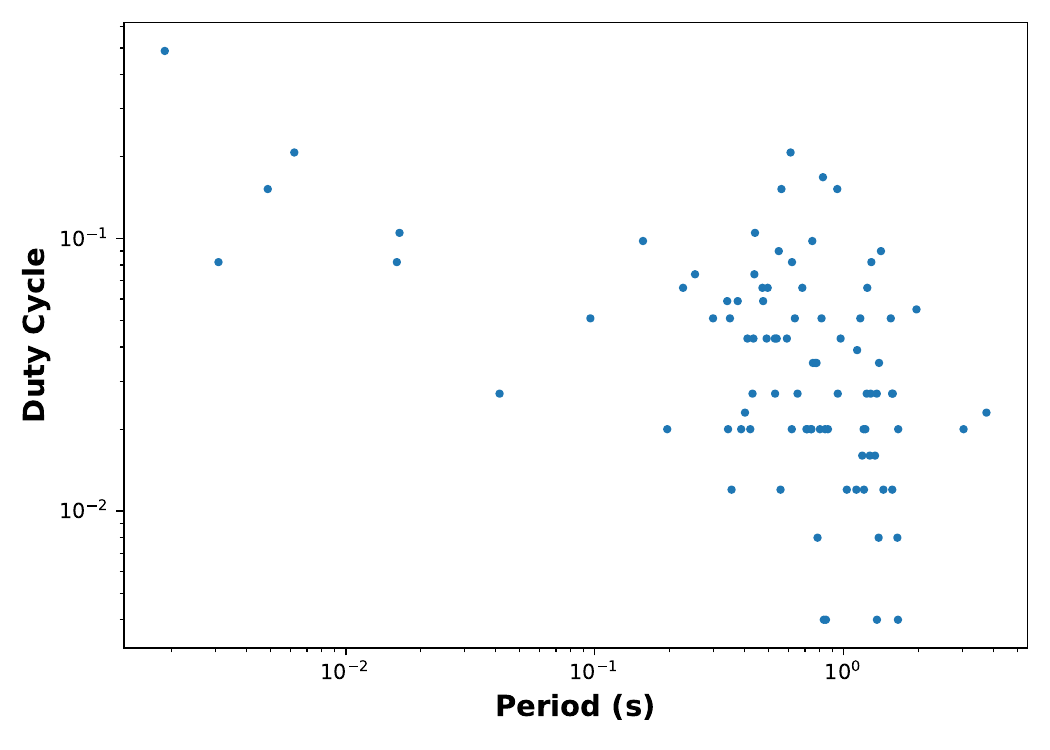}
\caption{Duty cycle (w$_{50}$/P) plotted against pulse period for all LWA pulsars.}
\label{fig:duty}
\end{figure}

 Previous studies have also shown that a broad inverse correlation exists between the pulse duty cycle (ratio of pulse width and period), and the period of the pulsar \citep[see, e.g.,][]{Lorimer1995}. For the LWA pulsar sample, a similar inverse correlation exists, and Figure \ref{fig:duty} shows this relation. For pulsars with periods $\sim$1\,sec we observe a duty cycle of $\sim$5$\%$, with a scatter of the same magnitude, whereas MSPs show a duty cycle larger by a factor of 4-10. The pulsar J1400-1431 does not follow the general inverse-correlation trend. This correlation directly affects the fraction of visible pulsars, has implications for pulsar population studies, and has been used to infer information about the magnetic and rotation axis inclination, as well as to constrain the beaming fraction \citep[e.g.,][]{Lyne1988}. 

\begin{figure*}[htbp!]
\includegraphics[width=\textwidth]{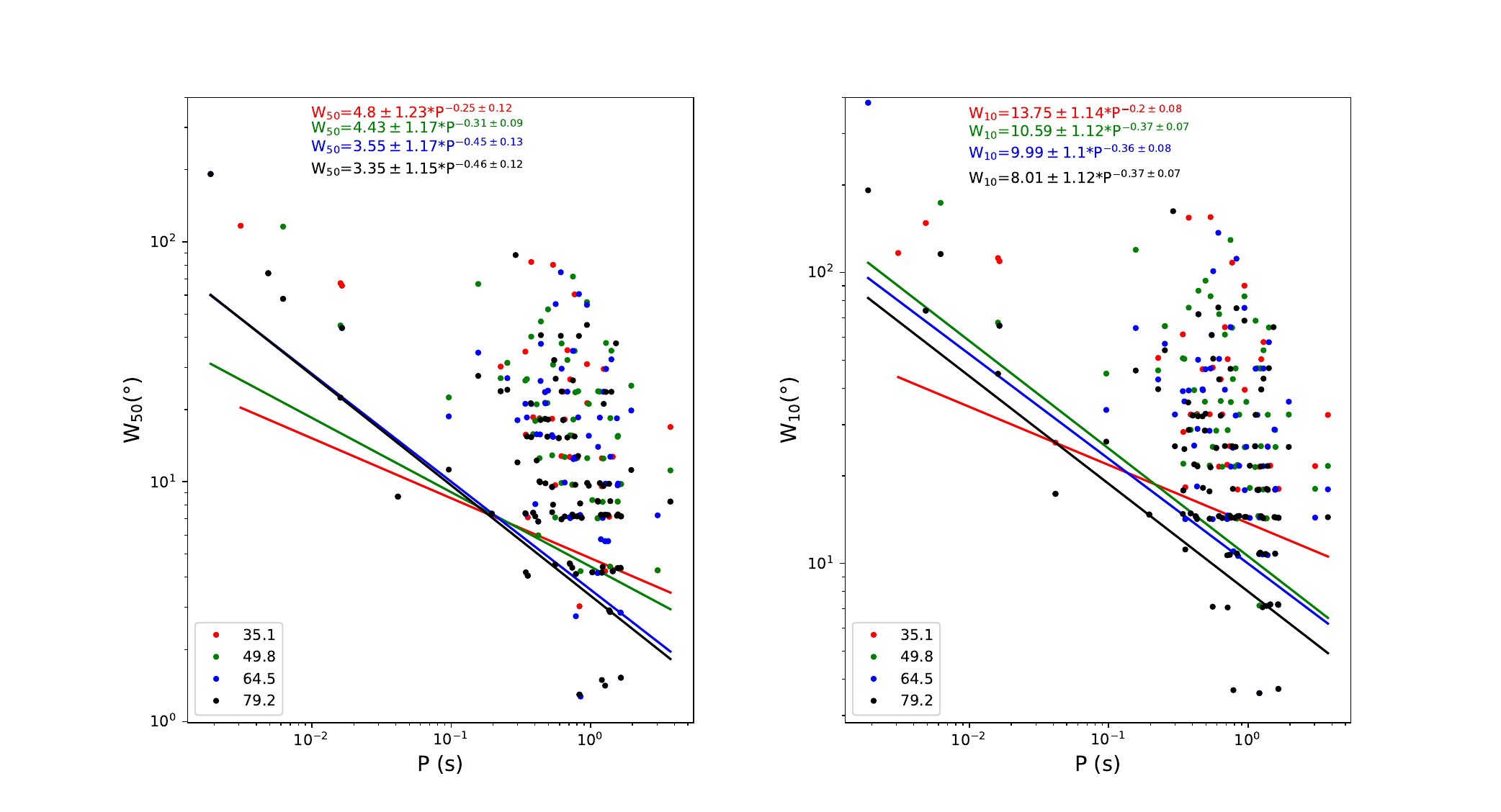}
\caption{The dots in figures shows the profile widths in degrees for respective pulse period in second. The colors represent the respective LWA frequency bands. The text in each subplot show the Lower boundary Line, which are plotted at solid lines for respective frequencies.}
\label{fig:LBL}
\end{figure*}

Previous studies have also shown the dependence of profile width in degrees on the period of the pulsar in seconds, and how this relation evolves with frequency, using a Lower boundary Line (LBL) of the form W$_{B}$$*$P$^{-d}$, where W$_{B}$ is the boundary width and P is the period of pulsar. Studies at higher frequencies have shown that the value of d is $\sim$ -0.5 \citep[see, e.g.,][]{skrzypczak,olszanskimitra}. We have applied a similar analysis to LWA pulsars, for W$_{50}$ and W$_{10}$ measurements, and estimated the parameter using quantile regression. For the case of W$_{50}$ we found the value of d $\sim$ $-0.4\pm0.1$, except for the case of our lowest frequency where it was $-0.25\pm0.12$. The value of d is consistent for three higher frequencies within error bars. The values at the two higher frequencies are similar to the value of -0.5 reported at higher frequencies. We also find the value of boundary width for these two frequencies to be $\sim$ $3.45\pm1.16$$^{\circ}$, consistent with the value of 2.37$^{\circ}$ at 327 MHz from \citet{skrzypczak}. The higher values are also consistent with previous results where lower frequencies have generally shown a higher boundary width. Similarly, for the case of W$_{10}$, we find an almost consistent value of d $\sim$ $-0.37\pm0.07$ for the three higher frequency bands, although slightly below the expected value of -0.5. However, in this case, we find the value of boundary widths consistent with the expectation of increasing with lower frequencies. If we assume the width increase to follow a power law dependence on frequency, we find the value for the power law index to be in the range of $-0.78$ to $-0.88$ with respect to the value at 327 MHz. The LBL analysis is summarized in Figure \ref{fig:LBL}





\subsection{Flux density, Spectra and Turnover}\label{sec:resultflux}
Table \ref{tab:fluxdensity} provides the mean flux densities of the LWA pulsars at one or more frequencies wherever possible. The flux density measurements along with other relevant quantities are only reported for 97 pulsars. We do not provide these measurements for three pulsars, since two are too weak for a meaningful measurement and one is only detected only in single pulses. For 29 of these pulsars, this is the first reported flux density measurement in this frequency range, while for 32 other pulsars, we present measurements at previously unreported frequencies. It is important to accurately determine flux densities in order to infer pulsar spectra as well as pulse luminosity and energetics. However, this is challenging due to two primary reasons, systematics related to the instruments, and observing time scales. The former can only be improved with extensive knowledge of telescope properties and how they change with several observing parameters, which evolve over time and need to be carefully monitored. In our case, we attribute 50$\%$ error to all our measurements due to these effects. Moreover, for pulsar observations at low frequencies flux densities can be underestimated due to pulse broadening caused by scattering. On the other hand, the observing time scales also matter. Pulsar flux densities can vary from a factor of two to an order of magnitude due to interstellar scintillation \citep[see, e.g.,][]{bell}. Given the frequency dependence, these effects are more prominent at low frequencies. In order to obtain accurate flux densities, observation longer than scintillation timescales of weeks to months is required. For our sample, these are obtained using profiles averaged over several epochs, separated by about three weeks, thus minimizing this effect.

Nevertheless, accurate flux density measurements have been taken in the past \citep[see, e.g.,][]{bilous2016,gnbcc350}. We make use of these measurements along with our results to obtain the spectra of these pulsars. 
We make use of the available data presented in other low-frequency pulsar catalogs \citep[][]{bilous2016,bilous2020,bondonneau,malofeev,atnf,gnbcc350,jankowski441psr,kravtsov,sanidaslotass,shrauner81.5MHz,zakharenko}, corresponding to our sample of pulsars. Using this data set, along with our spectral model and Equation \ref{eq:8}, as described in section \ref{sec:fluxspec}, we apply the least square fits to determine the spectral index and turnover frequency of the pulsar. 

We use all the flux density data available in the catalogs above except for \citet{gnbcc350}, where we discard data for the pulsars B0329+54, B0355+54, B1905+39, B1919+21, B2021+51, B2154+40, B2310+42, since the flux-density measurements from this catalog are not in alignment with measurements at nearby frequencies, assuming there is no abnormal change in spectral behavior. Using the rest of the data, we get a measurement of the spectral index and the curvature and use this value to determine the turnover frequency ($\nu_\mathrm{m}$). As described in section \ref{sec:fluxspec}, a reference frequency ($\nu_0$) is used for curve fitting, where $\nu_0$ is determined from the available data, and is given by equation \ref{eq:9}. For most pulsars, this value was 100 MHz, except for B0053+47, B0809+74, B1919+21, B1929+10, J0242+62, J0611+13, J1645+1012, and J1741+2758, in the 85-97 MHz range, for B0943+10, J1327+3423, J1741+2758, J1929+00 and J2227+30 $\sim$75 MHz and 45 MHz for J2208+4056. So, a uniform value of 100 MHz was used for $\nu_0$. 

As stated in section \ref{sec:fluxspec}, we also fit a linear power law to this data set. Based on the AIC value of the two fits, we find that a spectrum with low-frequency turnover is favored in about three-fourths of the pulsar sample. Figure \ref{fig:psrspectrum} shows the spectra of all pulsars in the sample, where red and blue points represent the measurements from this study and other catalogs respectively, and green and orange curves are the spectral fits to a quadratic (low-frequency turnover) and linear power law respectively. Since the spectral analysis was motivated to find the turnover at low frequencies, in 58 pulsars the range of frequencies used for spectral fitting limits to $\sim$ 1.4 GHz, however in 28 pulsars it goes up to 5 GHz, based on the values available from different collected catalogs presented here, as shown in Table \ref{tab:spectra}. 

\begin{figure*}[htbp!]
\includegraphics[width=\textwidth,scale=0.8]{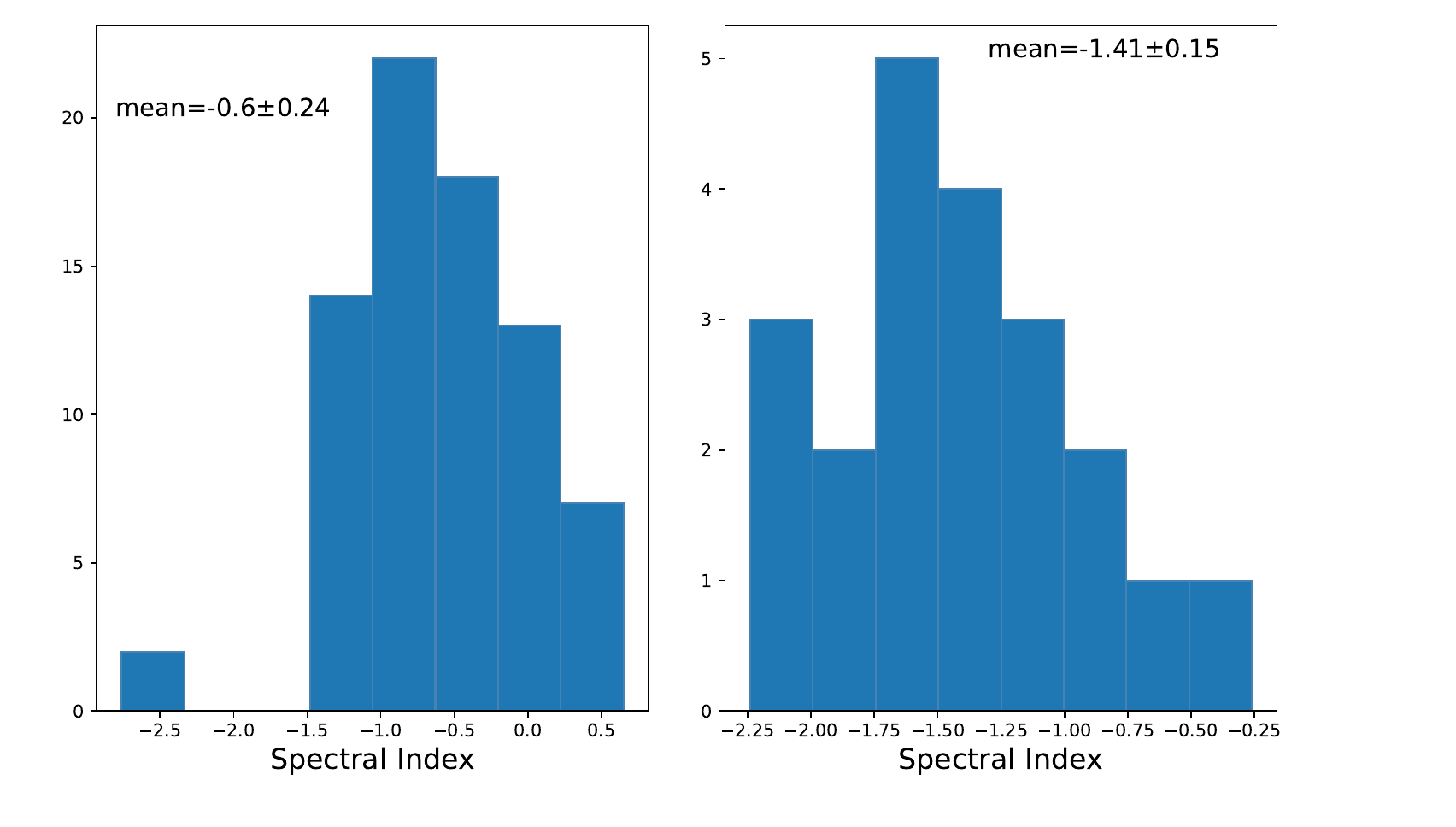}
\caption{Distribution of spectral indices for 97 LWA pulsars from best fit results.(\textit{Left}) Spectral index distribution at 100 MHz,
for pulsars which favor low-frequency turnover spectra. (\textit{Right}) Spectral index distribution from pulsars which favor linear spectra.}
\label{fig:spindex}
\end{figure*}

Most of our flux density measurements are in agreement with previous studies in the same frequency range, allowing for slight variability due to scintillation. We report a higher flux measurement than previous studies only for pulsar B0950+08 and B1633+24. 
The value of the spectral index and flux density at $\nu_0$, along with the curvature ($\beta$) from these fits, is given in Table \ref{tab:spectra}. This resulted in a mean spectral index of $-0.6\pm0.24$, from the low-frequency turnover spectra. The mean spectral index from the linear fit is $-1.41\pm0.15$, consistent with \citep[][]{Stovall}. Comparison of individual pulsar spectral indices with other low-frequency studies incorporating a low-frequency turnover in spectra \citep[][]{bilous2016,bilous2020} show reasonable agreement between the two with a smaller error in the current results by a factor of 2-4. Similarly, a comparison with values listed in the ATNF pulsar catalog, which used higher-frequency observations, shows a more negative value at higher frequencies. Comparing our mean linear spectral index with $\alpha=-1.41$, derived using observations near 1 GHz in \citet{bates}, and the weighted mean linear spectral index of $-1.6$ from \citet{jankowski441psr} from a wide range of frequencies, also indicates a steeper spectrum at higher frequencies. These results collectively provide evidence for a low-frequency turnover in spectra at around 100 MHz. Figure \ref{fig:spindex} shows the distribution of spectral index for all pulsars in our sample at 100 MHz.

Correlation between the spectral index and other pulsar properties such as period ($P$), spin-down rate ($\dot{P}$), characteristic age ($\tau$),  surface magnetic field ($B_\mathrm{s}$), and spin-down luminosity of the pulsar ($\dot{E}$) have been studied. \citet{Izvekova} found an increase in spectral index with period at low frequencies, whereas \citet{Maron} found no correlation with $P$, $\dot{P}$, or $\tau$. \citet{Lorimer1995} found a negative correlation between the spectral index and P, and a broad inverse correlation with $\tau$, and suggested that $\tau$ determines the value of the spectral index. On the other hand, \citet{Han} found the best correlation with $\dot{E}$. The correlation with $P$ and $\tau$ in \citet{jankowski441psr} is in agreement with \citet{Lorimer1995}, but they find the highest correlation to be with $\dot{E}$, $B_\mathrm{s}$ and $\dot{P}$. The Spearman correlation coefficient between the spectral index from the linear fit and these quantities for the LWA pulsar sample are listed in Table \ref{tab:spearman}. We find a positive correlation with $P$ and $\dot{P}$, while $\tau$ and $\dot{E}$ show a negative correlation. Although the correlation concerning $P$ is small, the p-value of 7$\%$ suggests that it is significant. Note that unlike \citet{Lorimer1995}, we find a positive correlation with $P$, not negative. In particular, we find the relation $\alpha=(-1.289\pm0.003)\log(P)-(0.112\pm0.001)$, whereas \citet{Izvekova} suggests $\alpha=0.7\log(P)+0.9$. The correlation with $\dot{E}$ agrees with \citet{jankowski441psr} and \citet{Han}, but both of these studies find a positive correlation instead of the negative correlation in our case. This might indicate some dependence of these correlations with observing frequency as well, but all these correlations are relatively weak and require further verification.

\begin{deluxetable}{ccc}[h!]
\tablecolumns{3}
\tabletypesize{\small}
\tablewidth{0pt}
\tablecaption{Spearman correlation coefficient between spectral index ($\alpha$) from linear fit and other quantities.\label{tab:spearman}}
\tablehead{
    \colhead{Correlation} & \colhead{r$_\mathrm{s}$} & \colhead{p-value} 
}

\startdata
$\alpha$-P & 0.19 & 0.07 \\ 
$\alpha$-$\dot{P}$ & 0.14 & 0.18 \\ 
$\alpha$-$\tau$ & -0.15 & 0.16 \\ 
$\alpha$-$\dot{E}$ & -0.15 & 0.15 \\ 
\enddata


\end{deluxetable}

In Table \ref{tab:spectra}, we also provide the calculated turnover frequency for all pulsars that favor low-frequency turnover spectra. Only two-thirds of the pulsar sample shows a constrained turnover frequency with errors smaller than its value, with a median value of 64$\pm$15 MHz. This result is lower than the value 130$\pm$80 MHz found by \citet{Izvekova}, and better constrained. Among the sample, 17 pulsars have a low-frequency turnover below 60 MHz, 31 pulsars have a turnover in the range of 60-160 MHz, and pulsars B0329+54, and B2021+51 show a turnover around 200 MHz. The poor constraints on turnover in the other 26 pulsars in our turnover spectra sample can be explained by a limited number of flux density measurements, bias in individual catalogs, as well as insufficient modeling of the spectra. Of particular interest are the spectra of six millisecond pulsars in our sample. Although they do indicate a preference for turnover, the measured value of $\beta$ is not significant suggesting no turnover in their spectra, in line with \citet{Kuzmin}. This suggests a difference between the two populations of normal and MSPs. However, given the small sample, it is difficult to establish this confidently. Moreover, recent studies of MSP spectra like \citet{rahulsharan} indicate a turnover at low-frequencies.

For the origin of the low-frequency turnovers in pulsar spectra, two possible scenarios exist: 1) origin via propagation through the interstellar medium, where pulsar emission gets preferentially absorbed at low frequencies by cold dense structures, leading to deviation from a simple power law; or 2) the features in spectra may be intrinsic to the pulsar emission mechanism, or may arise due to absorption processes close to the pulsar magnetosphere. \citet{Sieber} attributed low-frequency turnovers to two effects, namely synchrotron self-absorption near the emission region, and thermal absorption due to clouds along the LOS to the pulsar, whereas \citet{Malov} suggested free-free absorption. Verification of any such hypothesis would require some correlation of pulsar turnover frequency with positions in the sky or with dense structures along the LOS. We investigated this by looking for correlations between the turnover frequency, for LWA pulsars with well-constrained spectra, with Galactic coordinates as well as with DM (expected to be related to absorption along the LOS), but we found no such correlation. Among other possibilities, correlations with respect to previously detected structures in the Galaxy such as supernova remnants and molecular clouds need to be investigated. Similarly, investigations by \citet{jankowski441psr} for a sample of 40 pulsars revealed a partial correlation with spatial features, hence such a scenario is unlikely to fully explain the spectral turnover at low frequencies.

The other possibility requires that the turnover in spectra is closely tied to the pulsar emission mechanism. 
Our fitted turnover frequencies occur anywhere between 15-250 MHz (although most are around 100 MHz), rather than in a narrow range of frequencies. Given that pulsar emission is not well understood, it is possible that a wide range of turnover frequencies would be expected, if it depends on other pulsar properties such as spin period, characteristic age, etc. In particular \citet{Mfeev} found the relation $\nu_\mathrm{m}= 10^8P^{-0.5\pm0.2}$ for a sample of 39 pulsars and \citet{Izvekova} deduced the relation $\nu_\mathrm{m}= 100P^{-0.2}$ based on other correlations for a sample of 52 pulsars. We also looked for correlations between the turnover frequency and $P$, $\dot{P}$, $\tau$, and pulsar position in the $P$-$\dot{P}$ diagram, but no correlation was found. This may suggest that the turnover may be intrinsic but this conclusion needs tighter constraints.

Finally, the absence or presence of any correlation is dependent on other biases in observational data, namely, the influence of the ISM on low-frequency observations, statistical uncertainties, biases in individual catalog measurements, imperfect fitting of pulsar spectra, and the number of independent measurements available for spectral fitting. Some of these issues can be solved with observations with new and upcoming instruments and observational campaigns carefully designed to best determine the pulsar spectra and help to understand the associated emission. At the same time, the dependence of pulsar emission on observational frequency can not be ruled out, which could affect the observed flux density. 

\centerwidetable
\begin{deluxetable*}{cccccc}[!ht]
\tablecolumns{6}
\tabletypesize{\small}
\tablewidth{0pt}
\tablecaption{ New Timing Solutions for 5 pulsars\label{tab:partable}}
\tablehead{
    \colhead{PSR} & \colhead{J0242+62} & \colhead{J0611+30} & \colhead{J0815+4611} & \colhead{J1929+00} & \colhead{J2227+30} 
}
\startdata
Dataspan (yr) & 4.11 & 5.33 & 5.87 & 4.54 & 5.33 \\ 
Start Epoch (MJD) & 58373 & 57951 & 57718 & 57958 & 57950 \\ 
End Epoch (MJD) & 59876 & 59900 & 59865 & 59618 & 59897 \\ 
Timing Epoch (MJD) & 59124 & 58926 & 58792 & 58788 & 58923 \\ 
Number of TOAs & 636 & 262 & 188 & 344 & 692 \\ 
\hline
Right Ascension, $\alpha$ (J2000) & $02^{\mathrm{h}}\,42^{\mathrm{m}}\, 40\mathrm{s}.02(3)$ & $06^{\mathrm{h}}\,11^{\mathrm{m}}\,00\mathrm{s}.4(1)$ & $08^{\mathrm{h}}\,15^{\mathrm{m}}\,59\mathrm{s}.38(8)$ & $19^{\mathrm{h}}\,29^{\mathrm{m}}\,27\mathrm{s}.29(9)$ & $22^{\mathrm{h}}\,27^{\mathrm{m}}\,41\mathrm{s}.73(4)$ \\ 
Declination, $\delta$ (J2000) & $62\arcdeg\,56\arcmin\,41\farcs48(27)$ & $30\arcdeg\,13\arcmin\,58\arcsec(16)$ & $46\arcdeg\,11\arcmin\,51\arcsec(2)$ & $00\arcdeg\,31\arcmin\,23\arcsec(3)$ & $30\arcdeg\,38\arcmin\,21\farcs8(7)$ \\ 
Spin Frequency, $\nu$ (Hz) & 1.68993054312(5) & 0.70816896745(3) & 2.30286207906(5) & 0.85697025389(3) & 1.18707230283(1) \\ 
Spin Frequency Derivative, $\dot{\nu}$ (Hz s$^{-1}$)& -2.29418(7)$\times$10$^{-14}$ & -1.404(2)$\times$10$^{-15}$ & -1.9(1)$\times$10$^{-17}$ & -1.128(1)$\times$10$^{-15}$ & -1.6518(9)$\times$10$^{-15}$ \\ 
Spin Frequency Second Derivative, $\ddot{\nu}$ (Hz s$^{-2}$)& -7.3(8)$\times$10$^{-25}$ & ... & ... & ... & ... \\ 
Dispersion Measure (pc cm$^{-3}$) & 3.8202(1) & 45.27(2) & 11.2685(4) & 42.806(1) & 19.9611(2) \\ 
\enddata
\tablecomments{Timing parameters and data span for five LWA pulsars and the 1 $\sigma$ error bars are denoted with parenthesis on the last significant digit.}
\end{deluxetable*}

\subsection{Timing and Dispersion Measure}\label{sec:timeDM}
We also performed long-term timing of all our pulsars to improve existing parameters and the timing solution, as well as to accurately measure quantities such as DM and its variation over time. The following section describes the results of these efforts and their implications. 

\subsubsection{Long Term Timing and New Timing Solutions}
We perform the timing of all pulsars in our sample using the technique described in section \ref{sec:timing} and the data available in the LWA pulsar archive, to improve upon existing timing parameters, add new parameters to recently discovered pulsars, and measure DM time variations and frequency dependence. There was no notable change in pulsar RA or DEC, except for the case of J0051+0423, J0611+30, J1313+0931, J1929+00, and J2227+30, where the DEC changed by  60$\arcsec$, $-$180$\arcsec$, 60$\arcsec$, 300$\arcsec$, and 120$\arcsec$ respectively. 
Similarly, there was no significant change in the fractional pulse period of more than $10^{-7}$, except for J0242+62 and J0611+30, where we found fractional changes of 1.61 and $-$5.25 $\times 10^{-6}$, respectively. The value of mean DM from the previous measurements differed by $\sim 0.1\%$ for most pulsars, with the largest being 0.3 and 0.4 $\%$ for B0031-07 and B2045-16 respectively. Table \ref{tab:dmperiod} provides the period and DM value for LWA pulsars along with DM values from the LOFAR HBA study, and the ATNF values from high frequencies.

We provide new timing solutions for five pulsars, namely J0242+62, J0611+30, J0815+4611, J1929+00, and J2227+30, and the updated timing parameters are provided in Table \ref{tab:partable}. Most of these pulsars were recently discovered and did not have the derivative of spin frequency in the existing parameter files. All timing solutions were obtained using the JPL DE435 solar system ephemeris, and TT(BIPM2017) as the terrestrial time standard. 

\subsubsection{Pulsar DMs}
Thanks to observing at low frequencies where dispersion is largest, and our ability to measure DM by timing simultaneously at a range of frequencies with an octave of bandwidth, this survey provides more precise measurements of DM for these pulsars than any other observations. In particular, the median error on our measured DMs is 0.00018 pc cm$^{-3}$, compared to 0.0016 pc cm$^{-3}$ in the ATNF catalog for the same sample of pulsars. For the pulsars in our sample common with the LOFAR HBA census \citep[][]{bilous2016}, our median uncertainty is 0.00029 pc cm$^{-3}$ compared to 0.00053 pc cm$^{-3}$ for HBA. Even comparing with observations in the same frequency regime, median values for NenuFAR measurements of our sample \citep[][]{bondonneau} are 0.002 pc cm$^{-3}$, whereas our value for the matching sources is 0.00011 pc cm$^{-3}$. The median relative fractional change ($\delta DM/DM$) of our DM measurements with respect to the ATNF and HBA catalogs is $0.05\%$ and $0.04\%$, respectively. The ability to obtain precise DM measurements is of particular significance for the detection of low-frequency gravitational waves, which is greatly constrained by the ability to mitigate noise due to interstellar scattering \citep[][]{Cordshan}. Sensitive and precise measurements of DM at low frequencies can provide a path forward, by improving the models of propagation through the ISM \citep[][]{Lam}.

\subsubsection{DM variations}\label{sec:dmvar}
Since pulsars often have significant velocities and the ISM is an inhomogeneous medium by nature, the LOS to a given pulsar can sample quite different ISM conditions when compared at well-separated epochs. Systematic long-term monitoring campaigns of pulsars \citep[][]{keith} show that this can result in a slowly varying linear component of the DM, along with stochastic or periodic variations. Hence stochastic variations can be teased out of these measurements, considering that variations due to deterministic signals due to the LOS to the pulsar passing discrete structures can be corrected \citep[][]{Lammodel}. Additionally, DM variation can also arise due to stellar structures associated with pulsars, such as bow-shocks in the nebulae around young pulsars \citep[see, e.g.,][]{Ocker} which ionize nearby atomic gas. 

Inspecting the electron density fluctuation of pulsars \citet[][]{Armstrong} showed that the structure of density variation in the ISM has a steep red power spectrum leading to a prominent change in DM over long time scales. Hence, studying these variations elucidates the nature of turbulence in the ionized interstellar medium. Given the high precision in DM measurements achievable at low frequencies, long-term continuous monitoring of pulsars can be particularly useful in studying these effects and short-term DM variations. 

Using our long-term monitoring data, we have measured DM variation for all pulsars in our sample over timescales of several years, ranging up to a decade. Figure \ref{fig:dmtime} shows the variation in pulsar DM with respect to the fiducial values reported in Table \ref{tab:dmperiod} for LWA pulsars. The blue points indicate the value of excess DM (DMx) measured by pulsar timing, combining two or more frequency bands whenever possible, and the orange curve shows the angular separation with respect to the Sun (solar elongation) to indicate any correlation due to solar wind (SW) activity.

Some notable features include a strong and almost linear gradient of DM variation in pulsars B0149-16, B1237+25, and J1400-1431. The MSPs in the population, along with B0950+08, also indicate the presence of SW effects, while some other pulsars like B0329+54, B0450+55, B0628-28, B1133+16, B1642-03, etc., show slowly varying DM with some periodic or modulating effect.  Others like B0919+06 and B1822-09 show abnormal shapes with random DM variations. An interesting case is for the Guitar Nebula pulsar B2224+65\citep[][]{guitar}, which has a bow shock and is likely the origin of the observed short-term DM variations.

We use these data to study the variation in DM on long timescales. For this, we calculate the annual rate of change of DM by the relation below,

\begin{equation}\label{eq:14}
    \left|\frac{dDM}{dt}\right| = \left|\frac{\delta DM}{\delta MJD/365.25}\right|
\end{equation}

\noindent For discrete measurements, like comparing two census results, we use the above relation to compute the rate of change of DM, and the error is computed by the quadrature sum of the error of individual measurements divided by the time span in years. But for our DM time series values, we compute this directly by fitting a linear slope to the piecewise-binned data, accounting for the error in individual measurements. 
For a given binning size, we generate a distribution of DM slopes, and then after rejecting 3 sigma outliers, we use the mean and standard deviation of this distribution as the value of DM slope and error respectively, associated with that bin size. The process was repeated for different binning sizes from 6 months to 7 years, in 6-month increments. We found that the value of DM slope was consistent across pulsars for binning size of $\sim$1-5 years. Hence we choose a binning size of 3 years for our final calculations.
Since here we are only interested in ISM-related variations, we discard measurements within $<30^{\circ}$ of solar elongation for the slope fitting, to avoid misinterpretation due to solar effects. Both of these methods have inherent associated bias. In the former case, since the measurements are only available at two discrete points, and there is no information about the slow evolution of DM, the rate of change can be largely miscalculated depending on the shape of the DM time series. Furthermore, if one or both of these measurements have a large associated error, it will undermine the resulting value. While the DM time-series fitting method is not biased by these effects, since a large number of measurements are available and some information about the slow DM variations is known, a linear fit to the data may be an overestimation since, as shown before, for different pulsars in the sample deviation from simple slowly-varying or linear trends do exist and need to be accounted for. Nevertheless, the latter method has more leverage on DM variations and it is better suited for such analysis.
\begin{figure*}[htbp!]
    \centering
    \includegraphics[width=\textwidth]{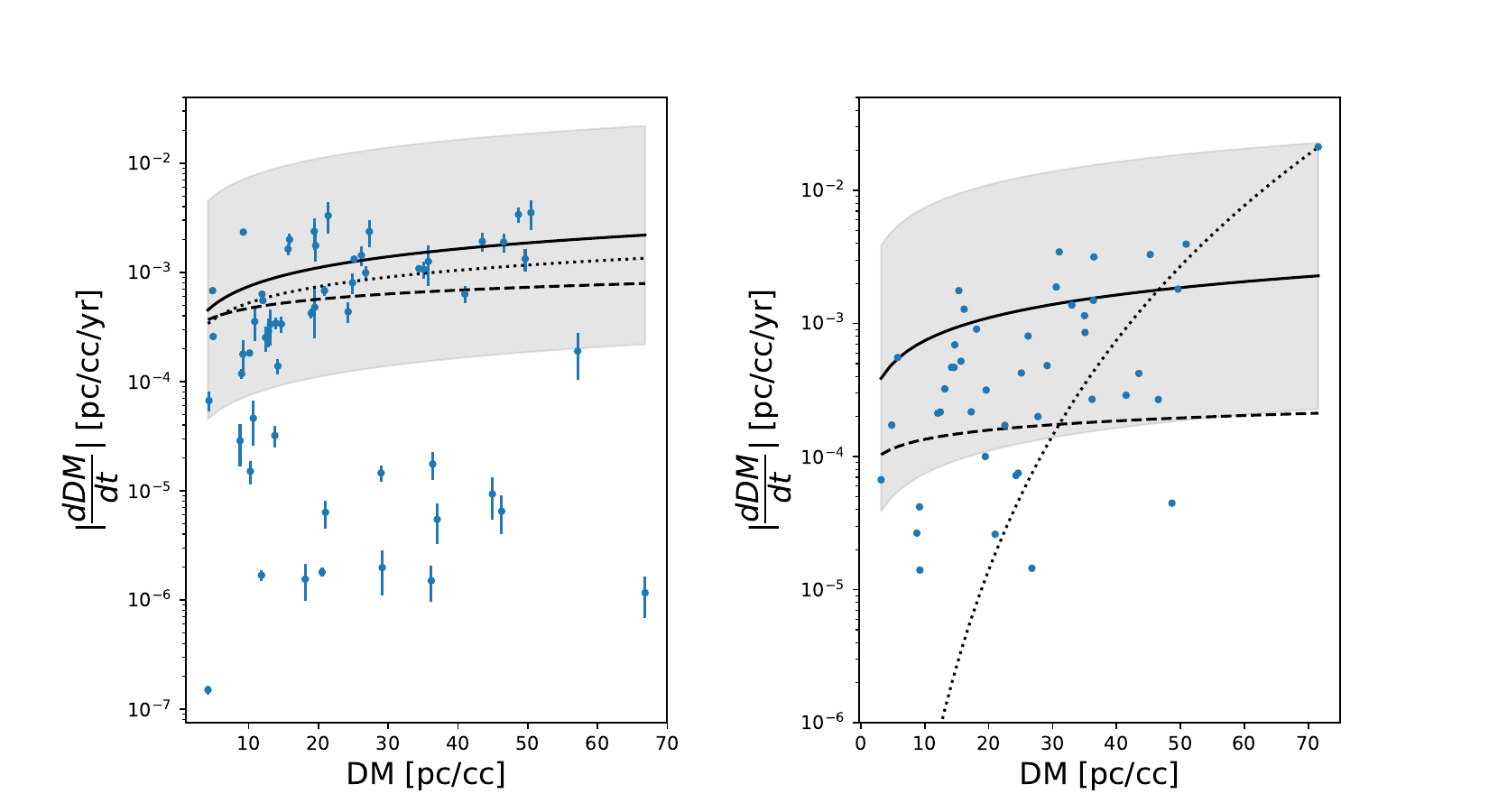}
    \caption{Dependence of DM variation on mean pulsar DMs. In both panels, blue points indicate the measured DM rate with errorbars, and the dashed, and dotted black curves indicate the best-fit results with and without using the errors as weights, and the solid black curve shows the relation from \citet{Hobbs} with an order of magnitude scatter in the shaded grey region. (\textit{Left}) Shows the results based on timing and DM time series data, (\textit{Right}) shows the results based on the difference between the LWA and HBA catalogs.}
    \label{fig:dmrate}
\end{figure*}

We compare our results with similar studies done in the past. 
In particular, \citet{Backer} proposed a wedge model for pulsar DM variations based on a sample of 13 pulsars in the range 2-200 pc cm$^{-3}$, where they found the relation $\left|\frac{dDM}{dt}\right| \sim \sqrt{DM}$. \citep[][]{Hobbs} measured the DM variation for a similar sample of about 100 pulsars, in the DM range 3-600 pc cm$^{-3}$, using the long-term timing of 6 to 34 years of data in some pulsars at frequencies 400-1400 MHz and deduced the relation $\left|\frac{dDM}{dt}\right| = 0.0002\times DM^{0.57\pm0.09}_{\mathrm{pc~cm}^{-3}}$ pc cm$^{-3}$ yr$^{-1}$. Both studies found an order-of-magnitude scatter around the fit. This spread could be partially due to differences in pulsar transverse velocities and inhomogeneities in the ISM on different scales as well as in different directions. \citet{bilous2016} performed an analysis based on the difference between their HBA census and ATNF values and found the relation $\left|\log\frac{dDM}{dt}\right| = \log A+ B \log DM$, with B=-0.1$\pm$0.1 and A = 0.03, indicating that fits based on discrete measurements are not that useful.

We applied both methods to measure the DM rate, using the slope of our DM time series and comparing DM census measurements with the HBA values as shown in Table \ref{tab:dmperiod}. Only measurements where the error on $\frac{dDM}{dt}$ was smaller than a factor of 2 were used, leaving 61 pulsars in the analysis. Figure \ref{fig:dmrate} shows the result from these analyses. Based on the time series data, we find the relation,
\begin{equation}
   \left|\frac{dDM}{dt}\right| = \tilde{A}\times DM^{\tilde{B}}
\end{equation}

\noindent where $\tilde{A}$ = (0.0002$\pm$0.0003) and (0.0002$\pm$0.0001), and $\tilde{B}$ = (0.27$\pm$0.34) and (0.5$\pm$0.26) for the unweighted and weighted fits, respectively. Qualitatively these results are similar to that of \citet{Hobbs}, albeit not very constraining.
However, we see two distributions of DM slopes in the left panel of Figure \ref{fig:dmrate}, above and below the value of $\sim$10$^{-4.5}$. We look for a correlation between these pulsars and their position on the sky but find no evidence to suggest its origin relating to the relative position of the pulsars with respect to the galactic plane or other discrete structures.
Similarly, evaluation of the census results yielded $\tilde{A}$ = (0.00008$\pm$0.00006), and $\tilde{B}$ = (0.23$\pm$0.27), which shows a weaker power-law dependence than in \citet{bilous2016}.

\begin{figure}
\centering
\includegraphics[width=0.45\textwidth]{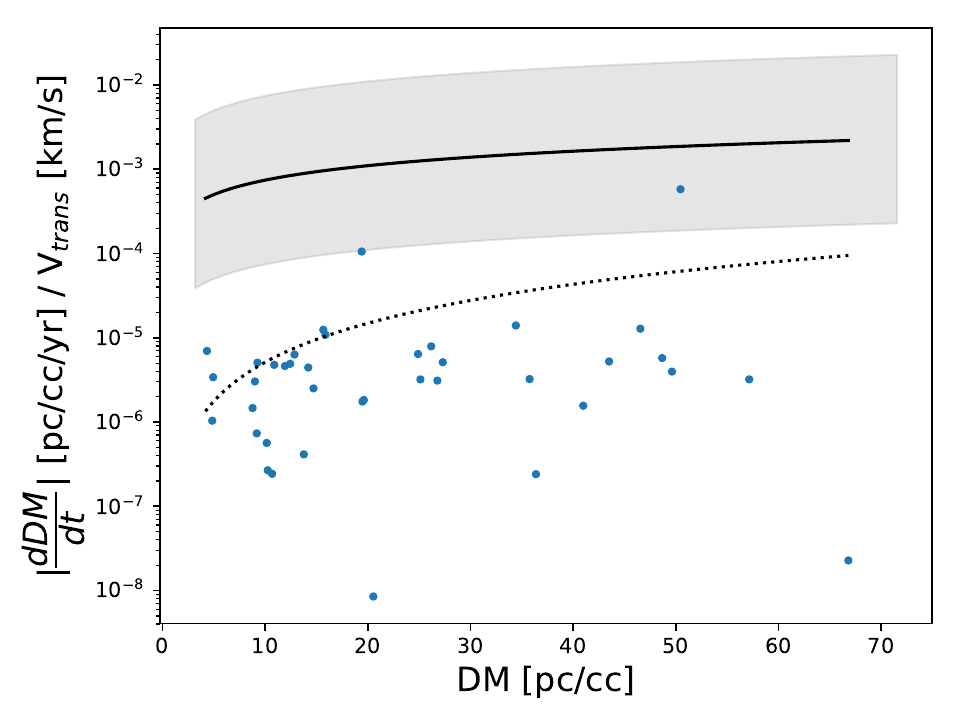}
\caption{Plot of the rate of change of DM, scaled by pulsar transverse velocity, with respect to pulsar DM. The dotted and solid curves and the grey region have the same meanings as in Figure \ref{fig:dmrate}.}
\label{fig:dmratewithvel}
\end{figure}

Additionally, we also evaluate the results from time-series data, along with transverse velocity estimates from ATNF to account for inherent bias due to pulsar transverse velocities. Since these do not have error estimates, only non-weighted analysis was applied. We found the relation

\begin{equation}
\begin{split}
    \left|\frac{dDM}{dt}\right|/V_\mathrm{Trans} = (1.5 \times 10^{-7}\pm7.7\times 10^{-7}) \\
    \times DM^{(1.53\pm1.34)}
\end{split}
\end{equation}

\noindent and the best-fit results are shown in Figure \ref{fig:dmratewithvel}. These results suggest consistency in DM variations at low and high frequencies, but it is hard to draw a definite conclusion since bias due to factors discussed above as well as due to profile evolution with frequency has not been properly accounted for.

\subsubsection{Solar Wind}
26 pulsars in our sample have a minimum solar elongation angle below 10$^{\circ}$, while another six passes within 15$^{\circ}$, where solar wind effects are most prominent and can clearly be detected in pulsar DM variations \citep[see, e.g.,][]{Kumar}. Among these, there are five MSPs, which have higher precision of DM measurement, while the others are normal pulsars.  We only detect clear variation due to the SW in four MSPs and one normal pulsar (B0950+08), while J0051+0423 also shows some evidence. Since solar wind effects are small and of the order of 10$^{-3}$ to 10$^{-4}$ pc cm$^{-3}$ \citep[][]{Tiburzi}, and the measurement error on DM variations obtained via pulsar timing depends on pulse width, shape and S/N of detection of the pulsar, the SW contribution is harder to detect in faint normal pulsars. A possibility remains that our sample is inadequate because pulsars are only close to the Sun for about 6-8 weeks per year. Nevertheless, we carried out dedicated campaigns for some normal pulsars, including B0525+21, B0823+26, B0834+06, B0919+06, and B043+10, which are all very bright in our frequency band. No clear detections of SW contributions were found. While B0943+10 and B0823+26 have reported mode-changing behavior which may limit our ability to make these detections, others do not indicate such behavior. Given the small amplitude of typical SW contributions, detections in these cases were expected assuming no large fluctuation due to ISM variations. However, Figure \ref{fig:dmtime} shows that is indeed not the case for these pulsars and these non-detections further support the turbulent nature of ISM along these LOS.
For pulsars where we do detect the SW effect, it is also possible to detect the change in its amplitude annually due to the solar cycle. These measurements provide an additional way to study the SW and solar activity on long time scales.

\subsubsection{Frequency- and time-dependent scattering and Pulsar B2217+47}
As stated in section \ref{sec:dmvar}, \citet{Armstrong} showed that the electron density power spectrum in the turbulent ISM has a steep spectrum, with more power at large scales. Since pulsars have finite velocities in the range of 100 km s$^{-1}$, long-term pulsar monitoring campaigns like PTAs, which require precision pulsar timing to detect gravitational waves, will be affected, and indeed noise-limited by this effect. Small-scale density fluctuation in the ISM can lead to chromaticity in DMs due to multipath propagation caused by scattering in the medium. While \citet{janssen} has proposed a way to mitigate these effects in PTA data by independently monitoring them at low frequencies and constructing DM measurements to correct for the variability at high frequencies, simulations suggest that low frequencies are unlikely to be useful for direct correction \citep{Cordesfdm}. Nonetheless, long-term low-frequency timing experiments provide the data necessary to study these effects and characterize them. These effects could not be studied until more recently after the advent of sensitive low-frequency instruments that began to monitor pulsars over long time scales. \citet{Donner} reported the first detection of this effect in B2217+47 and attributed it to extreme scattering events (``ESE") caused by compact lensing structures in the ISM. This was later challenged by \citet{Lam}, who showed it to be consistent with ISM turbulence and the presence of an enhanced scattering screen near the two ends along the LOS, and argued for the value of low-frequency observations to study these effects. More recently, \citet{Kaur} detected a frequency-dependent DM in J2241−5236 over a wide range of frequencies. 

\begin{figure}[htbp!]
\centering
\includegraphics[width=\linewidth]{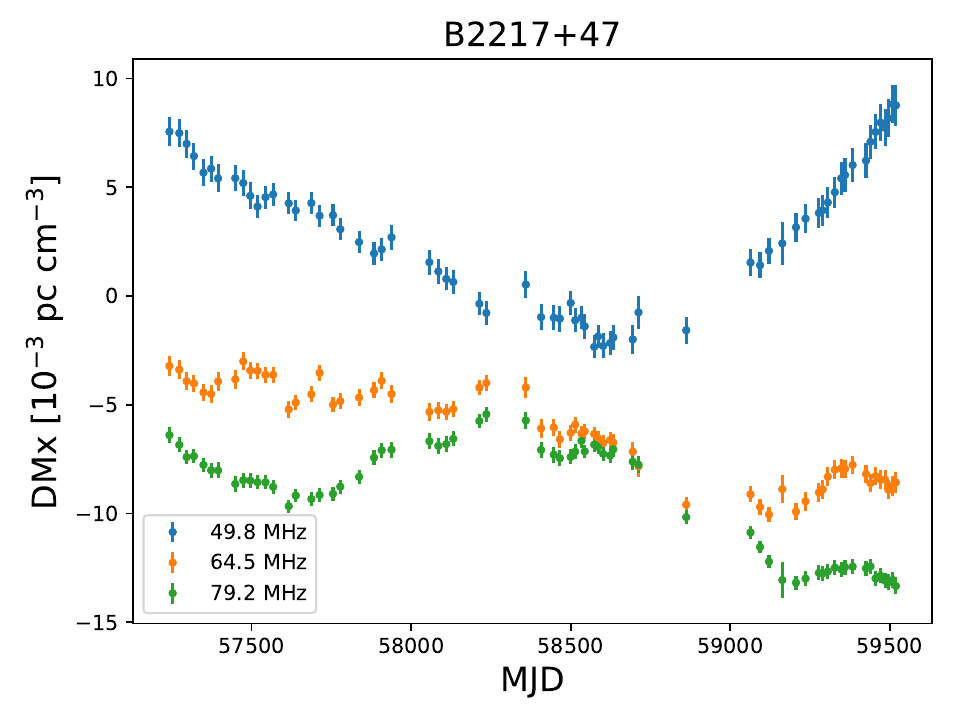}
\caption{Detection of frequency-dependent time-variable DM in pulsar B2217+47. The figure shows the measured DM excess in each subband, as indicated by the different labels.}
\label{fig:B2217}
\end{figure}

With our ability to measure DMs precisely at the low frequencies best suited to detect these effects, and continuous monitoring of a set of pulsars, LWA observations are well-placed to detect these frequency-dependent time-variable scattering events. As stated in section \ref{sec:timing}, we can time our pulsars independently in each sub-band of observation and use this to measure the DM from a narrow band of frequencies. We used this method to look for evidence of variable frequency-dependent DMs in our data. In particular, we report the detection of frequency-dependent DM in pulsar B2217+47, which varies with time. Figure \ref{fig:B2217} shows the DM excess in each subband, which changes with frequency and time. More importantly, the separation between each narrow band measurement evolves with time, similar to \citet{Donner}. This is the strongest detection of frequency-dependent time variable DM observed in pulsars, with the difference in DM measurements being about 5 times larger than reported in \citet{Donner}. Other than this, we also see evidence of frequency-dependent behavior in B0823+26, B1541+09, B1839+56, and B1929+10. \citet{Bansal} previously studied the scattering properties of B1839+56 for a smaller span of LWA data and found the index of scattering to be 2.70$\pm$0.16, deviating from a Kolmogorov spectrum. We leave a more detailed analysis of scattering and frequency-dependent behavior to future work.
\begin{figure*}[htbp!]
    \centering
    \includegraphics[width=\textwidth,height=21cm]{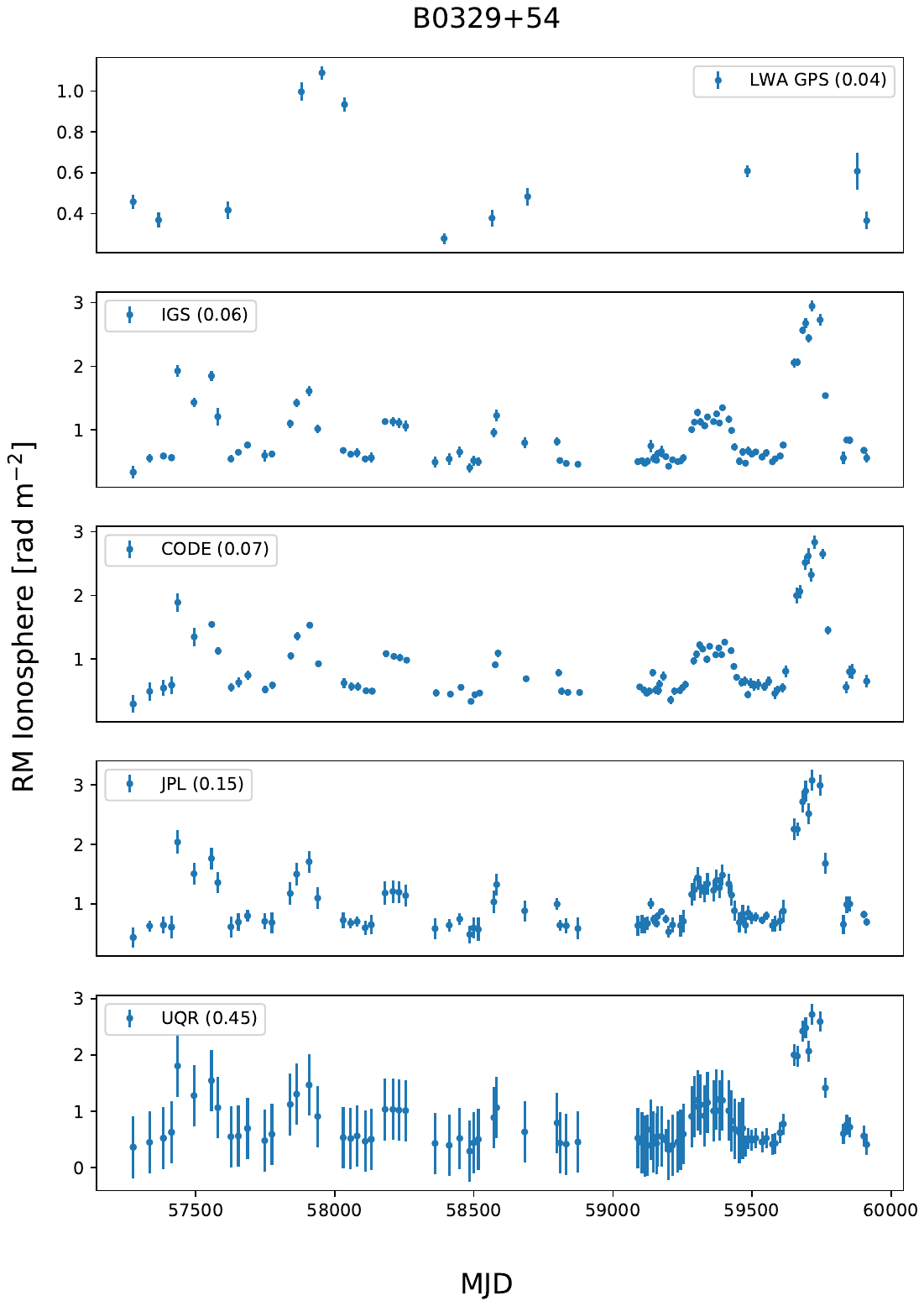}
    \caption{Calculated ionospheric RM corrections for pulsar B0329+54 on multiple epochs, obtained using different TEC maps and local LWA GPS measurements when available. The legend in each map shows which TEC measurement was used for calculating the ionospheric RM contribution and the value in parentheses is the corresponding value of mean error on all available measurements.}
    \label{fig:rmcorrections}
\end{figure*}

\subsection{Detecting Pulsar Polarization and its Variation}
Given the dependence of the amount of rotation of polarized signal due to the intervening ISM on observing wavelength, low frequencies are particularly well suited to perform sensitive searches and detect polarization and RM in pulsars. We searched about 50 pulsars in our sample to detect any polarized emission from these objects, while others were rejected as they were not sufficiently bright to perform any sensible search for polarization at the expected levels for pulsars. Among these objects, we detected the presence of linear and circular polarization in 27 pulsars.
Of these, 26 were reported previously at low or high frequencies, and one is a new detection. The method used to find the optimal RM and detect polarization is described in section \ref{sec:rm}. We used pulsar data at multiple frequencies and multiple epochs to detect any variation in the ISM magnetic field using the measurements.

\subsection{Choice of Ionospheric correction}
As described in section \ref{sec:rm}, the final value of pulsar RM due to Faraday rotation by ISM depends on the correction for Faraday rotation caused by Earth's ionosphere. More importantly, the ionospheric corrections (RM$_\mathrm{ion}$) depend on two quantities: the global magnetic field models (GMF) and the TEC contribution from the ionosphere, which together provide the correction term given by Equation \ref{eq:10}. While IGRF provides accurate GMF measurements, a few choices are available for TEC measurements. We can either choose from global TEC models such CODE (Center for Orbital Determination in Europe), IGS\footnote{\url{https://igs.org/wg/ionosphere/}}, JPL\footnote{\url{https://iono.jpl.nasa.gov/}} or UQR, or use local measurements if available. In our case, local measurements from a GPS receiver co-located with LWA (LWA GPS) with 10-sec sampling are available,  but were not continually operational, limiting us to provide corrections only for a handful of epochs. Measurements from global models can be used but there are inherent issues. First, these measurements are provided by interpolating between a discrete set of receivers, not accounting for all the spatial variations of the ionosphere. Also, the available data products vary between 15 min to 2 hours in duration, comparable to our observing time, providing insufficient measurements to get accurate corrections. Previous studies using LWA have shown that RM$_\mathrm{ion}$ shows strong diurnal variation, as expected, and can further vary on short time scales due to solar X-ray/EUV bursts and other phenomena. \citet[][]{Malins} demonstrated that GPS TEC measurements local to LWA can provide a suitable correction for the ionosphere. We compare the value of ionospheric corrections provided by these models against the available LWA GPS data to identify the best possible global model for studying long-term variation. Figure \ref{fig:rmcorrections} shows the result of this for pulsar B0329+54, where ionospheric corrections are plotted along with error bars for respective models over time. The legend in each panel indicates the TEC model and the mean of the error on all ionospheric correction measurements available for that model. After LWA GPS corrections, IGS performs the best and is used throughout this article for studying long-term variations. In fact, IGS TEC values interpolated to LWA1's location are dominated by a receiver located at Pie Town, NM, within 50 mi of LWA1, and hence the IGS values are equivalent to having a GPS receiver local to LWA1, but with a poorer time resolution of 2 hrs. In addition to the IGS values we also report on LWA GPS measurements wherever available.

\subsubsection{Measurements of Pulsar RM and ISM Magnetic Field}

For the 27 pulsars with polarization detections, we attempt to detect pulsar RM in all LWA frequency bands and at multiple epochs. Among the 27, 26 have reported values in the ATNF catalog, 11 overlap with low-frequency measurements from LOFAR \citep[][]{Sobey}, and 15 overlap with previous results using the LWA \citep[][]{Dike}. The resulting RM values from these detections are reported in Table \ref{tab:rm}, along with measured uncertainties, at one or more frequencies. We also provide the ionospheric RM correction using LWA GPS and IGS, as well as the LWA DM measured at that epoch. RM values from the ATNF catalog along with IGS model-corrected RM for the LWA data are also provided. Non-detection of pulsar RMs in all frequency bands is a consequence of multiple contributing factors, including the loss in sensitivity at low frequencies, as well as the lower brightness of pulsars at higher frequencies. The typical uncertainty on RM measurements is 0.01 rad m$^{-2}$, whereas ionospheric correction RM values have a typical uncertainty of 0.08 rad m$^{-2}$, for a combined uncertainty of the order of 0.1 rad m$^{-2}$.

The LWA values in Table \ref{tab:rm}, generally agree with high-frequency values from ATNF, with a median difference of $\pm$1 rad m$^{-2}$, where the median uncertainty in ATNF values is 0.1 rad m$^{-2}$. The largest deviation was for pulsar B0919+06, of 4.72 rad m$^{-2}$. Similarly, comparing with LOFAR results in \citet{Sobey}, the measured values of RM shows a mean difference of $-1.2$ rad m$^{-2}$, ranging from 0.02 rad m$^{-2}$ for B1839+56 to $-$2.45 rad m$^{-2}$ for B0823+26, but the difference in RM corrections are of the same order, leaving typical differences of $\pm$0.1 rad m$^{-2}$, which is comparable to the uncertainty in the measurements. The value of RM corrections for \citet{Sobey} is typically a factor of two larger than LWA corrections. Since the corrected values between LWA and LOFAR measurements are consistent with each other, it supports the idea that the ionospheric correction method used in \citet{Sobey} will overestimate the correction factor for LWA measurements. Since \citet{Dike} used the same method as \citet{Sobey} for correcting LWA pulsar RMs, we argue that our corrected RM measurements are more accurate, even though the measured RMs between this work and \citet{Dike} are consistent. 

\begin{figure*}[htbp!]
    \centering
    \includegraphics[width=0.9\textwidth,height=10cm]{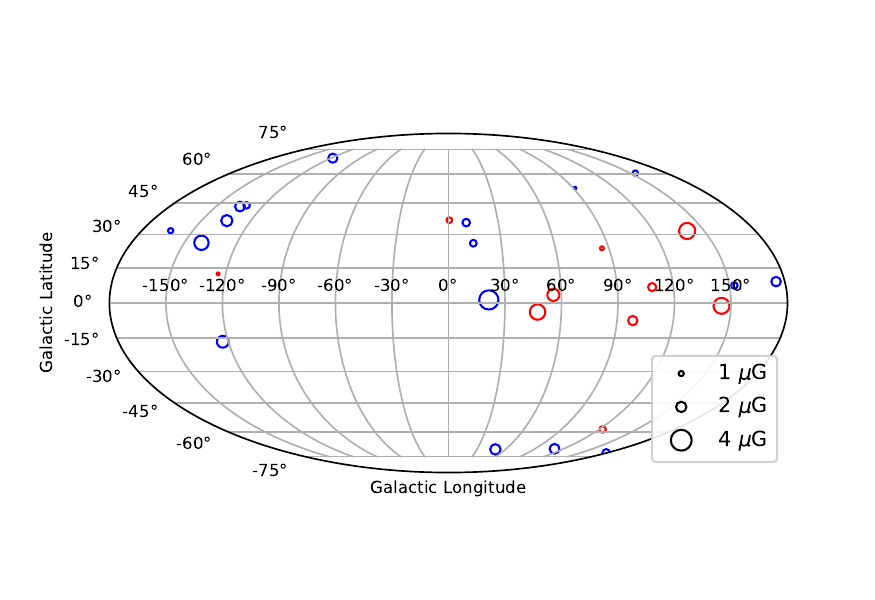}
    \caption{Calculated value of average Galactic magnetic field for all pulsars in our sample. The size of the circle corresponds to the absolute value of the magnetic field, and red and blue correspond to positive and negative respectively. The values range between $-$3 $\mu$G for B0809+74 and 4.26 $\mu$G for B1822-09.}
    \label{fig:magnetic}
\end{figure*}

With ionosphere-corrected RMs, we can calculate the parallel component of the average magnetic field of the ISM along the LOS to the pulsar, using the instantaneous value of the pulsar dispersion measure following Equation \ref{eq:5}. This value of the average ISM magnetic field ($B_{\parallel}$) using the RM corrected with the IGS model is also given in the last column of Table \ref{tab:rm}, along with measurement errors. Using this data, we plot the average value of $B_{\parallel}$ for all pulsars in our sample in Galactic coordinates which is shown in Figure \ref{fig:magnetic}. The map is qualitatively similar to those obtained in \citet{Sobey} using extragalactic sources and pulsar RMs, with most of our negative values near the  Galactic plane. The distances to these pulsars vary between 0.19\,kpc and 3.85\,kpc, per the ATNF catalog, with 20 pulsars within 1.25\,kpc and no correlation between B$_{\parallel}$ and distance. The region within $\sim$1\,kpc of the sun is thought to be between the spiral arms of the Galaxy and the magnetic field is mostly clockwise in three of the four galactic quadrants \citep[][]{hanmag}. This is also evident in our measurements as most of the negative magnetic field measurements are outside this distance range.  


\subsection{Polarization Profiles}
By correcting for the linear polarization using the measured RM values, we obtained 72 polarization profiles for 27 pulsars at multiple frequencies. Figure \ref{fig:psrprofrm} shows these polarization profiles at the epochs reference in Table \ref{tab:rm}. 

The bottom panel of each polarization plot shows the total intensity I, linearly polarized intensity P ($\sqrt{\mathrm{Q}^2+\mathrm{U}^2}$), circularly polarized intensity V, in black, orange, and cyan respectively, as a function of pulse phase in arbitrary units. The top panel in each case shows the change in PA for a fractional linear polarization of more than 30$\%$ (P/I $>0.3$).

Comparing our polarization profiles with those at higher frequencies in the EPN database shows that, in general, the fractional linear polarization for LWA pulsars at low frequencies is higher.
However, we see depolarization in the case of B0809+74 and B1839+56, which could be partly due to the large FWHM of the LWA beam (of the order of 2$^{\circ}$), and the low elevation of observations of B0809+74 at 49$^{\circ}$. Additionally, B0950+08 and B1929+10 both show high fractional linear polarization, close to 100$\%$. Both of these pulsars are nearby objects with DMs of 2.97030 pc cm$^{-3}$ and 3.18509 pc cm$^{-3}$, respectively, hence the detection of such strong linear polarization is not unusual. On the other hand, B2224+65 also shows $>60\%$ linear polarization but has a much larger DM, and the estimated distance is three times higher. However, this pulsar is associated with a bow-shock pulsar-wind nebula \citep[][]{guitar}, which could contribute to the polarization and could also explain the observed noise in the polarization profile of this pulsar.

\subsubsection{New Detection of Highly Circularly Polarized Emission in PSR J0051+0423}
Polarization searches have been previously conducted for J0051+0423, but no significant detections were made. In particular, \citet{kerr} made a marginal detection of polarization at 1400 MHz with an RM of $-$30.5 rad m$^{-2}$, whereas \citet{cng} did not detect any polarization from this source in their 400-800 MHz survey. We have a strong detection of polarization from this source in our brute-force search with an RM value of $-$4.49$\pm$0.02 rad m$^{-2}$ at multiple epochs. The polarized profile shown in Figure \ref{fig:J0051} shows an unusually high degree of circular polarization, $>50\%$, with almost no linear polarization. This is the highest degree of circular polarization among our entire sample. Since we do not detect such a large value in any other pulsar and the instrumental polarization of LWA is quite low, around $5\%$ for circular polarization, it is unlikely to be due to any instrumental effect. Moreover, we detect the same feature on multiple epochs, making it unlikely to be spurious in nature. The presence of strong circular polarization with almost no linear polarization also points towards the presence of a strong magnetoionic medium which can depolarize linear polarization while circular polarization remains unchanged. 

\begin{figure*}[htbp!]
    \centering
    \includegraphics[width=\textwidth,height=21cm]{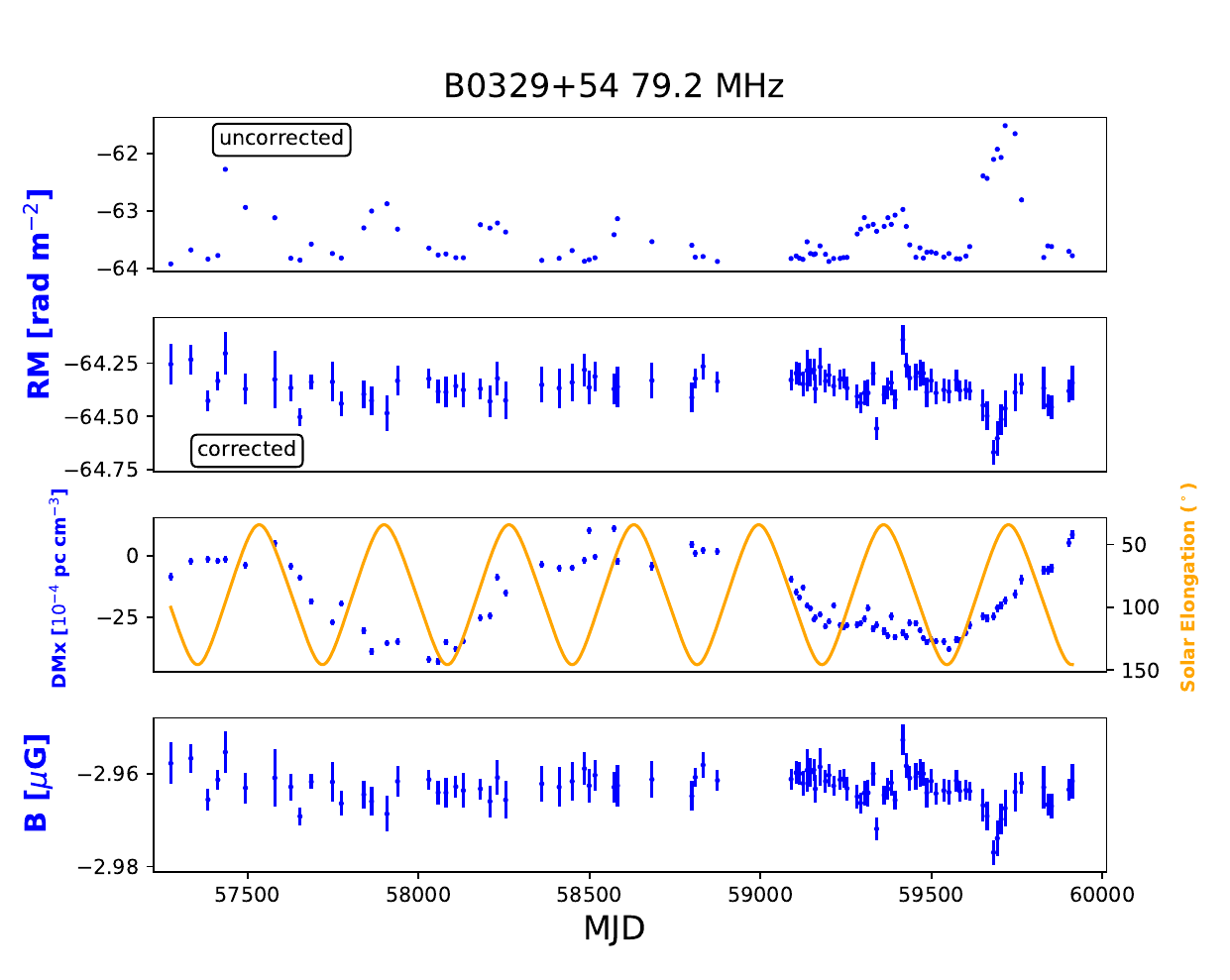}
    \caption{RM values for pulsar B0329+54 at 79.2 MHz at multiple epochs. The first two subplots show the measured RM from the polarization profiles, and after applying the ionospheric correction, respectively. The last two plots show the variation of dispersion measure obtained via pulsar timing and the calculated value of the average ISM magnetic field using Equation \ref{eq:5}.}
    \label{fig:composite}
\end{figure*}

\subsubsection{RM Variations and Fluctuations in the Average Magnetic Field of the ISM}
As pointed out earlier, using our long-term data set we can study the variation of pulsar RM over time in order to investigate the presence of any detectable variations in the ISM magnetic field. In general, pulsar RMs can be affected by the presence of supernova remnants or HII regions along the LOS \citep[see e.g.,][]{Mitra2003}. Using the same arguments, it is useful to study the variation of RM in pulsars to detect the presence of any discrete structure in the ISM. Since the LOS to the observer changes over time due to the high velocity of pulsars, studying RM variations allows us to sample ISM variations on very small scales. Assuming a typical pulsar velocity of 100 km s$^{-1}$ from \citet{Hobbsvel}, this amounts to variations on a scale of 2-200 AU, related to our observing cadence and total observing span. By studying the variation of RM in PTA data of MSPs observed at $\sim$1\,GHz, \citet{Wahl} found variation in the average magnetic field, in the form of linear trends as well as some periodicity, in five pulsars. 

Typical uncertainties in LWA RM measurements are of the order of 0.01 rad m$^{-2}$, with ionospheric correction uncertain at the level of 0.08 rad m$^{-2}$, hence the final uncertainty on corrected RMs is dominated by the latter. Figure \ref{fig:composite} shows the variation in RM for B0329+54, which is typical for our pulsar sample.  The first two panels indicate the values of measured RM before and after ionospheric correction, the third panel shows the measured DM variation in the pulsar using timing, and the last panel shows the calculated magnetic field over time using Equation \ref{eq:5}. It is worth repeating that although we see a large change in pulsar RM, of the order of 1-2 rad m$^{-2}$ in the top panel of Figure \ref{fig:composite}, the overall error on RM measurements is only of the order of 0.1 rad m$^{-2}$, and the large fluctuation seen in the top panel result from fluctuation in the ionospheric corrections. Since all our observations are planned to occur at transit, the observing UTC (universal coordinate time) changes with the epoch. Additionally, for long observing campaigns solar activity changes on timescales of months to years. Both of these factors will affect the RM contribution from the ionosphere and hence are the source of the large fluctuation in the first panel of Figure \ref{fig:composite}. However, we can clearly see that once these corrections are applied, the corrected RM value becomes almost flat with fluctuation of the order of 0.1 rad m$^{-2}$, as is evident from panel two of Figure \ref{fig:composite}. For all our pulsars where we have measurements at 10 or more epochs, Figure \ref{fig:rmtime} shows the variation in the ionosphere-corrected RM. Measurements from all frequencies are overplotted and are indicated by the legend at the bottom. Two things can be deduced from this plot: 1) in the case of completely overlapping measurements, only values at 79.2 MHz are visible; 2) there is intrinsic variation between the measurements at different frequencies. This is to be expected since RM measurements are derived using an S/N maximization technique, and lower frequencies generally have lower signal-to-noise as well as more RFI contamination.




Similarly, Figure \ref{fig:Btime} shows the variation of the average magnetic field calculated using Equation \ref{eq:5} using RM and DM measurements over time. 



\begin{figure}[htbp!]
    \includegraphics[width=0.45\textwidth]{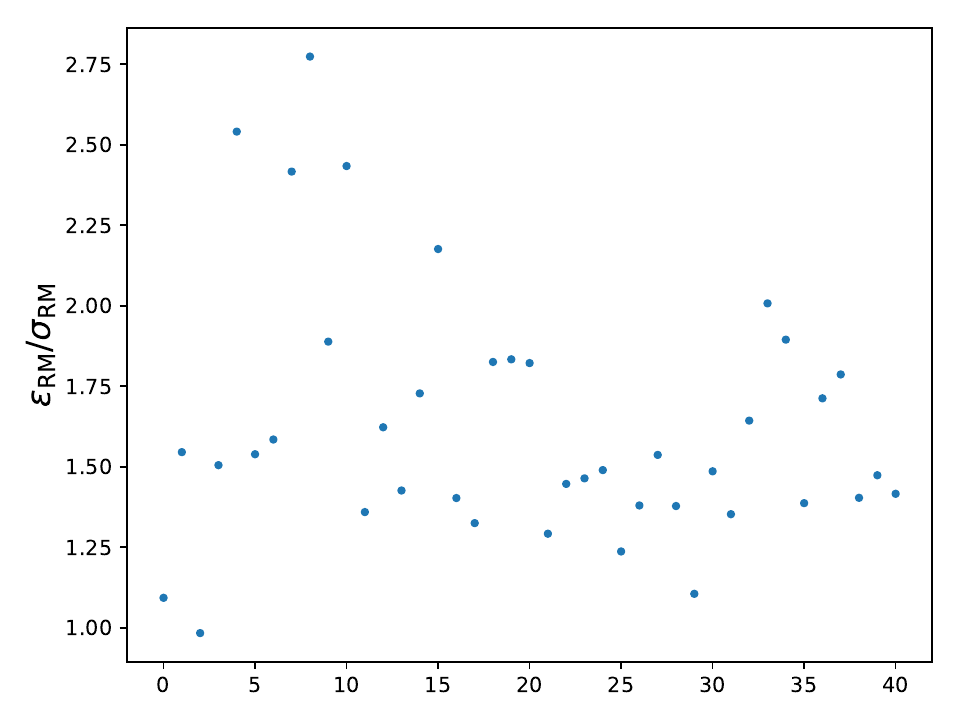}
    \caption{Ratio of root-mean-square variation of corrected RM ($\sigma_{\mathrm{RM}}$) values for all pulsars at all detected frequencies, and the corresponding value of mean error on ionosphere-corrected RMs.}
    \label{fig:deviations}
\end{figure}

To detect any variability in these measurements of potentially astrophysical origin, we applied purely linear trends to our time series data and compared it against a mean value, to test the statistical significance using an F-test. Although 12 pulsars in the sample statistically favored a linear trend at one or more frequencies, the slope of linear fits was of the order of $10^{-8} \mu$G yr$^{-1}$, which is close to zero. Hence it is reasonable to say that no statistically significant linear trend was found. Additionally, some of these pulsars also pass close to the Sun. Hence we applied a Lomb-Scargle periodogram analysis \citep[][]{Scargle} to detect any significant power on frequencies up to the time span of the data, but no statistically significant periodicity was detected. The absence of any strong variation is more evident from Figure \ref{fig:deviations}, which shows that the ratio of root-mean-square deviations of corrected RMs ($\sigma_{\mathrm{RM}}$) to the combined measurement uncertainties on RMs from detection and ionospheric corrections, is $\sim$1-2. 

This puts a limit of $\sim$20 nG on the variation in the average magnetic field on a scale of $\sim$100 AU. It is worth pointing out that our inability to detect any significant fluctuation in RM or $B_{\parallel}$ is currently limited by our ability to do ionospheric correction with better precision. As pointed out earlier, this can be improved by a factor of $\sim$2-5 by using LWA-localized GPS measurements for ionospheric corrections and the methods applied in \citet{Malins}. 

\section{Summary and Conclusions}\label{sec:summary}

In this section, we provide a summary of the most important results and concluding remarks.
\begin{itemize}
    \item We present the largest low-frequency pulsar catalog below 100 MHz, including 94 normal and 6-millisecond pulsars, along with the most sensitive measurements of some pulsar parameters, including temporal variations.
    \item Flux density measurements are reported for 97 sources including first and new measurements for 29 and 61 sources, respectively, in this frequency range. Measurements are provided over long-time scales to counter the effect of pulsar scintillation on flux measurements, but systematic errors are large due to the issue of limitations in low-frequency instrumental calibration.
    \item Multi-frequency narrow-band pulse profiles are shown, including 53 that are new detections in the specific frequency ranges, showing the effect of profile evolution with frequency due to effects both intrinsic and extrinsic to the pulsar.
    \item We detect 2 or more components in the pulse profiles of 32 pulsars, and provide pulse width measurements for all, which align with the expectations of the RFM model, with exceptions in a few cases.
    \item A study of turnover in pulsar spectra is presented where most pulsars favor a two-component model incorporating turnover, with a mean spectral index of -0.6$\pm$0.24, indicating the prevalence of this turnover effect at low frequencies. However, the mean spectral index from the linear model fit is -1.41$\pm$0.15, consistent with higher frequency measurements.
    \item Evaluating correlations between turnover frequency and other fundamental properties of the pulsar as well as its position in the sky yielded no significant correlations. While the former suggests that low-frequency turnover may not be due to intrinsic pulsar emission mechanism, the latter suggests they are not due to absorption in the ISM, as has been suggested. A possibility remains that these occur due to local pulsar environments such as pulsar wind nebula and supernova remnants as suggested by several authors before; however, this needs confirmation for a large sample of pulsars.
    \item Non-detection of any correlations may still be affected by intrinsic biases in various flux density measurements, hence more sensitive and systematic measurements are required to rule out these possibilities.
    \item Applying long-term timing to all our pulsars, we provide new timing solutions for 5 pulsars, namely J0242+62, J0611+30, J0815+4611, J1929+00, and J2227+30.
    \item We present a study of DM variation in these pulsars over a decade, with the most sensitive measurements of pulsar DM available: the median uncertainty is 2.9$\times$10$^{-4}$ pc cm$^{-3}$.
    \item Based on these measurements, we evaluate the model for electron density fluctuation in the ISM and find it to be consistent with prior studies, where the dependence between DM and DM rate-of-change for LWA observations is $|\frac{dDM}{dt}| = 0.0002\times DM^{0.5\pm0.26}$.
    \item We also report the strongest detection of frequency- and time-dependent DM in B2217+47, among only a few reported previously.
    \item Detection of polarization is reported in 27 pulsars, at multiple frequencies, along with the most sensitive measurements of pulsar RMs: the median uncertainty is 0.01 rad m$^{-2}$.
    \item A new detection of pulsar polarization is reported for J0051+0423: highly circular polarization with very low linear polarization, pointing towards magnetoionic interaction in the propagation path.
    \item The variation of pulsar RM and ISM magnetic field is studied for these targets, with no strong variations above the limit imposed by the ionospheric RM corrections. This suggests a limit to the variability in the ISM magnetic field of the order of 20nG on scales of $\sim$100 AU.
\end{itemize}

The results in this paper demonstrate the richness of pulsar and ISM physics that can be studied with low-frequency observations. They complement the higher-frequency observations, with strengths in different areas of pulsar parameter space. Future observations with existing, new, and upcoming facilities like LWA, LOFAR, NenuFAR, MWA, and the SKA will provide more robust measurements of these properties, and improve our understanding of pulsar emission, associated propagation effects, and the intervening medium. In particular, future studies involving pulsar scintillation and scattering can help improve our understanding of ISM processes. Improvements in low-frequency calibration and sensitivity will further improve upon the existing flux density measurements while expanding the study to more sources. Higher sensitivity would also allow us to perform analyses of single-pulse data, which encodes information about pulsar emission. Additionally, improvements to resolution will enable resolved studies of pulsar scattering screens for nearby objects.

\section{Acknowledgements}
We thank the referee for constructive suggestions. Construction of the LWA has been supported by the Office of Naval Research under Contract N00014-07-C-0147 and by the AFOSR. Support for operations and continuing development of the LWA1 is provided by the Air Force Research Laboratory and the National Science Foundation under grants AST-1835400 and AGS1708855. We would like to thank the UNM Center for Advanced Research Computing, supported in part by the National Science Foundation, for providing the high performance computing resources used in this work.


\bibliographystyle{aasjournal}
\bibliography{pulsarcatbib}

\appendix
\setcounter{figure}{0} 
\renewcommand{\thefigure}{A\arabic{figure}}
\renewcommand{\theHfigure}{A\arabic{figure}}
\renewcommand\theHtable{Appendix.\thetable}
\restartappendixnumbering

\section{LWA pulsars and their Dispersion Measures}
\startlongtable
\centerwidetable
\begin{deluxetable*}{cccccccc}
\tablecolumns{8}
\tabletypesize{\small}
\tablewidth{0pt}
\tablecaption{Period and Dispersion measure of Pulsars detected by the LWA \label{tab:dmperiod}}
\tablehead{
    \colhead{Name} & \colhead{Period} & \colhead{DM$_{\mathrm{ATNF}}$} & \colhead{DM$_{\mathrm{Epoch}}^{\mathrm{ATNF}}$} & \colhead{DM$_{\mathrm{LWA}}$} & \colhead{DM$_{\mathrm{Epoch}}^{\mathrm{LWA}}$} & \colhead{DM$_{\mathrm{LOFAR}}$} & \colhead{DM$_{\mathrm{Epoch}}^{\mathrm{LOFAR}}$}  \\ 
     \colhead{ } & \colhead{(s)} & \colhead{(pc cm$^{-3}$)} & \colhead{ } & \colhead{(pc cm$^{-3}$)} & \colhead{ } & \colhead{(pc cm$^{-3}$)} & \colhead{ }
}

\startdata
B0031-07 & 0.943 & 10.920(6) & 46635 & 10.88466(9) & 46635 & ... & ... \\ 
B0053+47 & 0.472 & 18.1400(13) & 49872 & 18.11842(13) & 49872 & 18.1354(13) & 56703 \\ 
B0105+65 & 1.2837 & 30.5500(14) & 50011 & 30.58307(114) & 50011 & 30.5482(14) & 56784 \\ 
B0136+57 & 0.2724 & 73.81141(76) & 49289 & 73.811(20) & 57758 & 73.81141 (76) & 56753 \\ 
B0138+59 & 1.2229 & 34.930(4) & 49293 & 34.92012(19) & 49293 & ... & ... \\ 
B0149-16 & 0.8327 & 11.93000(4) & 48227 & 11.92794(1) & 48227 & ... & ... \\ 
B0301+19 & 1.3876 & 15.66000(35) & 49289 & 15.66734(59) & 49289 & 15.65677(35) & 56703 \\ 
B0320+39 & 3.0321 & 26.19000(93) & 49290 & 26.17342(156) & 49290 & 26.18975(93) & 56703 \\ 
B0329+54 & 0.7145 & 26.7600(1) & 46473 & 26.76516(2) & 46473 & ... & ... \\ 
B0355+54 & 0.1564 & 57.1400(3) & 54096 & 57.14860(21) & 49616 & ... & ... \\ 
B0447-12 & 0.438 & 37.04(1) & 49338 & 37.03022(96) & 49338 & ... & ... \\ 
B0450-18 & 0.5489 & 39.900(3) & 49289 & 39.89792(83) & 49289 & ... & ... \\ 
B0450+55 & 0.3407 & 14.59000(15) & 49910 & 14.59881(6) & 49910 & 14.59002(15) & 56772 \\ 
B0525+21 & 3.7455 & 50.8700(13) & 54200 & 50.89705(34) & 54200 & 50.8695(13) & 56747 \\ 
B0628-28 & 1.2444 & 34.420(1) & 46603 & 34.41520(10) & 46603 & ... & ... \\ 
B0655+64 & 0.1957 & 8.77000(27) & 48806 & 8.774450(1) & 48806 & 8.77387(27) & 56772 \\ 
B0809+74 & 1.2922 & 5.75000(48) & 49162 & 5.762190(6) & 49162 & 5.75066(48) & 56747 \\ 
B0818-13 & 1.2381 & 40.940(3) & 48904 & 40.98380(33) & 48904 & ... & ... \\ 
B0820+02 & 0.8649 & 23.730(6) & 49281 & 23.78146(55) & 49281 & ... & ... \\ 
B0823+26 & 0.5307 & 19.48000(18) & 46450 & 19.47915(1) & 46450 & 19.47633(18) & 56747 \\ 
B0834+06 & 1.2738 & 12.8600(4) & 48721 & 12.86134(0) & 48721 & ... & ... \\ 
B0906-17 & 0.4016 & 15.880(2) & 48737 & 15.87306(17) & 48737 & ... & ... \\ 
B0917+63 & 1.568 & 13.15000(18) & 49687 & 13.14800(55) & 49687 & 13.15423(18) & 56747 \\ 
B0919+06 & 0.4306 & 27.3000(5) & 55140 & 27.29982(1) & 55140 & ... & ... \\ 
B0943+10 & 1.0977 & 15.3200(9) & 48483 & 15.33080(1) & 54226 & 15.31845(90) & 56779 \\ 
B0950+08 & 0.2531 & 2.97000(8) & 46375 & 2.970300(0) & 46375 & ... & ... \\ 
B1112+50 & 1.6564 & 9.19000(26) & 49334 & 9.187190(17) & 49334 & 9.18634(26) & 56747 \\ 
B1133+16 & 1.1879 & 4.84000(34) & 46407 & 4.845510(0) & 46407 & 4.84066(34) & 56687 \\ 
B1237+25 & 1.3824 & 9.25000(53) & 46531 & 9.251980(2) & 46531 & 9.25159(53) & 56689 \\ 
B1257+12 & 0.0062 & 10.17000(3) & 49750 & 10.15332(0) & 57220 & ... & ... \\ 
B1322+83 & 0.0062 & 13.31624(76) & 48889 & 13.324(20) & 57758 & 13.31624(76) & 56687 \\
B1508+55 & 0.7397 & 19.6200(3) & 49904 & 19.61303(1) & 49904 & 19.6189(13) & 56687 \\ 
B1530+27 & 1.1248 & 14.690(16) & 49666 & 14.70432(17) & 49666 & 14.691(16) & 56703 \\ 
B1540-06 & 0.7091 & 18.3(2) & 49423 & 18.37955(17) & 49423 & ... & ... \\ 
B1541+09 & 0.7484 & 34.9800(16) & 48716 & 35.00107(32) & 48716 & 34.9758(16) & 56780 \\ 
B1600-27 & 0.7783 & 46.200(16) & 49911 & 46.25939(170) & 49911 & ... & ... \\ 
B1604-00 & 0.4218 & 10.6800(1) & 46973 & 10.68148(1) & 46973 & ... & ... \\ 
B1612+07 & 1.2068 & 21.3900(3) & 49897 & 21.40453(11) & 49897 & ... & ... \\ 
B1633+24 & 0.4905 & 24.2700(44) & 48736 & 24.26552(40) & 48736 & 24.2671(44) & 56748 \\ 
B1642-03 & 0.3877 & 35.7600(8) & 46515 & 35.75697(6) & 46515 & ... & ... \\ 
B1702-19 & 0.299 & 22.910(3) & 48733 & 22.92196(51) & 48733 & ... & ... \\ 
B1706-16 & 0.6531 & 24.890(1) & 46993 & 24.89079(9) & 46993 & ... & ... \\ 
B1717-16 & 1.5656 & 44.83(3) & 49686 & 44.93165(174) & 49686 & ... & ... \\ 
B1717-29 & 0.6204 & 42.64(3) & 49863 & 42.61806(189) & 49863 & ... & ... \\ 
B1737+13 & 0.8031 & 48.67000(43) & 48262 & 48.66926(29) & 48262 & 48.66823(43) & 56687 \\ 
B1747-46 & 0.7424 & 20.4(4) & 57600 & 20.53728(13) & 46800 & ... & ... \\ 
B1749-28 & 0.5626 & 50.370(8) & 46483 & 50.45210(83) & 46483 & ... & ... \\ 
B1821+05 & 0.7529 & 66.780(3) & 48713 & 66.80706(27) & 48713 & ... & ... \\ 
B1822-09 & 0.769 & 19.3800(9) & 57600 & 19.40154(11) & 54262 & ... & ... \\ 
B1831-04 & 0.2901 & 79.310(8) & 49714 & 79.51473(165) & 49714 & ... & ... \\ 
B1839+56 & 1.6529 & 26.77000(17) & 48717 & 26.77195(2) & 48717 & 26.77163(17) & 56789 \\ 
B1842+14 & 0.3755 & 41.49000(61) & 49362 & 41.49135(11) & 49362 & 41.48555(61) & 56703 \\ 
B1857-26 & 0.6122 & 37.990(5) & 48891 & 37.99183(111) & 48891 & ... & ... \\ 
B1905+39 & 1.2358 & 30.970(14) & 48713 & 31.04227(923) & 48713 & 30.966(14) & 56789 \\ 
B1911-04 & 0.8259 & 89.39(1) & 46634 & 89.42068(367) & 46634 & ... & ... \\ 
B1918+26 & 0.7855 & 27.71000(82) & 49912 & 27.71256(22) & 49912 & 27.70882(82) & 56748 \\ 
B1919+21 & 1.3373 & 12.44000(63) & 48999 & 12.43938(1) & 48999 & 12.44399(63) & 56788 \\ 
B1929+10 & 0.2265 & 3.18000(16) & 57600 & 3.185090(3) & 46523 & 3.18321(16) & 56781 \\ 
B1940-12 & 0.9724 & 28.920(15) & 48717 & 28.98699(156) & 48717 & ... & ... \\ 
B1944+17 & 0.4406 & 16.1400(73) & 48790 & 16.16371(94) & 48790 & 16.1356(73) & 56801 \\ 
B2016+28 & 0.558 & 14.2000(6) & 46384 & 14.19723(4) & 46384 & 14.1839(13) & 56781 \\ 
B2020+28 & 0.3434 & 24.63000(18) & 49692 & 24.63254(6) & 49692 & 24.63109(18) & 56748 \\ 
B2021+51 & 0.5292 & 22.55000(56) & 46640 & 22.54491(119) & 46640 & 22.54968(56) & 56789 \\ 
B2022+50 & 0.3726 & 32.99000(37) & 49910 & 33.01410(133) & 49910 & 32.98817(37) & 56789 \\ 
B2043-04 & 1.5469 & 35.80(1) & 48739 & 35.74060(987) & 48739 & ... & ... \\ 
B2045-16 & 1.9616 & 11.460(5) & 46423 & 11.50060(231) & 46423 & ... & ... \\ 
B2053+21 & 0.8152 & 36.35000(29) & 49726 & 36.37842(129) & 49726 & 36.34963(29) & 56773 \\ 
B2110+27 & 1.2029 & 25.11000(18) & 48741 & 25.12042(8) & 48741 & 25.11106(18) & 56784 \\ 
B2152-31 & 1.03 & 14.85(5) & 48714 & 14.87062(24) & 48714 & ... & ... \\ 
B2154+40 & 1.5253 & 71.1200(22) & 49277 & 71.55517(283) & 49277 & 71.1239(22) & 56687 \\ 
B2217+47 & 0.5385 & 43.5000(5) & 46599 & 43.49784(4) & 46599 & 43.4862(60) & 56687 \\ 
B2224+65 & 0.6825 & 36.44000(51) & 54420 & 36.46412(124) & 54420 & 36.44362(51) & 56784 \\ 
B2303+30 & 1.5759 & 49.5800(12) & 48714 & 49.62452(90) & 48714 & 49.5845(12) & 56773 \\ 
B2306+55 & 0.4751 & 46.54000(37) & 48717 & 46.54498(87) & 48717 & 46.53905(37) & 56785 \\ 
B2310+42 & 0.3494 & 17.28000(33) & 48241 & 17.28194(38) & 48241 & 17.27693(33) & 56687 \\ 
B2315+21 & 1.4446 & 20.86959(31) & 48716 & 20.87579(24) & 48716 &  20.86959(31)& 56687 \\ 
B2327-20 & 1.6436 & 8.460(2) & 49878 & 8.455940(4) & 49878 & ... & ... \\ 
B2334+61 & 0.4954 & 58.410(15) & 54521 & 58.45541(85) & 54521 & ... & ... \\ 
J0030+0451 & 0.0049 & 4.33000(11) & 55664 & 4.332480(0) & 54250 & ... & ... \\ 
J0034-0534 & 0.0019 & 13.77000(4) & 55000 & 13.76526(0) & 55000 & ... & ... \\ 
J0051+0423 & 0.3547 & 13.93(9) & 49800 & 13.92669(2) & 49800 & ... & ... \\ 
J0242+62 & 0.5917 & 3.92(16) & 54100 & 3.820240(12) & 59124 & ... & ... \\ 
J0459-0210 & 1.1331 & 21.02(3) & 50845 & 21.05038(80) & 50845 & ... & ... \\ 
J0611+30 & 1.4121 & 45.2600(16) & 49717 & 45.27457(183) & 58926 & 45.2551(16) & 56772 \\ 
J0613+3731 & 0.6192 & 18.990(12) & 56000 & 18.97839(9) & 56000 & ... & ... \\ 
J0815+4611 & 0.4342 & 11.2700(3) & 57662 & 11.26831(39) & 58792 & ... & ... \\ 
J1022+1001 & 0.0165 & 10.2600(8) & 55636 & 10.25366(0) & 50250 & ... & ... \\ 
J1313+0931 & 0.8489 & 12.04000(4) & 50984 & 12.04398(31) & 50984 & 12.040623(43) & 56779 \\ 
J1327+3423 & 0.0415 & ... & ... & 4.183920(1) & 56800 & ... & ... \\ 
J1400-1431 & 0.0031 & 4.93(0) & 56960 & 4.933250(0) & 56960 & ... & ... \\ 
J1645+1012 & 0.4109 & 36.17000(19) & 48957 & 36.17699(71) & 48957 & 36.17129(19) & 56687 \\ 
J1741+2758 & 1.3607 & 29.14000(11) & 51797 & 29.15145(17) & 51797 & 29.14487(11) & 56773 \\ 
J1758+3030 & 0.9473 & 35.0700(14) & 49800 & 35.08378(45) & 49800 & 35.0674(14) & 56784 \\ 
J1929+00 & 1.1669 & 42.95(7) & 49717 & 42.80258(114) & 58788 & ... & ... \\ 
J2043+2740 & 0.0961 & 21.02000(15) & 49773 & 21.02114(4) & 49773 & 21.02064(15) & 56784 \\ 
J2145-0750 & 0.0161 & 9.0000(13) & 55657 & 9.004510(4) & 54197 & ... & ... \\ 
J2208+4056 & 0.637 & 11.840(9) & 56375 & 11.85126(12) & 56086 & ... & ... \\ 
J2227+30 & 0.8424 & ... & ... & 19.96113(23) & 58923 & ... & ... \\ 
J2234+2114 & 1.3587 & 35.08(9) & 49079 & 35.35754(248) & 49079 & 35.31(24) & 56703 \\ 
\enddata
\tablenotetext{a}{$\Delta \phi_{12}$ is phase difference between the leading and lagging component. Similarly A$_{12}$ is the ratio of leading to lagging component.}
\tablenotetext{b}{\textit{W}$_{50}^1$ and \textit{W}$_{50}^2$ are the full-width at half max values for the leading and lagging component respectively.}
\end{deluxetable*}

\section{Component Analysis}
\startlongtable
\centerwidetable
\begin{deluxetable*}{cccccccccccc}
\tablecolumns{12}
\tabletypesize{\small}
\tablewidth{0pt}
\tablecaption{Component Spacing and Widths in pulse longitude, for 27 pulsars with two components
in the LWA1 frequency band.\label{tab:components}}
\tablehead{
    \colhead{Name} & \colhead{$\nu$} & \colhead{$\Delta \phi_{12}$} & \colhead{A${12}$} & \colhead{\textit{W}$_{50}^1$} & \colhead{\textit{W}$_{50}^2$} & \colhead{Name} & \colhead{$\nu$} & \colhead{$\Delta \phi_{12}$} & \colhead{A${12}$} & \colhead{\textit{W}$_{50}^1$} & \colhead{\textit{W}$_{50}^2$} \\
    \colhead{ } & \colhead{(MHz)} & \colhead{(phase)} & \colhead{(phase)} & \colhead{(phase)} & \colhead{ } & \colhead{ } & \colhead{(MHz)} & \colhead{(phase)} & \colhead{(phase)} & \colhead{(phase)} & \colhead{ } 
}

\startdata
B0031-07 & 35.1 & 0.115(2) & 0.074(4) & 0.082(4) & 0.4(8) & B1530+27 & 79.2 & 0.033(1) & 0.027(4) & 0.012(4) & 1.27(2) \\
B0031-07 & 49.8 & 0.094(2) & 0.082(4) & 0.074(4) & 0.7(8) & B1604-00 & 35.1 & 0.02(0) & 0.012(4) & 0.02(4) & 1.12(2) \\
B0031-07 & 64.5 & 0.084(2) & 0.066(4) & 0.074(4) & 0.85(9) & B1604-00 & 49.8 & 0.02(0) & 0.012(4) & 0.012(4) & 1.5(1) \\
B0031-07 & 79.2 & 0.071(2) & 0.059(4) & 0.082(4) & 0.82(7) & B1604-00 & 64.5 & 0.019(0) & 0.012(4) & 0.012(4) & 1.23(1) \\
B0138+59 & 49.8 & 0.046(7) & 0.191(20) & 0.02(4) & 1.33(3) & B1604-00 & 79.2 & 0.019(0) & 0.012(4) & 0.012(4) & 1.16(2) \\
B0138+59 & 64.5 & 0.054(5) & 0.145(12) & 0.02(4) & 0.76(2) & B1905+39 & 79.2 & 0.059(14) & 0.012(27) & 0.02(20) & 0.46(2) \\
B0138+59 & 79.2 & 0.049(7) & 0.129(12) & 0.012(4) & 0.85(2) & B1918+26 & 64.5 & 0.987(5) & 0.02(12) & 0.012(4) & 0.29(1) \\
B0149-16 & 35.1 & 0.023(0) & 0.012(4) & 0.02(4) & 0.97(1) & B2016+28 & 35.1 & 0.018(5) & 0.02(4) & 0.09(12) & 0.38(3) \\
B0149-16 & 49.8 & 0.022(0) & 0.004(4) & 0.012(4) & 1.16(1) & B2016+28 & 49.8 & 0.025(2) & 0.02(4) & 0.066(4) & 1.42(2) \\
B0149-16 & 64.5 & 0.021(0) & 0.004(4) & 0.012(4) & 1.32(1) & B2016+28 & 64.5 & 0.019(2) & 0.012(4) & 0.074(4) & 1.74(2) \\
B0149-16 & 79.2 & 0.02(0) & 0.004(4) & 0.012(4) & 1.38(1) & B2016+28 & 79.2 & 0.018(2) & 0.012(4) & 0.059(4) & 1.86(1) \\
B0301+19 & 49.8 & 0.062(1) & 0.02(4) & 0.027(4) & 0.86(2) & B2020+28 & 49.8 & 0.051(21) & 0.02(43) & 0.02(20) & 0.33(2) \\
B0301+19 & 64.5 & 0.056(1) & 0.02(4) & 0.02(4) & 1.16(2) & B2020+28 & 64.5 & 9.047(23) & 0.02(51) & 0.02(12) & 0.34(2) \\
B0301+19 & 79.2 & 0.054(1) & 0.012(4) & 0.012(4) & 1.28(1) & B2020+28 & 79.2 & 0.046(18) & 0.02(43) & 0.012(12) & 0.38(2) \\
B0320+39 & 35.1 & 0.024(3) & 0.012(4) & 0.035(4) & 0.61(1) & B2045-16 & 49.8 & 0.04(30) & 0.035(59) & 0.02(43) & 1.6(2) \\
B0320+39 & 64.5 & 0.018(0) & 0.012(4) & 0.012(4) & 1.5(2) & B2045-16 & 64.5 & 0.04(23) & 0.02(27) & 0.043(51) & 0.39(3) \\
B0525+21 & 35.1 & 0.066(1) & 0.027(4) & 0.027(4) & 1.29(3) & B2045-16 & 79.2 & 0.037(16) & 0.012(12) & 0.027(35) & 0.63(2) \\
B0525+21 & 49.8 & 0.062(0) & 0.02(4) & 0.02(4) & 1.41(2) & B2152-31 & 49.8 & 0.025(12) & 0.012(20) & 0.012(20) & 1.14(1) \\
B0525+21 & 64.5 & 0.058(0) & 0.012(4) & 0.012(4) & 1.3(2) & B2152-31 & 64.5 & 0.023(17) & 0.012(20) & 0.012(35) & 1.91(1) \\
B0525+21 & 79.2 & 0.055(0) & 0.012(4) & 0.012(4) & 1.37(1) & B2152-31 & 79.2 & 0.012(34) & 0.004(12) & 0.043(66) & 0.26(1) \\
B0809+74 & 35.1 & 0.074(0) & 0.043(4) & 0.051(4) & 0.51(6) & B2217+47 & 35.1 & 0.127(11) & 0.113(4) & 0.254(20) & 0.58(14) \\
B0809+74 & 49.8 & 0.061(0) & 0.043(4) & 0.059(4) & 0.55(6) & B2217+47 & 49.8 & 0.072(7) & 0.066(4) & 0.145(12) & 0.97(8) \\
B0809+74 & 64.5 & 0.037(0) & 0.027(4) & 0.074(4) & 0.2(7) & B2217+47 & 64.5 & 0.038(3) & 0.035(4) & 0.082(4) & 1.14(4) \\
B0809+74 & 79.2 & 0.022(0) & 0.02(4) & 0.066(4) & 0.14(6) & B2217+47 & 79.2 & 0.017(2) & 0.02(4) & 0.043(4) & 1.61(2) \\
B0820+02 & 49.8 & 1.064(12) & 0.02(4) & 0.066(35) & 1.69(1) & B2327-20 & 35.1 & 0.012(0) & 0.004(4) & 0.004(4) & 2.05(1) \\
B0820+02 & 64.5 & 0.046(2) & 0.02(4) & 0.02(4) & 2.21(1) & B2327-20 & 49.8 & 0.011(0) & 0.004(4) & 0.004(4) & 1.54(1) \\
B0906-17 & 49.8 & 0.006(82) & 0.051(90) & 0.035(215) & 3.45(5) & B2327-20 & 64.5 & 0.01(0) & 0.004(4) & 0.004(4) & 1.15(1) \\
B0906-17 & 64.5 & 0.037(53) & 0.02(27) & 0.043(121) & 1.22(2) & B2327-20 & 79.2 & 0.01(10) & 0.004(12) & 0.004(20) & 0.89(1)\\
B0906-17 & 79.2 & 0.025(68) & 0.012(27) & 0.059(113) & 0.56(2) & J0051+0423 & 49.8 & 0.045(2) & 0.012(4) & 0.02(4) & 4.74(1) \\
B0917+63 & 35.1 & 0.033(14) & 0.012(20) & 0.012(27) & 1.34(1) & J0051+0423 & 64.5 & 0.044(1) & 0.012(4) & 0.012(4) & 3.44(1) \\
B0917+63 & 49.8 & 0.032(10) & 0.012(20) & 0.012(20) & 1.02(1) & J0051+0423 & 79.2 & 0.042(1) & 0.012(4) & 0.012(4) & 3.62(1) \\
B0917+63 & 64.5 & 0.03(9) & 0.004(12) & 0.012(20) & 1.07(1) & J0459-0210 & 49.8 & 0.03(7) & 0.043(4) & 0.129(12) & 0.48(5) \\
B0943+10B & 35.1 & 0.047(18) & 0.02(20) & 0.02(35) & 2.87(2) & J0459-0210 & 64.5 & 0.007(2) & 0.027(4) & 0.09(4) & 0.34(4) \\
B0943+10B & 79.2 & 0.032(21) & 0.02(20) & 0.02(51) & 3.45(2) & J1313+0931 & 49.8 & 0.018(30) & 0.012(74) & 0.012(20) & 0.42(1) \\
B0943+10Q & 35.1 & 0.037(21) & 0.027(43) & 0.027(27) & 0.67(3) & J1313+0931 & 64.5 & 0.003(54) & 0.027(137) & 0.004(12) & 1.06(1) \\
B0943+10Q & 49.8 & 0.033(23) & 0.027(43) & 0.027(35) & 0.56(3) & J1741+2758 & 79.2 & 0.013(0) & 0.004(4) & 0.004(4) & 0.64(1) \\
B0943+10Q & 64.5 & 0.028(27) & 0.02(43) & 0.035(43) & 0.29(3) & J2227+30 & 35.1 & 0.037(1) & 0.012(4) & 0.02(4) & 1.16(1) \\
B1530+27 & 35.1 & 0.049(2) & 0.02(4) & 0.02(4) & 0.4(2) & J2227+30 & 49.8 & 0.032(1) & 0.012(4) & 0.02(4) & 1.05(2) \\
B1530+27 & 49.8 & 0.043(1) & 0.02(4) & 0.02(4) & 0.63(2) & J2227+30 & 64.5 & 0.032(1) & 0.012(4) & 0.012(4) & 1.21(1) \\
B1530+27 & 64.5 & 0.037(1) & 0.027(4) & 0.012(4) & 0.8(1) & J2227+30 & 79.2 & 0.028(1) & 0.012(4) & 0.012(4) & 1.13(1) \\
\enddata
\tablenotetext{a}{$\Delta \phi_{12}$ is phase difference between the leading and lagging component. Similarly A$_{12}$ is the ratio of leading to lagging component.}
\tablenotetext{b}{\textit{W}$_{50}^1$ and \textit{W}$_{50}^2$ are the full-width at half max values for the leading and lagging component respectively.}
\tablenotetext{c}{Term in brackets represent errors obtained from Gaussian component fitting process. For $\Delta \phi_{12}$ and A$_{12}$, it is the quadrature error.}
\end{deluxetable*}

\section{LWA pulsar flux density and average profiles}

\startlongtable
\begin{deluxetable*}{c@{\hskip 0.5in}c@{\hskip 0.5in}c@{\hskip 0.5in}c@{\hskip 0.2in}c@{\hskip 0.5in}c@{\hskip 0.5in}c@{\hskip 0.5in}c}
\tablecolumns{8}
\tabletypesize{\small}
\tablewidth{0pt}
\tablecaption{ Mean Flux Density (S$_\nu$) Measurements for Pulsars Detected by the LWA at Multiple Observing Frequencies ($\nu$)\label{tab:fluxdensity}}
\tablehead{
    \colhead{Name} & \colhead{$\nu$} & \colhead{T} & \colhead{\textit{W}$_{50}$} & \colhead{\textit{W}$_{10}$} & \colhead{S/N} & \colhead{SEFD} & \colhead{S$_{\nu}$}  \\ 
     \colhead{ } & \colhead{(MHz)} & \colhead{(min)} & \colhead{(phase)} & \colhead{(phase)} & \colhead{ } & \colhead{(kJy)} & \colhead{(mJy)}     
}
\startdata
B0031-07 & 35.1 & 300 & 0.081 & 0.236 & 616.93 & 14.87 & 3350(1675) \\ 
B0031-07 & 49.8 & 180 & 0.147 & 0.217 & 732.24 & 11.9 & 5758(2879) \\ 
B0031-07 & 64.5 & 180 & 0.143 & 0.198 & 718.6 & 11.9 & 5565(2782) \\ 
B0031-07 & 79.2 & 180 & 0.118 & 0.179 & 628.06 & 11.9 & 4342(2171) \\ 
\hline
B0053+47 & 35.1 & 360 & 0.031 & 0.061 & 172.2 & 9.27 & 461(230) \\ 
B0053+47 & 49.8 & 600 & 0.028 & 0.052 & 224.29 & 7.42 & 350(175) \\ 
B0053+47 & 64.5 & 180 & 0.031 & 0.052 & 108.99 & 7.42 & 330(165) \\ 
B0053+47 & 79.2 & 420 & 0.024 & 0.042 & 67.2 & 7.42 & 116(58) \\ 
\hline
B0105+65 & 49.8 & 120 & 0.035 & 0.064 & 21.69 & 9.88 & 67(33) \\ 
B0105+65 & 64.5 & 240 & 0.035 & 0.077 & 51.16 & 9.88 & 112(56) \\ 
B0105+65 & 79.2 & 240 & 0.026 & 0.051 & 39.02 & 9.88 & 73(36) \\ 
\hline
B0138+59 & 49.8 & 540 & 0.028 & 0.159 & 212.02 & 8.89 & 256(128) \\ 
B0138+59 & 64.5 & 600 & 0.024 & 0.049 & 240.94 & 8.89 & 257(128) \\ 
B0138+59 & 79.2 & 240 & 0.015 & 0.037 & 160.49 & 8.89 & 209(104) \\
\hline
B0149-16 & 35.1 & 300 & 0.007 & 0.033 & 229.49 & 18.85 & 462(231) \\ 
B0149-16 & 49.8 & 120 & 0.003 & 0.033 & 195.05 & 15.08 & 350(175) \\ 
B0149-16 & 64.5 & 300 & 0.003 & 0.025 & 258.97 & 15.08 & 294(147) \\ 
B0149-16 & 79.2 & 300 & 0.003 & 0.025 & 162.15 & 15.08 & 184(92) \\
\hline
B0301+19 & 49.8 & 720 & 0.071 & 0.125 & 81.46 & 7.47 & 108(54) \\ 
B0301+19 & 64.5 & 1080 & 0.049 & 0.097 & 83.93 & 7.47 & 74(37) \\ 
B0301+19 & 79.2 & 1200 & 0.032 & 0.069 & 55.85 & 7.47 & 38(19) \\ 
\hline
B0320+39 & 35.1 & 300 & 0.036 & 0.182 & 58.38 & 8.37 & 64(32) \\ 
B0320+39 & 49.8 & 600 & 0.036 & 0.152 & 160.45 & 6.7 & 99(49) \\ 
B0320+39 & 64.5 & 540 & 0.061 & 0.121 & 212.31 & 6.7 & 180(90) \\ 
\hline
B0329+54 & 35.1 & 180 & 0.053 & 0.1 & 255.05 & 10.18 & 1127(563) \\
B0329+54 & 49.8 & 90 & 0.031 & 0.057 & 206.29 & 8.14 & 773(386) \\ 
B0329+54 & 64.5 & 90 & 0.014 & 0.029 & 191.26 & 8.14 & 483(241) \\ 
B0329+54 & 79.2 & 90 & 0.009 & 0.014 & 224.93 & 8.14 & 438(219) \\ 
\hline
B0355+54 & 49.8 & 840 & 0.029 & 0.052 & 156.83 & 8.1 & 429(214) \\ 
B0355+54 & 64.5 & 1200 & 0.015 & 0.028 & 225.35 & 8.1 & 358(179) \\ 
B0355+54 & 79.2 & 1200 & 0.012 & 0.02 & 243.64 & 8.1 & 332(166) \\ 
\hline
B0447-12 & 64.5 & 720 & 0.032 & 0.061 & 89.01 & 13.61 & 263(131) \\ 
B0447-12 & 79.2 & 360 & 0.022 & 0.039 & 36.01 & 13.61 & 123(61) \\ 
\hline
B0450-18 & 64.5 & 840 & 0.049 & 0.093 & 108.07 & 15.72 & 404(202) \\ 
B0450-18 & 79.2 & 1200 & 0.049 & 0.093 & 127.12 & 15.72 & 352(176) \\ 
\hline
B0450+55 & 35.1 & 360 & 0.033 & 0.058 & 108.99 & 10.35 & 256(128) \\ 
B0450+55 & 49.8 & 600 & 0.025 & 0.048 & 178.8 & 8.28 & 107(53) \\ 
B0450+55 & 64.5 & 600 & 0.02 & 0.037 & 147.16 & 8.28 & 380(190) \\ 
B0450+55 & 79.2 & 360 & 0.007 & 0.014 & 83.4 & 8.28 & 374(187) \\ 
\hline
B0525+21 & 35.1 & 600 & 0.176 & 0.337 & 168.18 & 9.05 & 284(142) \\ 
B0525+21 & 49.8 & 240 & 0.116 & 0.225 & 214.8 & 7.24 & 370(185) \\ 
B0525+21 & 64.5 & 120 & 0.086 & 0.187 & 146.96 & 7.24 & 307(153) \\ 
B0525+21 & 79.2 & 180 & 0.086 & 0.15 & 133.86 & 7.24 & 228(114) \\ 
\hline
B0628-28 & 35.1 & 360 & 0.102 & 0.174 & 360.92 & 28.76 & 3371(1685) \\ 
B0628-28 & 49.8 & 180 & 0.082 & 0.162 & 427.35 & 23.01 & 4017(2008) \\ 
B0628-28 & 64.5 & 180 & 0.082 & 0.137 & 442.72 & 23.01 & 4162(2081) \\ 
B0628-28 & 79.2 & 180 & 0.073 & 0.137 & 392.49 & 23.01 & 3475(1737) \\ 
\hline
B0655+64 & 35.1 & 360 & 0.004 & 0.008 & 91.48 & 11.9 & 169(84) \\ 
B0655+64 & 49.8 & 120 & 0.004 & 0.008 & 121.78 & 9.53 & 311(155) \\
B0655+64 & 64.5 & 360 & 0.004 & 0.008 & 198.92 & 9.53 & 294(147) \\ 
B0655+64 & 79.2 & 180 & 0.004 & 0.008 & 150.15 & 9.53 & 314(157) \\
\hline 
B0809+74 & 35.1 & 180 & 0.106 & 0.207 & 924.84 & 14.56 & 6183(3091) \\ 
B0809+74 & 49.8 & 180 & 0.136 & 0.194 & 1539.0 & 11.65 & 9437(4718) \\ 
B0809+74 & 64.5 & 180 & 0.106 & 0.168 & 1383.74 & 11.65 & 7404(3702) \\ 
B0809+74 & 79.2 & 180 & 0.085 & 0.155 & 1069.03 & 11.65 & 5087(2543) \\ 
\hline
B0818-13 & 49.8 & 420 & 0.043 & 0.087 & 102.48 & 14.03 & 275(137) \\ 
B0818-13 & 64.5 & 480 & 0.033 & 0.05 & 133.53 & 14.03 & 293(146) \\
B0818-13 & 79.2 & 480 & 0.025 & 0.05 & 145.82 & 14.03 & 275(137) \\
\hline 
B0820+02 & 49.8 & 960 & 0.017 & 0.078 & 30.98 & 9.87 & 29(14) \\ 
B0820+02 & 64.5 & 720 & 0.017 & 0.052 & 52.27 & 9.87 & 56(28) \\ 
B0820+02 & 79.2 & 120 & 0.017 & 0.035 & 17.68 & 9.87 & 46(23) \\ 
\hline
B0823+26 & 35.1 & 180 & 0.027 & 0.048 & 325.18 & 8.54 & 989(494) \\
B0823+26 & 49.8 & 180 & 0.019 & 0.032 & 416.94 & 6.84 & 834(417) \\ 
B0823+26 & 64.5 & 180 & 0.014 & 0.021 & 486.44 & 6.84 & 851(425) \\
B0823+26 & 79.2 & 180 & 0.011 & 0.021 & 394.77 & 6.84 & 592(296) \\
\hline 
B0834+06 & 35.1 & 90 & 0.015 & 0.051 & 416.48 & 11.45 & 1142(571) \\ 
B0834+06 & 49.8 & 90 & 0.02 & 0.038 & 805.06 & 9.16 & 2044(1022) \\
B0834+06 & 64.5 & 90 & 0.02 & 0.038 & 1278.55 & 9.16 & 3246(1623) \\ 
B0834+06 & 79.2 & 90 & 0.005 & 0.025 & 927.01 & 9.16 & 1169(584)\\
\hline 
B0906-17 & 49.8 & 360 & 0.02 & 0.036 & 85.78 & 15.63 & 337(168) \\ 
B0906-17 & 64.5 & 600 & 0.009 & 0.036 & 170.7 & 15.63 & 344(172) \\ 
B0906-17 & 79.2 & 360 & 0.008 & 0.036 & 85.7 & 15.63 & 207(103) \\
\hline
B0917+63 & 35.1 & 840 & 0.036 & 0.078 & 31.58 & 11.6 & 39(19) \\ 
B0917+63 & 49.8 & 1080 & 0.036 & 0.078 & 49.65 & 9.28 & 44(22) \\ 
B0917+63 & 64.5 & 120 & 0.019 & 0.047 & 21.36 & 9.28 & 41(20) \\ 
\hline
B0919+06 & 35.1 & 60 & 0.022 & 0.039 & 269.02 & 11.36 & 1886(943) \\ 
B0919+06 & 49.8 & 60 & 0.015 & 0.026 & 227.84 & 9.09 & 1050(525) \\
B0919+06 & 64.5 & 60 & 0.012 & 0.022 & 171.18 & 9.09 & 690(345) \\ 
B0919+06 & 79.2 & 60 & 0.012 & 0.017 & 122.28 & 9.09 & 492(246) \\ 
\hline
B0943+10B & 35.1 & 60 & 0.022 & 0.077 & 622.91 & 10.78 & 2552(1276) \\ 
B0943+10B & 49.8 & 60 & 0.022 & 0.066 & 843.5 & 8.62 & 2765(1382) \\ 
B0943+10B & 64.5 & 60 & 0.022 & 0.077 & 814.29 & 8.62 & 2670(1335) \\ 
B0943+10B & 79.2 & 60 & 0.022 & 0.066 & 399.67 & 8.62 & 1310(655) \\ 
\hline
B0943+10Q & 35.1 & 60 & 0.06 & 0.099 & 254.06 & 10.78 & 1758(879) \\ 
B0943+10Q & 49.8 & 60 & 0.06 & 0.088 & 286.17 & 8.62 & 1584(792) \\ 
B0943+10Q & 64.5 & 60 & 0.056 & 0.077 & 260.33 & 8.62 & 1385(692) \\ 
B0943+10Q & 79.2 & 60 & 0.047 & 0.088 & 194.88 & 8.62 & 948(474) \\
\hline
B0950+08 & 35.1 & 90 & 0.022 & 0.046 & 1258.16 & 11.12 & 9329(4664) \\ 
B0950+08 & 49.8 & 90 & 0.022 & 0.046 & 1263.79 & 8.9 & 7498(3749) \\ 
B0950+08 & 64.5 & 90 & 0.019 & 0.04 & 1440.15 & 8.9 & 7875(3937) \\
B0950+08 & 79.2 & 90 & 0.017 & 0.038 & 1688.32 & 8.9 & 8681(4340) \\ 
\hline
B1112+50 & 35.1 & 240 & 0.045 & 0.083 & 55.72 & 9.57 & 118(59) \\ 
B1112+50 & 49.8 & 180 & 0.045 & 0.066 & 98.25 & 7.66 & 192(96) \\ 
B1112+50 & 64.5 & 360 & 0.033 & 0.066 & 116.18 & 7.66 & 138(69) \\ 
B1112+50 & 79.2 & 120 & 0.033 & 0.066 & 51.67 & 7.66 & 106(53) \\ 
\hline
B1133+16 & 35.1 & 90 & 0.024 & 0.048 & 596.84 & 9.85 & 1824(912) \\
B1133+16 & 49.8 & 90 & 0.024 & 0.048 & 830.96 & 7.88 & 2033(1016) \\ 
B1133+16 & 64.5 & 90 & 0.019 & 0.059 & 647.93 & 7.88 & 1415(707) \\ 
B1133+16 & 79.2 & 90 & 0.024 & 0.059 & 703.61 & 7.88 & 1721(860) \\
\hline
B1237+25 & 35.1 & 600 & 0.017 & 0.097 & 182.26 & 8.73 & 147(73) \\ 
B1237+25 & 49.8 & 360 & 0.017 & 0.069 & 199.97 & 6.99 & 167(83) \\ 
B1237+25 & 64.5 & 300 & 0.011 & 0.041 & 274.62 & 6.99 & 205(102) \\ 
B1237+25 & 79.2 & 300 & 0.011 & 0.083 & 159.66 & 6.99 & 119(59) \\ 
\hline
B1257+12 & 49.8 & 1080 & 0.002 & 0.003 & 196.27 & 8.26 & 605(302) \\ 
B1257+12 & 64.5 & 1200 & 0.001 & 0.002 & 242.28 & 8.26 & 608(304) \\ 
B1257+12 & 79.2 & 1200 & 0.001 & 0.002 & 168.1 & 8.26 & 432(216) \\ 
\hline
B1508+55 & 35.1 & 90 & 0.032 & 0.052 & 175.13 & 10.3 & 830(415) \\ 
B1508+55 & 49.8 & 90 & 0.02 & 0.03 & 182.84 & 8.24 & 545(272) \\ 
B1508+55 & 64.5 & 90 & 0.015 & 0.022 & 192.16 & 8.24 & 491(245) \\ 
B1508+55 & 79.2 & 90 & 0.009 & 0.022 & 213.3 & 8.24 & 421(210) \\ 
\hline
B1530+27 & 35.1 & 300 & 0.022 & 0.079 & 61.19 & 8.43 & 87(43) \\ 
B1530+27 & 49.8 & 540 & 0.022 & 0.067 & 138.38 & 6.74 & 118(59) \\ 
B1530+27 & 64.5 & 300 & 0.013 & 0.079 & 90.48 & 6.74 & 80(40) \\ 
B1530+27 & 79.2 & 300 & 0.026 & 0.079 & 92.83 & 6.74 & 114(57) \\ 
\hline
B1540-06 & 35.1 & 180 & 0.025 & 0.043 & 75.81 & 14.6 & 323(161) \\ 
B1540-06 & 49.8 & 600 & 0.014 & 0.028 & 160.93 & 11.68 & 226(113) \\ 
B1540-06 & 64.5 & 300 & 0.014 & 0.028 & 158.68 & 11.68 & 315(157) \\ 
B1540-06 & 79.2 & 120 & 0.009 & 0.021 & 87.32 & 11.68 & 211(105) \\
\hline
B1541+09 & 49.8 & 540 & 0.149 & 0.269 & 242.75 & 8.67 & 931(465) \\ 
B1541+09 & 64.5 & 480 & 0.073 & 0.135 & 247.16 & 8.67 & 664(332) \\ 
B1541+09 & 79.2 & 300 & 0.055 & 0.097 & 329.75 & 8.67 & 962(481) \\ 
\hline
B1600-27 & 49.8 & 1200 & 0.051 & 0.093 & 54.91 & 21.78 & 189(94) \\ 
B1600-27 & 64.5 & 840 & 0.027 & 0.054 & 82.15 & 21.78 & 242(121) \\ 
B1600-27 & 79.2 & 840 & 0.021 & 0.039 & 73.93 & 21.78 & 190(95) \\ 
\hline
B1604-00 & 35.1 & 360 & 0.007 & 0.017 & 423.47 & 12.94 & 759(379) \\ 
B1604-00 & 49.8 & 180 & 0.007 & 0.017 & 377.4 & 10.35 & 765(382) \\ 
B1604-00 & 64.5 & 180 & 0.008 & 0.017 & 289.68 & 10.35 & 658(329) \\ 
B1604-00 & 79.2 & 360 & 0.008 & 0.017 & 282.52 & 10.35 & 454(227) \\ 
\hline
B1612+07 & 35.1 & 180 & 0.042 & 0.072 & 39.66 & 11.17 & 129(64) \\ 
B1612+07 & 49.8 & 540 & 0.014 & 0.024 & 64.09 & 8.94 & 56(28) \\ 
B1612+07 & 64.5 & 600 & 0.014 & 0.012 & 78.46 & 8.94 & 65(32) \\ 
B1612+07 & 79.2 & 600 & 0.005 & 0.012 & 75.75 & 8.94 & 36(18) \\ 
\hline
B1633+24 & 49.8 & 120 & 0.029 & 0.049 & 42.06 & 7.03 & 139(69) \\ 
B1633+24 & 64.5 & 240 & 0.021 & 0.039 & 89.89 & 7.03 & 178(89) \\ 
B1633+24 & 79.2 & 240 & 0.021 & 0.039 & 92.76 & 7.03 & 184(92) \\ 
\hline
B1642-03 & 35.1 & 240 & 0.02 & 0.035 & 62.59 & 13.67 & 263(131) \\ 
B1642-03 & 49.8 & 600 & 0.017 & 0.031 & 235.88 & 10.94 & 460(230) \\ 
B1642-03 & 64.5 & 360 & 0.008 & 0.016 & 206.96 & 10.94 & 351(175) \\ 
B1642-03 & 79.2 & 360 & 0.008 & 0.016 & 231.46 & 10.94 & 393(196) \\ 
\hline
B1702-19 & 64.5 & 600 & 0.015 & 0.027 & 31.97 & 16.31 & 101(50) \\ 
B1702-19 & 79.2 & 840 & 0.01 & 0.021 & 40.69 & 16.31 & 89(44) \\ 
\hline
B1706-16 & 35.1 & 420 & 0.033 & 0.059 & 169.48 & 18.95 & 748(374) \\ 
B1706-16 & 49.8 & 360 & 0.023 & 0.039 & 289.06 & 15.16 & 907(453) \\ 
B1706-16 & 64.5 & 360 & 0.018 & 0.026 & 285.53 & 15.16 & 783(391) \\ 
B1706-16 & 79.2 & 360 & 0.013 & 0.026 & 174.26 & 15.16 & 410(205) \\ 
\hline
B1717-16 & 49.8 & 480 & 0.067 & 0.125 & 48.83 & 15.1 & 147(73) \\ 
B1717-16 & 64.5 & 360 & 0.042 & 0.078 & 48.42 & 15.1 & 132(66) \\ 
B1717-16 & 79.2 & 480 & 0.019 & 0.047 & 35.91 & 15.1 & 56(28) \\ 
\hline
B1717-29 & 49.8 & 960 & 0.065 & 0.124 & 65.03 & 24.01 & 355(177) \\
B1717-29 & 64.5 & 840 & 0.051 & 0.087 & 104.11 & 24.01 & 531(265) \\ 
B1717-29 & 79.2 & 840 & 0.041 & 0.074 & 91.65 & 24.01 & 416(208) \\ 
\hline
B1737+13 & 49.8 & 180 & 0.028 & 0.048 & 63.39 & 8.18 & 151(75) \\ 
B1737+13 & 64.5 & 360 & 0.016 & 0.032 & 100.91 & 8.18 & 128(64) \\ 
B1737+13 & 79.2 & 300 & 0.016 & 0.032 & 103.49 & 8.18 & 143(71) \\ 
\hline
B1747-46 & 35.1 & 240 & 0.02 & 0.03 & 34.83 & 110.51 & 853(426) \\ 
B1747-46 & 49.8 & 420 & 0.015 & 0.03 & 68.22 & 88.43 & 867(433) \\ 
B1747-46 & 64.5 & 360 & 0.015 & 0.03 & 67.71 & 88.43 & 929(464) \\ 
B1747-46 & 79.2 & 180 & 0.015 & 0.03 & 24.85 & 88.43 & 482(241) \\ 
\hline
B1749-28 & 64.5 & 600 & 0.086 & 0.158 & 238.21 & 22.54 & 1913(956) \\ 
B1749-28 & 79.2 & 600 & 0.042 & 0.079 & 270.73 & 22.54 & 1452(726) \\ 
\hline
B1821+05 & 49.8 & 420 & 0.038 & 0.075 & 100.37 & 9.2 & 215(107) \\ 
B1821+05 & 64.5 & 540 & 0.026 & 0.045 & 154.3 & 9.2 & 239(119) \\ 
B1821+05 & 79.2 & 600 & 0.015 & 0.03 & 146.67 & 9.2 & 162(81) \\ 
\hline
B1822-09 & 35.1 & 180 & 0.129 & 0.231 & 134.23 & 15.7 & 1455(727) \\ 
B1822-09 & 49.8 & 240 & 0.075 & 0.138 & 182.25 & 12.57 & 1004(502) \\ 
B1822-09 & 64.5 & 120 & 0.027 & 0.054 & 156.09 & 12.57 & 703(351) \\ 
B1822-09 & 79.2 & 360 & 0.033 & 0.054 & 282.23 & 12.57 & 816(408) \\ 
\hline
B1831-04 & 79.2 & 1080 & 0.071 & 0.131 & 140.84 & 11.19 & 564(282) \\ 
\hline
B1839+56 & 35.1 & 360 & 0.033 & 0.066 & 179.54 & 10.5 & 292(146) \\ 
B1839+56 & 49.8 & 180 & 0.02 & 0.033 & 211.83 & 8.4 & 301(150) \\ 
B1839+56 & 64.5 & 180 & 0.007 & 0.017 & 231.89 & 8.4 & 189(94) \\ 
B1839+56 & 79.2 & 180 & 0.007 & 0.017 & 134.14 & 8.4 & 109(54) \\ 
\hline
B1842+14 & 35.1 & 420 & 0.086 & 0.161 & 192.43 & 9.96 & 1054(527) \\ 
B1842+14 & 49.8 & 600 & 0.042 & 0.079 & 310.35 & 7.97 & 743(371) \\ 
B1842+14 & 64.5 & 360 & 0.022 & 0.041 & 334.79 & 7.97 & 726(363) \\ 
B1842+14 & 79.2 & 300 & 0.016 & 0.03 & 283.92 & 7.97 & 571(285) \\ 
\hline
B1857-26 & 64.5 & 720 & 0.127 & 0.233 & 202.1 & 20.68 & 1640(820) \\ 
B1857-26 & 79.2 & 1200 & 0.069 & 0.129 & 206.94 & 20.68 & 909(454) \\ 
\hline
B1905+39 & 79.2 & 480 & 0.033 & 0.074 & 37.56 & 6.72 & 39(19) \\ 
\hline
B1911-04 & 64.5 & 600 & 0.139 & 0.256 & 128.42 & 11.24 & 546(273) \\ 
B1911-04 & 79.2 & 1200 & 0.093 & 0.173 & 203.76 & 11.24 & 486(243) \\ 
\hline
B1918+26 & 49.8 & 360 & 0.009 & 0.024 & 36.74 & 6.81 & 29(14) \\ 
B1918+26 & 64.5 & 600 & 0.006 & 0.024 & 62.25 & 6.81 & 32(16) \\ 
B1918+26 & 79.2 & 480 & 0.009 & 0.008 & 42.04 & 6.81 & 29(14) \\ 
\hline
B1919+21 & 35.1 & 60 & 0.027 & 0.053 & 400.81 & 9.05 & 1379(689) \\ 
B1919+21 & 49.8 & 90 & 0.021 & 0.053 & 867.92 & 7.24 & 1742(871) \\ 
B1919+21 & 64.5 & 90 & 0.021 & 0.04 & 1278.11 & 7.24 & 2565(1282) \\ 
B1919+21 & 79.2 & 90 & 0.027 & 0.04 & 1064.82 & 7.24 & 2394(1197) \\
\hline
B1929+10 & 35.1 & 120 & 0.019 & 0.032 & 107.38 & 10.56 & 638(319) \\ 
B1929+10 & 49.8 & 540 & 0.017 & 0.029 & 350.69 & 8.45 & 743(371) \\
B1929+10 & 64.5 & 300 & 0.015 & 0.027 & 340.16 & 8.45 & 909(454) \\
B1929+10 & 79.2 & 180 & 0.015 & 0.025 & 235.38 & 8.45 & 812(406) \\ 
\hline
B1940-12 & 49.8 & 240 & 0.057 & 0.097 & 31.04 & 13.54 & 140(70) \\ 
B1940-12 & 64.5 & 720 & 0.042 & 0.068 & 70.55 & 13.54 & 155(77) \\ 
B1940-12 & 79.2 & 600 & 0.026 & 0.039 & 56.5 & 13.54 & 107(53) \\ 
\hline
B1944+17 & 49.8 & 600 & 0.057 & 0.106 & 121.1 & 7.62 & 298(149) \\ 
B1944+17 & 64.5 & 960 & 0.046 & 0.088 & 145.95 & 7.62 & 253(126) \\
B1944+17 & 79.2 & 720 & 0.05 & 0.088 & 80.72 & 7.62 & 168(84) \\ 
\hline
B2016+28 & 35.1 & 240 & 0.015 & 0.073 & 99.86 & 8.34 & 184(92) \\ 
B2016+28 & 49.8 & 600 & 0.011 & 0.039 & 337.56 & 6.67 & 270(135) \\
B2016+28 & 64.5 & 360 & 0.007 & 0.022 & 300.6 & 6.67 & 240(120) \\ 
B2016+28 & 79.2 & 360 & 0.007 & 0.011 & 286.69 & 6.67 & 229(114) \\ 
\hline
B2020+28 & 35.1 & 600 & 0.015 & 0.027 & 45.27 & 8.3 & 67(33) \\ 
B2020+28 & 49.8 & 540 & 0.007 & 0.021 & 136.54 & 6.64 & 115(57) \\ 
B2020+28 & 64.5 & 480 & 0.007 & 0.017 & 150.37 & 6.64 & 134(67) \\ 
B2020+28 & 79.2 & 480 & 0.004 & 0.017 & 157.22 & 6.64 & 108(54) \\
\hline
B2021+51 & 64.5 & 600 & 0.023 & 0.042 & 56.76 & 7.83 & 79(39) \\ 
B2021+51 & 79.2 & 960 & 0.014 & 0.026 & 66.11 & 7.83 & 57(28) \\ 
\hline
B2022+50 & 79.2 & 720 & 0.022 & 0.037 & 84.53 & 7.69 & 125(62) \\
\hline
B2043-04 & 64.5 & 720 & 0.079 & 0.124 & 41.73 & 11.16 & 82(41) \\ 
B2043-04 & 79.2 & 360 & 0.031 & 0.062 & 22.56 & 11.16 & 39(19) \\
\hline
B2045-16 & 49.8 & 240 & 0.137 & 0.177 & 70.71 & 14.94 & 385(192) \\ 
B2045-16 & 64.5 & 360 & 0.108 & 0.196 & 83.72 & 14.94 & 327(163) \\ 
B2045-16 & 79.2 & 360 & 0.061 & 0.137 & 61.04 & 14.94 & 177(88) \\
\hline
B2053+21 & 49.8 & 360 & 0.054 & 0.106 & 58.17 & 7.21 & 121(60) \\ 
B2053+21 & 64.5 & 360 & 0.042 & 0.073 & 64.38 & 7.21 & 117(58) \\ 
B2053+21 & 79.2 & 240 & 0.029 & 0.057 & 33.22 & 7.21 & 60(30) \\
\hline
B2110+27 & 35.1 & 540 & 0.032 & 0.06 & 131.93 & 8.4 & 163(81) \\ 
B2110+27 & 49.8 & 180 & 0.024 & 0.048 & 155.01 & 6.72 & 228(114) \\ 
B2110+27 & 64.5 & 360 & 0.024 & 0.036 & 179.32 & 6.72 & 187(93) \\ 
B2110+27 & 79.2 & 300 & 0.014 & 0.036 & 159.39 & 6.72 & 140(70) \\
\hline
B2152-31 & 49.8 & 240 & 0.024 & 0.052 & 35.51 & 25.9 & 187(93) \\ 
B2152-31 & 64.5 & 480 & 0.012 & 0.041 & 53.5 & 25.9 & 143(71) \\ 
B2152-31 & 79.2 & 360 & 0.012 & 0.062 & 51.25 & 25.9 & 158(79) \\
\hline
B2154+40 & 79.2 & 360 & 0.16 & 0.275 & 74.18 & 6.74 & 186(93) \\
\hline
B2217+47 & 35.1 & 300 & 0.12 & 0.232 & 774.94 & 9.27 & 4581(2290) \\ 
B2217+47 & 49.8 & 90 & 0.046 & 0.124 & 748.13 & 7.42 & 3699(1849) \\ 
B2217+47 & 64.5 & 90 & 0.023 & 0.07 & 687.17 & 7.42 & 2348(1174) \\ 
B2217+47 & 79.2 & 90 & 0.012 & 0.032 & 665.16 & 7.42 & 1645(822) \\
\hline
B2224+65 & 35.1 & 120 & 0.067 & 0.123 & 61.77 & 12.22 & 468(234) \\ 
B2224+65 & 49.8 & 60 & 0.061 & 0.116 & 131.7 & 9.78 & 1077(538) \\ 
B2224+65 & 64.5 & 120 & 0.045 & 0.075 & 139.01 & 9.78 & 679(339) \\ 
B2224+65 & 79.2 & 60 & 0.029 & 0.048 & 68.86 & 9.78 & 379(189) \\ 
\hline
B2303+30 & 49.8 & 480 & 0.068 & 0.11 & 45.63 & 6.48 & 59(29) \\ 
B2303+30 & 64.5 & 840 & 0.043 & 0.079 & 75.09 & 6.48 & 57(28) \\ 
B2303+30 & 79.2 & 480 & 0.032 & 0.063 & 56.62 & 6.48 & 49(24) \\ 
\hline
B2306+55 & 64.5 & 480 & 0.028 & 0.052 & 64.02 & 8.3 & 125(62) \\ 
B2306+55 & 79.2 & 240 & 0.013 & 0.024 & 40.36 & 8.3 & 74(37) \\
\hline
B2310+42 & 49.8 & 840 & 0.026 & 0.049 & 116.74 & 6.96 & 163(81) \\ 
B2310+42 & 64.5 & 720 & 0.018 & 0.035 & 83.39 & 6.96 & 103(51) \\ 
B2310+42 & 79.2 & 1080 & 0.015 & 0.024 & 103.96 & 6.96 & 96(48) \\
\hline
B2315+21 & 35.1 & 360 & 0.051 & 0.087 & 34.14 & 9.05 & 63(31) \\ 
B2315+21 & 49.8 & 360 & 0.017 & 0.029 & 37.17 & 7.24 & 32(16) \\ 
B2315+21 & 64.5 & 540 & 0.017 & 0.029 & 38.18 & 7.24 & 27(13) \\ 
B2315+21 & 79.2 & 360 & 0.017 & 0.029 & 44.61 & 7.24 & 38(19) \\
\hline
B2327-20 & 35.1 & 240 & 0.013 & 0.033 & 51.32 & 20.92 & 128(64) \\ 
B2327-20 & 49.8 & 540 & 0.013 & 0.033 & 122.54 & 16.74 & 163(81) \\ 
B2327-20 & 64.5 & 540 & 0.013 & 0.033 & 97.23 & 16.74 & 129(64) \\ 
B2327-20 & 79.2 & 240 & 0.02 & 0.033 & 38.18 & 16.74 & 93(46) \\
\hline
B2334+61 & 49.8 & 240 & 0.072 & 0.129 & 100.04 & 9.17 & 502(251) \\ 
B2334+61 & 64.5 & 600 & 0.033 & 0.064 & 170.61 & 9.17 & 350(175) \\ 
B2334+61 & 79.2 & 540 & 0.025 & 0.045 & 134.08 & 9.17 & 252(126) \\
\hline
J0030+0451 & 35.1 & 360 & 0.001 & 0.002 & 136.61 & 11.67 & 885(442) \\ 
J0030+0451 & 49.8 & 1200 & 0.001 & 0.001 & 346.04 & 9.34 & 814(407) \\ 
J0030+0451 & 64.5 & 1200 & 0.001 & 0.001 & 265.7 & 9.34 & 625(312) \\ 
J0030+0451 & 79.2 & 480 & 0.001 & 0.001 & 152.8 & 9.34 & 535(267) \\ 
\hline
J0034-0534 & 49.8 & 360 & 0.001 & 0.002 & 2032.51 & 11.44 & 29334(14667) \\ 
J0034-0534 & 64.5 & 360 & 0.001 & 0.002 & 2080.19 & 11.44 & 25250(12625) \\ 
J0034-0534 & 79.2 & 360 & 0.001 & 0.001 & 1142.78 & 11.44 & 12041(6020) \\
\hline
J0051+0423 & 35.1 & 360 & 0.007 & 0.018 & 105.51 & 11.77 & 192(96) \\ 
J0051+0423 & 49.8 & 180 & 0.004 & 0.014 & 94.15 & 9.42 & 150(75) \\ 
J0051+0423 & 64.5 & 300 & 0.004 & 0.014 & 97.25 & 9.42 & 120(60) \\ 
J0051+0423 & 79.2 & 480 & 0.004 & 0.011 & 74.84 & 9.42 & 73(36) \\
\hline
J0242+62 & 35.1 & 960 & 0.025 & 0.041 & 143.17 & 11.65 & 235(117) \\ 
J0242+62 & 49.8 & 1200 & 0.025 & 0.047 & 237.69 & 9.32 & 279(139) \\ 
J0242+62 & 64.5 & 480 & 0.025 & 0.041 & 141.28 & 9.32 & 262(131) \\ 
J0242+62 & 79.2 & 840 & 0.025 & 0.041 & 73.92 & 9.32 & 103(51) \\
\hline
J0459-0210 & 49.8 & 600 & 0.075 & 0.215 & 214.15 & 10.67 & 511(255) \\ 
J0459-0210 & 64.5 & 600 & 0.044 & 0.147 & 202.71 & 10.67 & 366(183) \\ 
J0459-0210 & 79.2 & 420 & 0.026 & 0.102 & 103.6 & 10.67 & 170(85) \\ 
\hline
J0611+30 & 49.8 & 360 & 0.138 & 0.254 & 74.1 & 7.02 & 186(93) \\ 
J0611+30 & 64.5 & 1200 & 0.127 & 0.226 & 183.79 & 7.02 & 241(120) \\ 
J0611+30 & 79.2 & 1200 & 0.093 & 0.184 & 103.46 & 7.02 & 114(57) \\
\hline
J0613+3731 & 35.1 & 480 & 0.022 & 0.037 & 81.12 & 8.15 & 118(59) \\ 
J0613+3731 & 49.8 & 600 & 0.017 & 0.025 & 158.88 & 6.52 & 145(72) \\ 
J0613+3731 & 64.5 & 600 & 0.012 & 0.025 & 130.82 & 6.52 & 102(51) \\ 
J0613+3731 & 79.2 & 600 & 0.012 & 0.025 & 110.13 & 6.52 & 86(43) \\
\hline
J0815+4611 & 49.8 & 360 & 0.022 & 0.039 & 48.65 & 7.24 & 88(44) \\ 
J0815+4611 & 64.5 & 360 & 0.019 & 0.035 & 50.42 & 7.24 & 84(42) \\ 
J0815+4611 & 79.2 & 240 & 0.012 & 0.026 & 25.18 & 7.24 & 40(20) \\ 
\hline
J1022+1001 & 35.1 & 1200 & 0.003 & 0.005 & 109.49 & 10.75 & 323(161) \\ 
J1022+1001 & 49.8 & 600 & 0.002 & 0.003 & 136.59 & 8.6 & 338(169) \\ 
J1022+1001 & 64.5 & 480 & 0.002 & 0.003 & 77.34 & 8.6 & 214(107) \\ 
J1022+1001 & 79.2 & 480 & 0.002 & 0.003 & 87.12 & 8.6 & 241(120) \\ 
\hline
J1313+0931 & 49.8 & 360 & 0.01 & 0.034 & 28.71 & 8.67 & 29(14) \\ 
J1313+0931 & 64.5 & 360 & 0.003 & 0.025 & 37.37 & 8.67 & 22(11) \\ 
J1313+0931 & 79.2 & 1080 & 0.003 & 0.0 & 158.91 & 8.67 & 54(27) \\
\hline
J1327+3423 & 35.1 & 240 & 0.001 & 0.003 & 64.59 & 7.86 & 128(64) \\ 
J1327+3423 & 49.8 & 300 & 0.001 & 0.002 & 192.97 & 6.29 & 275(137) \\ 
J1327+3423 & 64.5 & 480 & 0.001 & 0.002 & 221.69 & 6.29 & 218(109) \\ 
J1327+3423 & 79.2 & 540 & 0.001 & 0.002 & 207.14 & 6.29 & 192(96) \\
\hline
J1400-1431 & 35.1 & 480 & 0.001 & 0.001 & 350.43 & 17.87 & 3511(1755) \\ 
J1400-1431 & 49.8 & 600 & 0.0 & 0.001 & 405.38 & 14.3 & 1944(972) \\ 
J1400-1431 & 64.5 & 480 & 0.0 & 0.0 & 305.21 & 14.3 & 1227(613) \\ 
J1400-1431 & 79.2 & 360 & 0.0 & 0.0 & 147.81 & 14.3 & 686(343) \\ 
\hline
J1645+1012 & 49.8 & 1200 & 0.024 & 0.045 & 47.93 & 8.57 & 61(30) \\ 
J1645+1012 & 64.5 & 720 & 0.018 & 0.029 & 91.6 & 8.57 & 127(63) \\ 
J1645+1012 & 79.2 & 480 & 0.014 & 0.025 & 68.32 & 8.57 & 104(52) \\ 
\hline
J1741+2758 & 35.1 & 480 & 0.027 & 0.054 & 33.21 & 8.4 & 37(18) \\ 
J1741+2758 & 49.8 & 1200 & 0.011 & 0.054 & 121.19 & 6.72 & 43(21) \\ 
J1741+2758 & 64.5 & 1200 & 0.005 & 0.0 & 256.54 & 6.72 & 65(32) \\ 
J1741+2758 & 79.2 & 960 & 0.011 & 0.027 & 71.09 & 6.72 & 28(14) \\
\hline
J1758+3030 & 35.1 & 600 & 0.056 & 0.104 & 46.46 & 8.16 & 79(39) \\ 
J1758+3030 & 49.8 & 420 & 0.033 & 0.066 & 120.83 & 6.53 & 151(75) \\ 
J1758+3030 & 64.5 & 540 & 0.026 & 0.047 & 106.44 & 6.53 & 102(51) \\
J1758+3030 & 79.2 & 420 & 0.026 & 0.038 & 75.58 & 6.53 & 82(41) \\ 
\hline
J1929+00 & 49.8 & 480 & 0.077 & 0.152 & 56.21 & 10.14 & 142(71) \\ 
J1929+00 & 64.5 & 960 & 0.06 & 0.105 & 100.48 & 10.14 & 157(78) \\ 
J1929+00 & 79.2 & 1080 & 0.032 & 0.058 & 90.89 & 10.14 & 96(48) \\
\hline
J2043+2740 & 49.8 & 180 & 0.006 & 0.012 & 53.94 & 6.74 & 148(74) \\ 
J2043+2740 & 64.5 & 240 & 0.005 & 0.009 & 45.05 & 6.74 & 93(46) \\ 
J2043+2740 & 79.2 & 240 & 0.003 & 0.007 & 41.36 & 6.74 & 70(35) \\ 
\hline
J2145-0750 & 35.1 & 720 & 0.003 & 0.005 & 219.31 & 15.04 & 1139(569) \\ 
J2145-0750 & 49.8 & 1200 & 0.002 & 0.003 & 448.38 & 12.04 & 1100(550) \\ 
J2145-0750 & 64.5 & 720 & 0.001 & 0.002 & 360.57 & 12.04 & 996(498) \\ 
J2145-0750 & 79.2 & 600 & 0.001 & 0.002 & 271.15 & 12.04 & 821(410) \\ 
\hline
J2208+4056 & 35.1 & 600 & 0.042 & 0.076 & 269.94 & 8.49 & 512(256) \\ 
J2208+4056 & 49.8 & 600 & 0.032 & 0.064 & 345.93 & 6.79 & 458(229) \\ 
J2208+4056 & 64.5 & 600 & 0.032 & 0.057 & 238.35 & 6.79 & 315(157) \\ 
J2208+4056 & 79.2 & 360 & 0.032 & 0.057 & 96.59 & 6.79 & 165(82) \\ 
\hline
J2227+30 & 35.1 & 1200 & 0.017 & 0.042 & 52.54 & 8.14 & 36(18) \\ 
J2227+30 & 49.8 & 1200 & 0.019 & 0.051 & 92.11 & 6.51 & 54(27) \\ 
J2227+30 & 64.5 & 1200 & 0.017 & 0.034 & 44.93 & 6.51 & 24(12) \\ 
J2227+30 & 79.2 & 1200 & 0.019 & 0.034 & 37.78 & 6.51 & 22(11) \\
\hline
J2234+2114 & 64.5 & 480 & 0.037 & 0.068 & 33.86 & 7.3 & 38(19) \\ 
J2234+2114 & 79.2 & 600 & 0.037 & 0.068 & 26.22 & 7.3 & 26(13) \\ 
\enddata
\tablenotetext{a}{$\nu$, T and S/N are the center frequency, total integration time and signal-to-noise ratio of the pulse profile}
\tablenotetext{b}{\textit{W}$_{50}$ and \textit{W}$_{10}$ are the full-width at half max and at 10 $\%$.}
\tablenotetext{c}{S$_{\nu}$ is the flux density calculated using \textit{W}$_{50}$.}
\end{deluxetable*}
\clearpage
\begin{figure*}[htbp!]
\centering
\includegraphics[width=\textwidth,height=21cm]{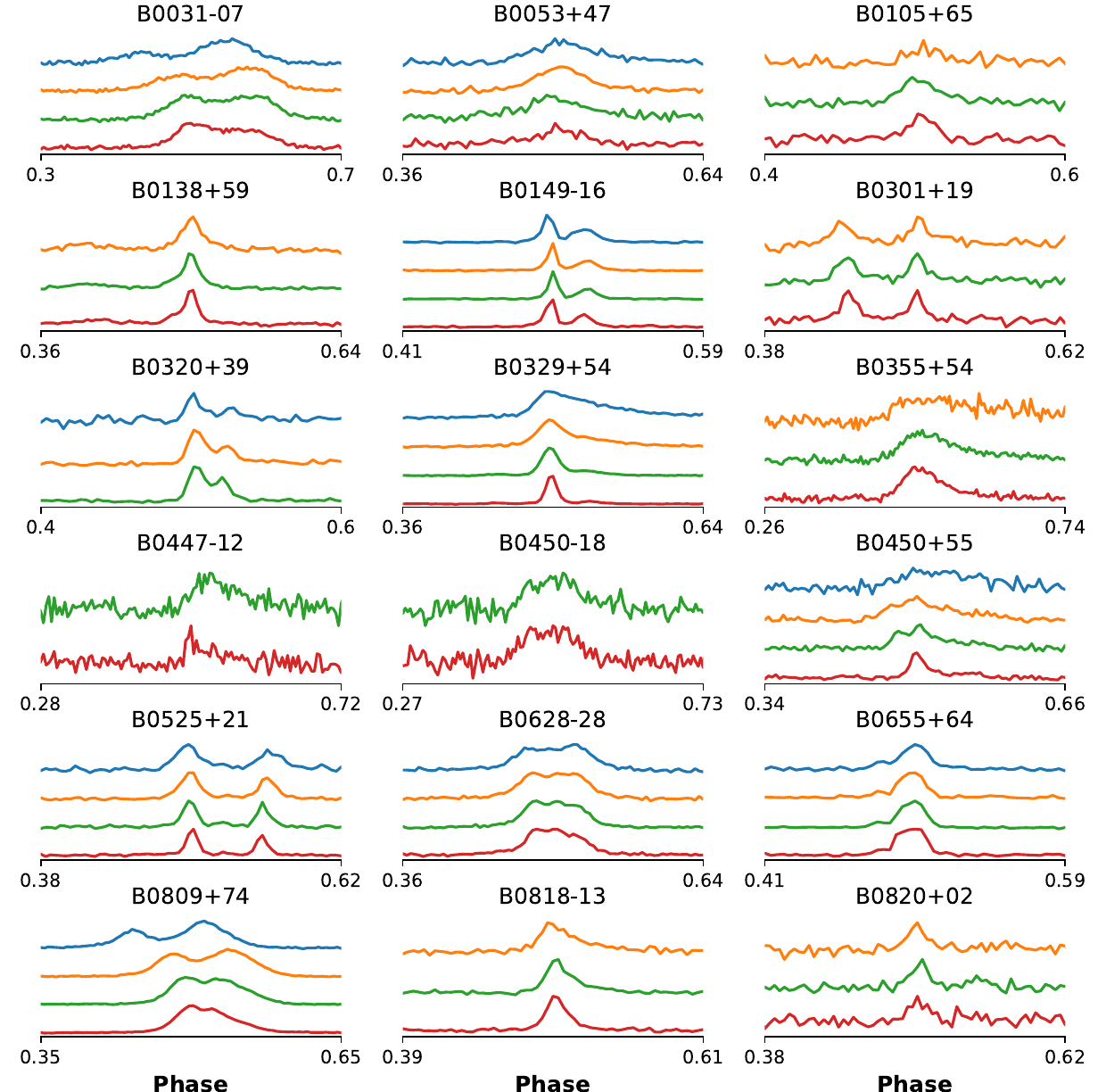}
\caption{Average integrated pulse profiles, zooming in on the region of emission, for pulsars detected by LWA at multiple observing frequencies. The marker of each subplots horizontal axis indicates the phase range shown. The different colors cyan, orange, green and red correspond to center frequencies of 35.1, 49.8, 64.5 and 79.2 MHz respectively.}
\label{fig:psrprof}
\end{figure*}
\begin{figure*}[htbp!]
\centering
\includegraphics[width=\textwidth,height=22cm]{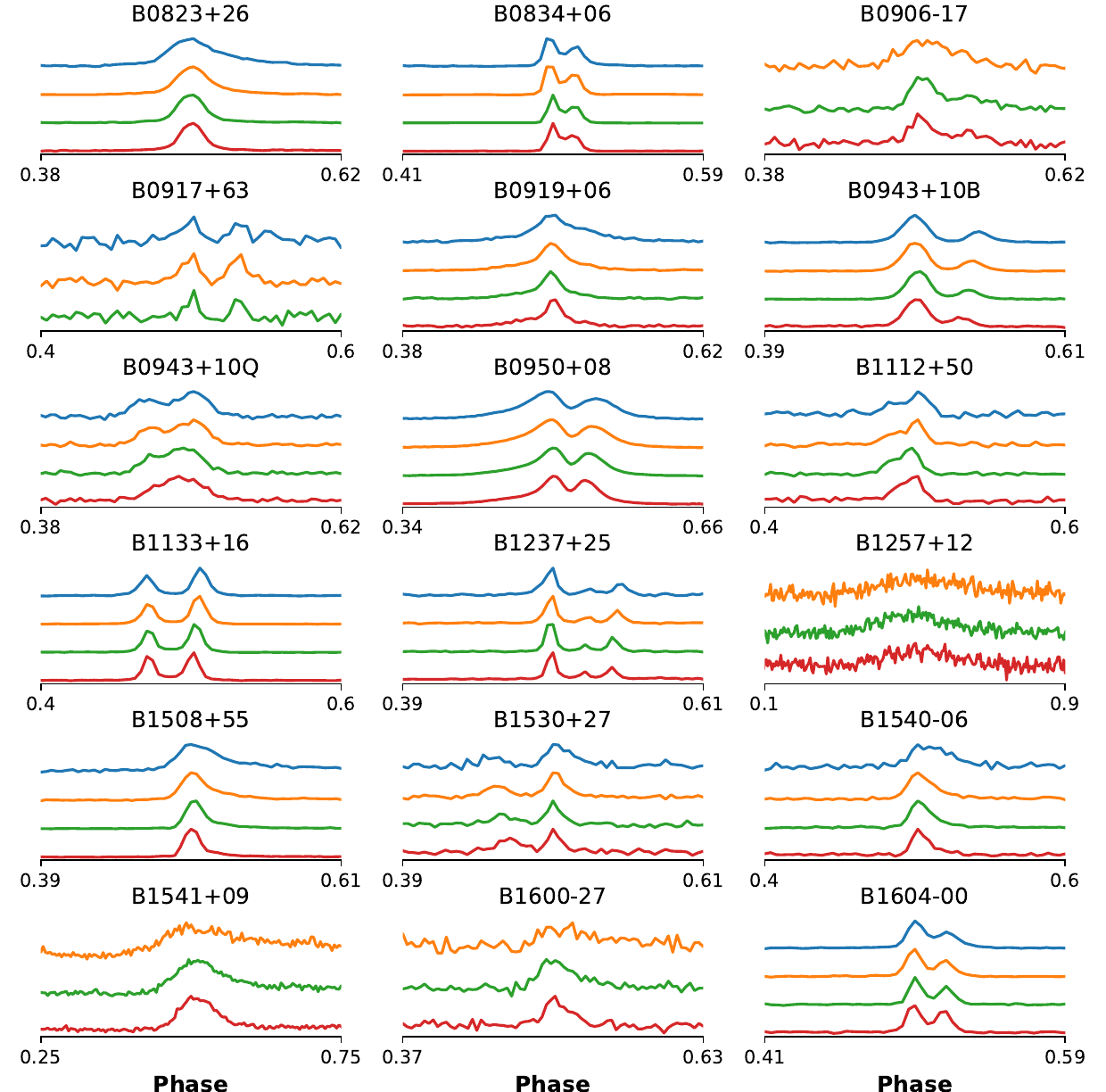}
\caption{Fig \ref{fig:psrprof} Continued}
\end{figure*}
\begin{figure*}[htbp!]
\centering
\includegraphics[width=\textwidth,height=22cm]{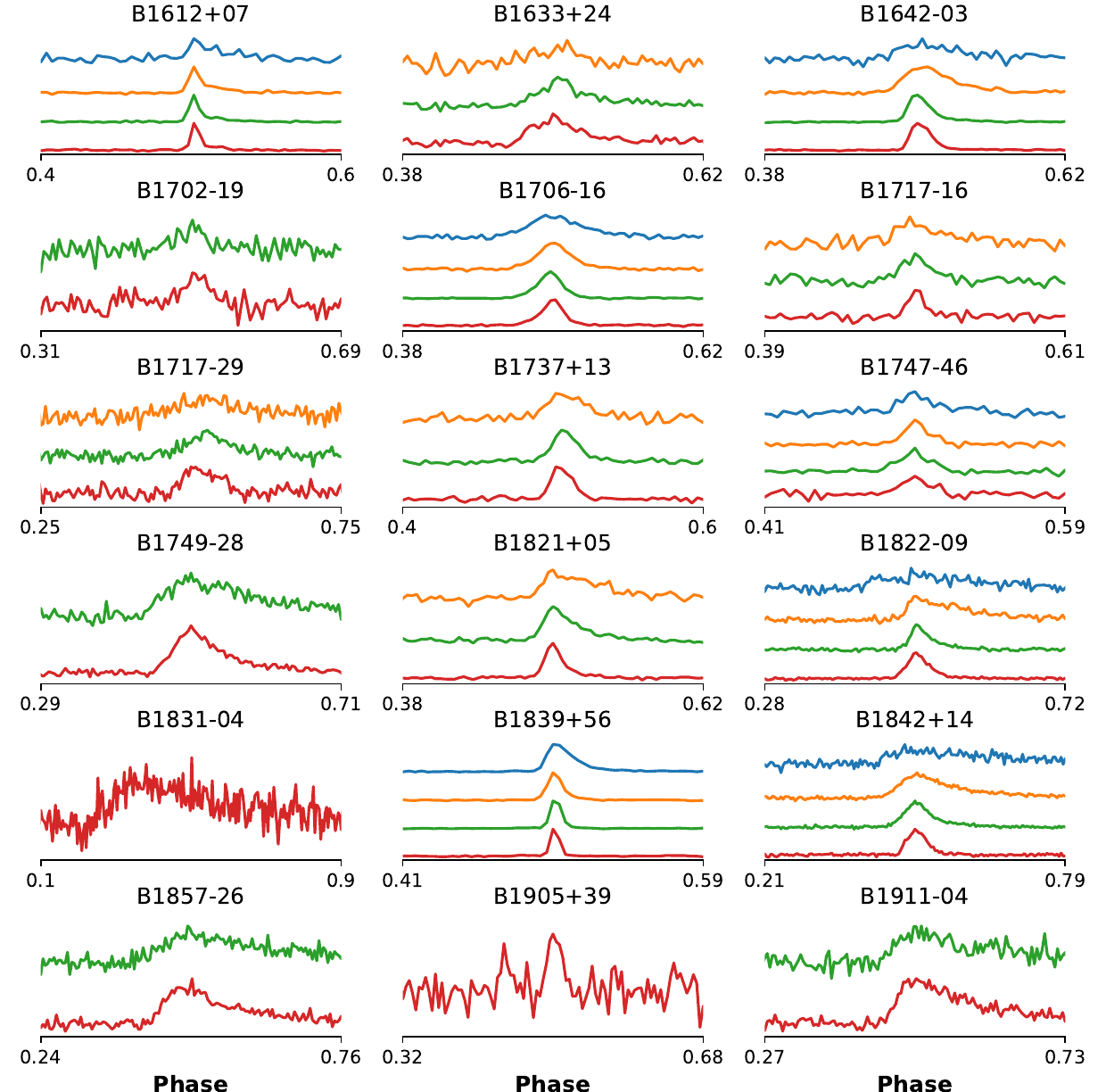}
\caption{Fig \ref{fig:psrprof} Continued}
\end{figure*}
\begin{figure*}[htbp!]
\centering
\includegraphics[width=\textwidth,height=22cm]{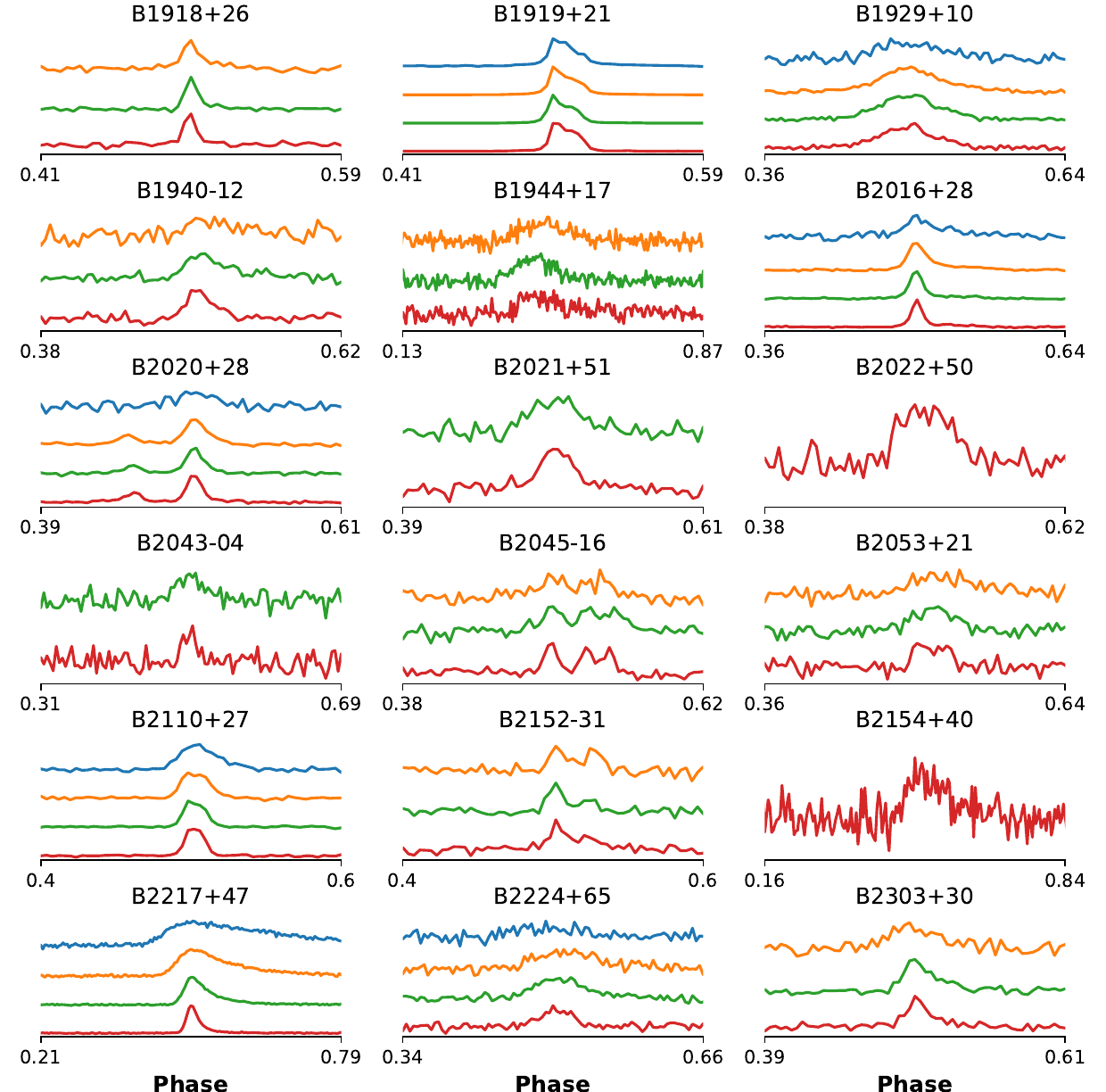}
\caption{Fig \ref{fig:psrprof} Continued}
\end{figure*}
\begin{figure*}[htbp!]
\centering
\includegraphics[width=\textwidth,height=22cm]{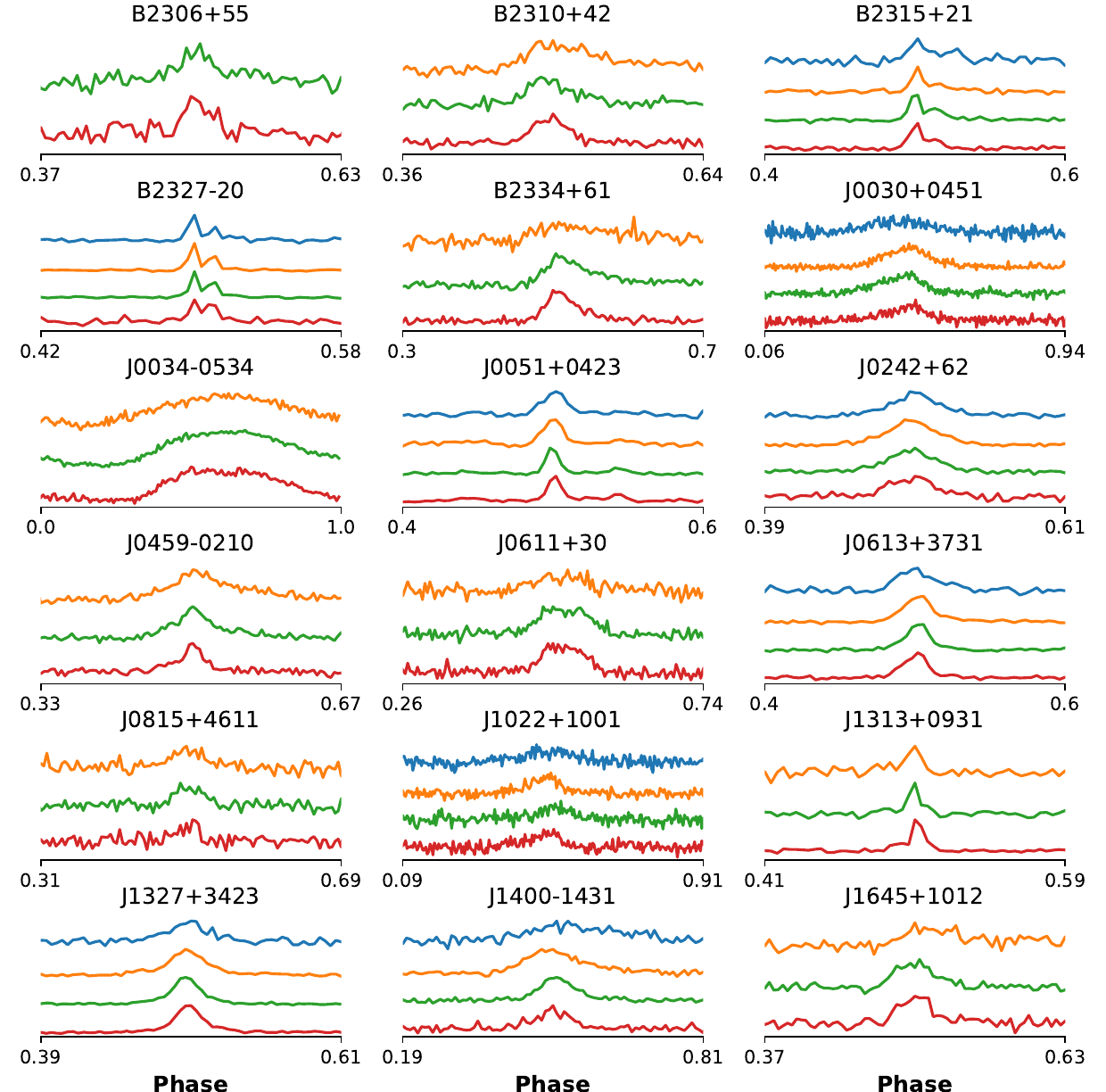}
\caption{Fig \ref{fig:psrprof} Continued}
\end{figure*}
\begin{figure*}[htbp!]
\centering
\includegraphics[width=\textwidth,height=22cm]{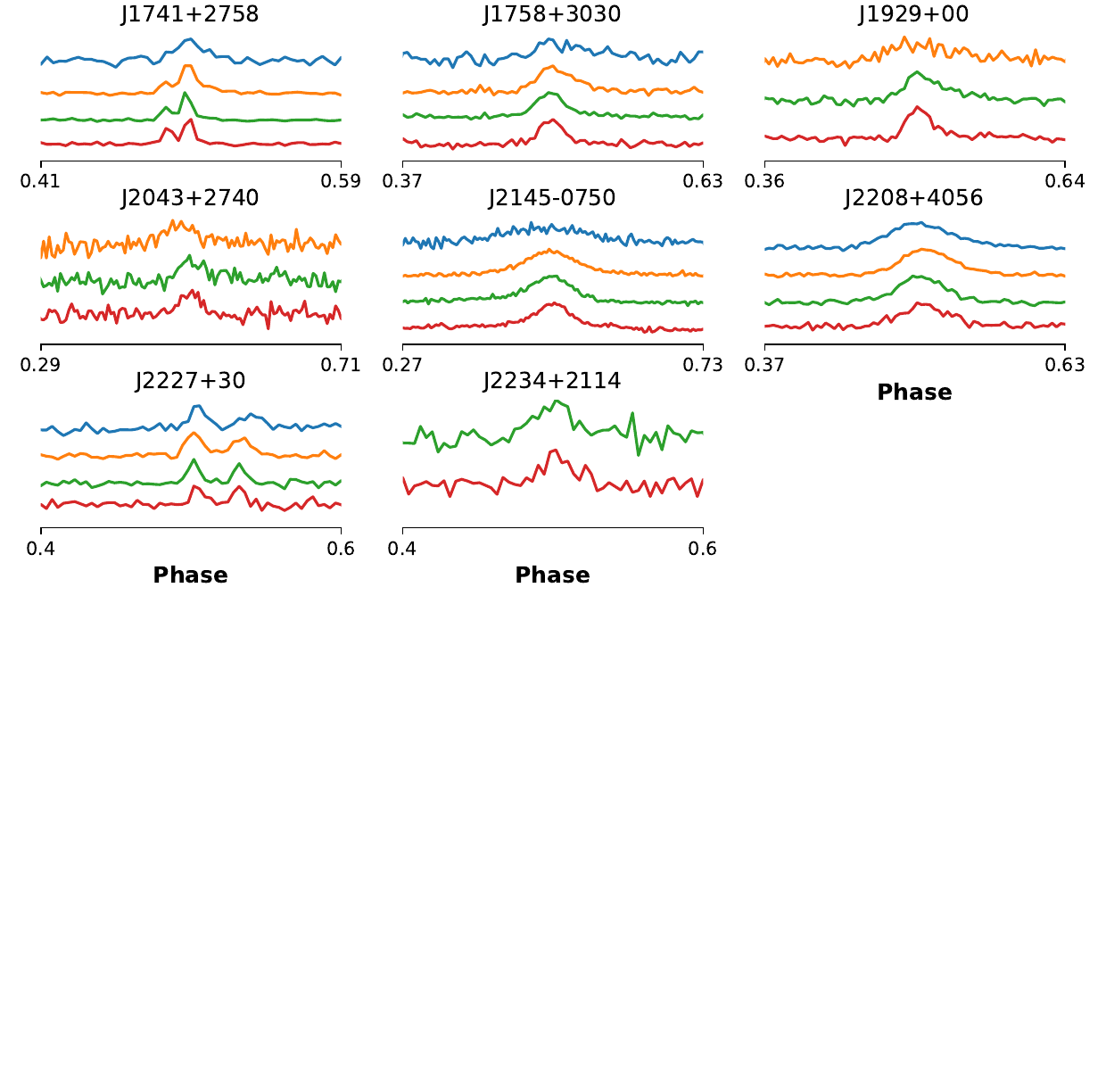}
\caption{Fig \ref{fig:psrprof} Continued}
\end{figure*}
\clearpage

\section{LWA Pulsar Spectra}
\startlongtable
\begin{deluxetable*}{c@{\hskip 0.2in}c@{\hskip 0.2in}c@{\hskip 0.2in}c@{\hskip 0.2in}c@{\hskip 0.2in}c@{\hskip 0.2in}c@{\hskip 0.2in}c@{\hskip 0.2in}c}
\tablecolumns{9}
\tabletypesize{\small}
\tablewidth{0pt}
\tablecaption{ \label{tab:spectra}}
\tablehead{
    \colhead{Name} & \colhead{$\alpha$} & \colhead{$\beta$} & \colhead{$\nu_0$} & \colhead{S$_0$} & \colhead{$\nu_{\mathrm{m}}$} &  \colhead{$\alpha_{L}$} & \colhead{$\nu_{R}$} & \colhead{References}\\
    \colhead{} & \colhead{} & \colhead{} & \colhead{(MHz)} & \colhead{(mJy)} & \colhead{(MHz)} &\colhead{} & \colhead{(GHz)} & \colhead{}
}

\startdata
B0031-07 & ... & ... & ... & ... & ... & -2.24(27) & 0.035-1.4& 4, 5  \\
B0053+47 & -1.23(16) & -0.78(18) & 100 & 75(22) & 45(9) & -1.34(24) & 0.02-0.6& 1,  2,  3,  4,  5,  6,  9, 11  \\
B0105+65 & -0.99(13) & -0.26(6) & 100 & 60(5) & 14(7) & -1.48(10) & 0.05-1.4& 1, 3, 4, 5, 6, 9  \\
B0138+59 & -0.7(11) & -0.28(5) & 100 & 207(19) & 28(8) & -1.25(10) & 0.05-1.4& 3, 5, 9  \\
B0149-16 & -1.19(10) & -0.16(4) & 100 & 147(7) & 2(1) & -1.56(7) & 0.035-1.4& 5  \\
B0301+19 & -0.01(24) & -0.18(11) & 100 & 40(9) & 97(64) & -0.32(15) & 0.025-1.4& 1, 2, 4, 5, 8  \\
B0320+39 & -0.27(12) & -0.67(7) & 100 & 145(17) & 81(7) & -1.29(17) & 0.025-1.4& 1,  2,  3,  4,  5,  6,  9, 11  \\
B0329+54 & 0.65(19) & -0.51(8) & 100 & 1577(316) & 189(39) & -0.31(23) & 0.025-5.0& 3,  4,  5, 10, 11  \\
B0355+54 & ... & ... & ... & ... & ... & -0.84(9) & 0.05-5.0& 3, 5, 9  \\
B0447-12 & ... & ... & ... & ... & ... & -1.6(9) & 0.064-1.4& 4, 5  \\
B0450-18 & ... & ... & ... & ... & ... & -1.0(17) & 0.064-1.4& 5  \\
B0450+55 & -0.65(17) & -0.07(7) & 100 & 108(18) & 0(0) & -0.81(10) & 0.025-5.0& 1,  2,  3,  4,  5,  6,  9, 10, 11  \\
B0525+21 & -0.64(18) & -0.23(7) & 100 & 203(34) & 24(13) & -1.13(13) & 0.035-5.0& 1, 3, 4, 5, 9  \\
B0628-28 & ... & ... & ... & ... & ... & -1.35(24) & 0.035-1.4& 5, 7  \\
B0655+64 & -1.31(24) & -0.39(14) & 100 & 105(21) & 18(12) & -1.89(17) & 0.035-1.4& 1,  2,  3,  5,  6,  9, 10  \\
B0809+74 & -1.03(20) & -0.27(11) & 100 & 880(270) & 14(12) & -1.44(13) & 0.02-1.4& 1,  2,  3,  4,  5,  6, 10, 11  \\
B0818-13 & -0.1(23) & -0.48(8) & 100 & 255(45) & 90(21) & -1.31(18) & 0.05-5.0& 5, 7  \\
B0820+02 & -0.27(42) & -0.28(19) & 100 & 50(13) & 61(49) & -0.84(20) & 0.05-1.4& 4, 5, 6, 7, 9  \\
B0823+26 & -0.47(16) & -0.32(8) & 100 & 335(66) & 47(14) & -0.9(14) & 0.02-5.0& 1,  2,  3,  4,  5,  6,  9, 10, 11  \\
B0834+06 & -0.42(14) & -0.7(6) & 100 & 1215(105) & 74(7) & -1.61(33) & 0.02-1.4& 3,  4,  5,  7, 10, 11  \\
B0906-17 & -1.31(34) & -0.04(16) & 100 & 110(30) & 0(0) & -1.38(18) & 0.05-1.4& 4, 5, 7  \\
B0917+63 & -0.62(13) & -0.34(8) & 100 & 23(3) & 40(11) & -1.03(15) & 0.025-1.4& 1, 2, 4, 5, 6, 8, 9  \\
B0919+06 & -0.87(16) & -0.22(10) & 100 & 261(61) & 13(12) & -1.12(12) & 0.02-1.4& 3,  4,  5,  6,  7, 10, 11  \\
B0943+10 & -1.47(15) & -1.55(16) & 100 & 712(183) & 62(4) & -1.51(41) & 0.02-0.6& 1,  2,  3,  4,  5,  9, 10, 11  \\
B0950+08 & 0.15(20) & -0.43(6) & 100 & 2347(273) & 119(27) & -1.06(23) & 0.02-5.0& 3,  4,  5,  7,  9, 10, 11  \\
B1112+50 & -0.37(13) & -0.53(11) & 100 & 55(11) & 70(9) & -0.77(16) & 0.02-1.4& 1,  2,  3,  4,  5,  6,  9, 11  \\
B1133+16 & 0.25(17) & -0.68(7) & 100 & 1919(211) & 120(15) & -0.65(34) & 0.02-5.0& 1,  2,  3,  4,  5,  7,  9, 10, 11  \\
B1237+25 & 0.39(9) & -0.45(4) & 100 & 154(17) & 154(16) & -0.47(16) & 0.02-5.0& 1,  2,  3,  4,  5,  9, 10, 11  \\
B1257+12 & -0.43(63) & -0.76(26) & 100 & 232(79) & 75(31) & -2.18(23) & 0.05-1.4& 4, 5, 6, 9  \\
B1508+55 & -0.33(19) & -0.36(5) & 100 & 314(65) & 63(17) & -1.38(20) & 0.02-5.0& 1,  2,  3,  4,  5,  6, 10, 11  \\
B1530+27 & -0.43(10) & -0.49(6) & 100 & 68(6) & 64(7) & -1.08(14) & 0.025-1.4& 1,  2,  3,  4,  5,  9, 11  \\
B1540-06 & -0.71(20) & -0.4(11) & 100 & 161(25) & 41(14) & -1.36(14) & 0.035-1.4& 4, 5  \\
B1541+09 & -0.9(16) & -0.31(7) & 100 & 465(63) & 23(9) & -1.53(13) & 0.05-1.4& 1,  2,  3,  5,  9, 10  \\
B1600-27 & -0.93(33) & -0.23(11) & 100 & 119(35) & 13(15) & -1.52(20) & 0.05-1.4& 5  \\
B1604-00 & -0.8(12) & -0.5(5) & 100 & 464(44) & 44(6) & -1.83(19) & 0.035-1.4& 4, 5  \\
B1612+07 & -0.68(14) & -0.55(10) & 100 & 68(9) & 53(8) & -1.27(16) & 0.025-1.4& 3,  4,  5,  9, 11  \\
B1633+24 & -0.63(15) & -0.59(9) & 100 & 70(9) & 58(8) & -1.4(17) & 0.025-1.4& 1,  2,  3,  4,  5,  9, 11  \\
B1642-03 & -0.26(54) & -0.56(16) & 100 & 2460(939) & 79(38) & -2.09(14) & 0.035-5.0& 4, 5  \\
B1702-19 & ... & ... & ... & ... & ... & -1.21(13) & 0.064-5.0& 4, 5  \\
B1706-16 & -1.24(44) & -0.18(17) & 100 & 545(156) & 3(10) & -1.65(18) & 0.035-1.4& 5  \\
B1717-16 & -0.92(14) & -0.26(6) & 100 & 75(8) & 17(8) & -1.44(12) & 0.05-1.4& 5, 7  \\
B1717-29 & -1.15(10) & -0.29(3) & 100 & 265(27) & 13(3) & -2.19(13) & 0.05-1.4& 5  \\
B1737+13 & -0.46(9) & -0.35(4) & 100 & 94(6) & 51(7) & -1.14(11) & 0.05-1.4& 1, 2, 3, 4, 5, 9  \\
B1747-46 & -0.65(26) & -0.49(9) & 100 & 810(117) & 51(14) & -1.93(24) & 0.035-1.4& 5, 7  \\
B1749-28 & -0.64(41) & -0.43(15) & 100 & 4660(1032) & 47(25) & -1.84(11) & 0.064-5.0& 5, 7  \\
B1821+05 & -1.08(28) & -0.16(11) & 100 & 128(29) & 3(7) & -1.44(13) & 0.05-1.4& 5, 9  \\
B1822-09 & ... & ... & ... & ... & ... & -2.04(25) & 0.035-5.0& 5  \\
B1831-04 & ... & ... & ... & ... & ... & -1.05(5) & 0.079-1.4& 5  \\
B1839+56 & -0.55(13) & -0.32(9) & 100 & 94(17) & 42(13) & -0.87(13) & 0.02-1.4& 1,  2,  3,  4,  5,  6,  9, 10, 11  \\
B1842+14 & ... & ... & ... & ... & ... & -1.92(6) & 0.035-1.4& 1, 2, 3, 5, 9  \\
B1857-26 & -0.53(28) & -0.3(9) & 100 & 549(88) & 41(22) & -1.43(15) & 0.064-5.0& 5, 7  \\
B1905+39 & 0.55(17) & -0.65(6) & 100 & 38(5) & 152(20) & -1.29(28) & 0.079-1.4& 1, 5, 9  \\
B1911-04 & -0.17(45) & -0.53(16) & 100 & 335(114) & 85(36) & -1.61(22) & 0.064-1.4& 5, 7  \\
B1918+26 & -0.41(11) & -0.54(9) & 100 & 28(1) & 68(8) & -1.04(11) & 0.05-0.6& 1, 4, 5, 6, 9  \\
B1919+21 & -0.07(30) & -0.88(15) & 100 & 3188(653) & 96(16) & -1.02(45) & 0.02-1.4& 1,  2,  3,  4,  5, 10, 11  \\
B1929+10 & 0.18(14) & -0.34(4) & 100 & 379(39) & 130(27) & -0.61(20) & 0.02-5.0& 1,  2,  3,  4,  5, 11  \\
B1940-12 & -1.42(44) & -0.2(15) & 100 & 138(48) & 2(5) & -1.99(16) & 0.05-1.4& 5  \\
B1944+17 & -0.31(21) & -0.15(5) & 100 & 83(18) & 35(27) & -0.83(14) & 0.025-5.0& 1,  5,  9, 11  \\
B2016+28 & 0.04(26) & -0.4(7) & 100 & 443(97) & 105(34) & -1.18(22) & 0.025-5.0& 1,  2,  3,  4,  5,  6,  9, 10, 11  \\
B2020+28 & 0.08(19) & -0.37(9) & 100 & 163(28) & 111(28) & -0.47(18) & 0.025-5.0& 1, 2, 3, 4, 5, 6, 8, 9  \\
B2021+51 & 0.56(14) & -0.31(4) & 100 & 65(8) & 246(62) & -0.54(16) & 0.05-5.0& 1, 3, 4, 5, 9  \\
B2022+50 & ... & ... & ... & ... & ... & -1.27(11) & 0.06-1.4& 1, 2, 4, 5, 6, 9  \\
B2043-04 & -0.0(30) & -0.57(11) & 100 & 56(12) & 100(0) & -1.46(23) & 0.064-1.4& 4, 5  \\
B2045-16 & ... & ... & ... & ... & ... & -0.57(19) & 0.05-1.4& 5, 7  \\
B2053+21 & ... & ... & ... & ... & ... & -1.17(5) & 0.05-0.6& 1, 5, 6, 9  \\
B2110+27 & -0.79(8) & -0.38(4) & 100 & 117(7) & 35(5) & -1.45(11) & 0.025-1.4& 1,  3,  4,  5,  6,  9, 10, 11  \\
B2152-31 & -1.22(18) & -0.26(5) & 100 & 100(17) & 9(5) & -2.03(14) & 0.05-1.4& 5  \\
B2154+40 & -0.04(26) & -0.34(8) & 100 & 193(43) & 94(35) & -1.08(16) & 0.079-1.4& 1, 4, 5, 9  \\
B2217+47 & -1.4(16) & -0.28(10) & 100 & 1085(137) & 8(7) & -1.8(10) & 0.035-1.4& 1,  2,  3,  4,  5,  6,  9, 10  \\
B2224+65 & -1.48(16) & -0.06(7) & 100 & 178(24) & 0(0) & -1.6(8) & 0.035-1.4& 1,  2,  3,  5,  6,  9, 10  \\
B2303+30 & 0.02(24) & -0.57(10) & 100 & 59(11) & 101(21) & -1.12(21) & 0.05-1.4& 1, 2, 3, 5, 6, 9  \\
B2306+55 & -0.5(45) & -0.48(17) & 100 & 153(56) & 59(29) & -1.64(28) & 0.05-1.4& 1, 2, 3, 5, 6, 9  \\
B2310+42 & 0.34(11) & -0.47(6) & 100 & 122(12) & 143(17) & -0.37(15) & 0.025-1.4& 1,  2,  3,  4,  5,  9, 11  \\
B2315+21 & -0.17(15) & -0.55(10) & 100 & 50(6) & 85(11) & -0.85(16) & 0.025-1.4& 1, 2, 3, 4, 5, 6, 9, 11  \\
B2327-20 & -0.6(21) & -0.3(6) & 100 & 118(26) & 36(14) & -1.45(19) & 0.035-5.0& 5  \\
B2334+61 & ... & ... & ... & ... & ... & -1.65(9) & 0.05-1.4& 4, 5, 6, 9  \\
J0030+0451 & ... & ... & ... & ... & ... & -1.63(6) & 0.035-1.4& 3, 5, 6  \\
J0034-0534 & 0.23(140) & -1.23(62) & 100 & 337(178) & 109(62) & -2.39(55) & 0.05-1.4& 5, 6  \\
J0051+0423 & -0.49(30) & -0.53(15) & 100 & 70(22) & 62(19) & -1.46(17) & 0.025-1.4& 3,  5,  6,  9, 11  \\
J0242+62 & ... & ... & ... & ... & ... & -1.42(24) & 0.035-0.35& 6, 9  \\
J0459-0210 & ... & ... & ... & ... & ... & -2.04(4) & 0.05-1.4& 4, 5, 6  \\
J0611+30 & -1.47(33) & -0.97(40) & 100 & 67(16) & 46(16) & -1.97(33) & 0.05-0.4& 1, 2, 3, 5, 9  \\
J0613+3731 & -0.94(28) & -0.56(16) & 100 & 49(16) & 43(14) & -1.66(30) & 0.025-1.4& 5, 6, 8, 9  \\
J0815+4611 & ... & ... & ... & ... & ... & -1.75(16) & 0.05-0.4& 5, 6, 9  \\
J1022+1001 & -0.97(18) & -0.1(5) & 100 & 101(18) & 0(0) & -1.32(7) & 0.035-5.0& 4, 5, 9  \\
J1313+0931 & -0.85(19) & -0.4(5) & 100 & 24(4) & 34(9) & -2.2(17) & 0.05-1.4& 1, 2, 3, 5, 9  \\
J1327+3423 & ... & ... & ... & ... & ... & -0.26(41) & 0.035-0.135& 9  \\
J1400-1431 & -2.75(68) & 0.08(32) & 100 & 152(89) & 0(0) & -2.59(17) & 0.035-1.4& 5, 6  \\
J1645+1012 & -0.85(17) & -1.27(16) & 100 & 85(9) & 71(5) & -2.1(16) & 0.05-0.4& 1, 2, 3, 4, 5, 9  \\
J1741+2758 & -0.72(8) & -0.77(9) & 100 & 35(3) & 62(4) & -1.33(13) & 0.025-0.4& 1,  3,  4,  5,  9, 11  \\
J1758+3030 & -0.63(21) & -0.43(16) & 100 & 55(9) & 48(17) & -1.11(14) & 0.035-0.9& 1, 2, 3, 4, 5, 6, 9  \\
J1929+00 & ... & ... & ... & ... & ... & -1.57(24) & 0.05-0.15& 9  \\
J2043+2740 & -0.04(35) & -1.2(34) & 100 & 159(29) & 98(14) & -0.88(40) & 0.05-0.4& 1, 4, 5, 6, 9  \\
J2145-0750 & ... & ... & ... & ... & ... & -1.47(9) & 0.035-5.0& 4, 5, 7  \\
J2208+4056 & -2.76(38) & -1.62(33) & 100 & 129(20) & 42(8) & -1.14(52) & 0.025-0.15& 8  \\
J2227+30 & -1.09(44) & -0.83(37) & 100 & 22(5) & 51(20) & -0.3(35) & 0.025-0.35& 6, 8, 9  \\
J2234+2114 & ... & ... & ... & ... & ... & -1.58(19) & 0.064-0.4& 1, 4, 5, 6, 9  \\
\enddata
\tablerefs{\citep[(1)][]{bilous2016},\citep[(2)][]{bilous2020},\citep[(3)][]{bondonneau},\citep[(4)][]{malofeev},\citep[(5)][]{atnf},\citep[(6)][]{gnbcc350},\citep[(7)][]{jankowski441psr},\citep[(8)][]{kravtsov},\citep[(9)][]{sanidaslotass},\citep[(10)][]{shrauner81.5MHz},\citep[(11)][]{zakharenko}}
\tablecomments{We also give the reference frequency($\nu_0$) used for spectral fitting, flux density (S$_0$) at the reference frequency, calculated turnover frequency($\nu_\mathrm{m}$), spectral index from linear fit ($\alpha_{L}$) and the list of references used for fitting the pulsar spectrum. Turnover and associated parameters are only reported for pulsars which prefer it over the linear model.}
\tablenotetext{a}{$\alpha$ and $\beta$ are the spectral index and curvature parameter of the low frequency turnover fit.}
\tablenotetext{b}{$\nu_\mathrm{R}$ is the frequency range of the data used for spectral fitting.}
\end{deluxetable*}

\clearpage

\begin{figure*}[htbp!]
\centering
\includegraphics[width=\textwidth,height=21cm]{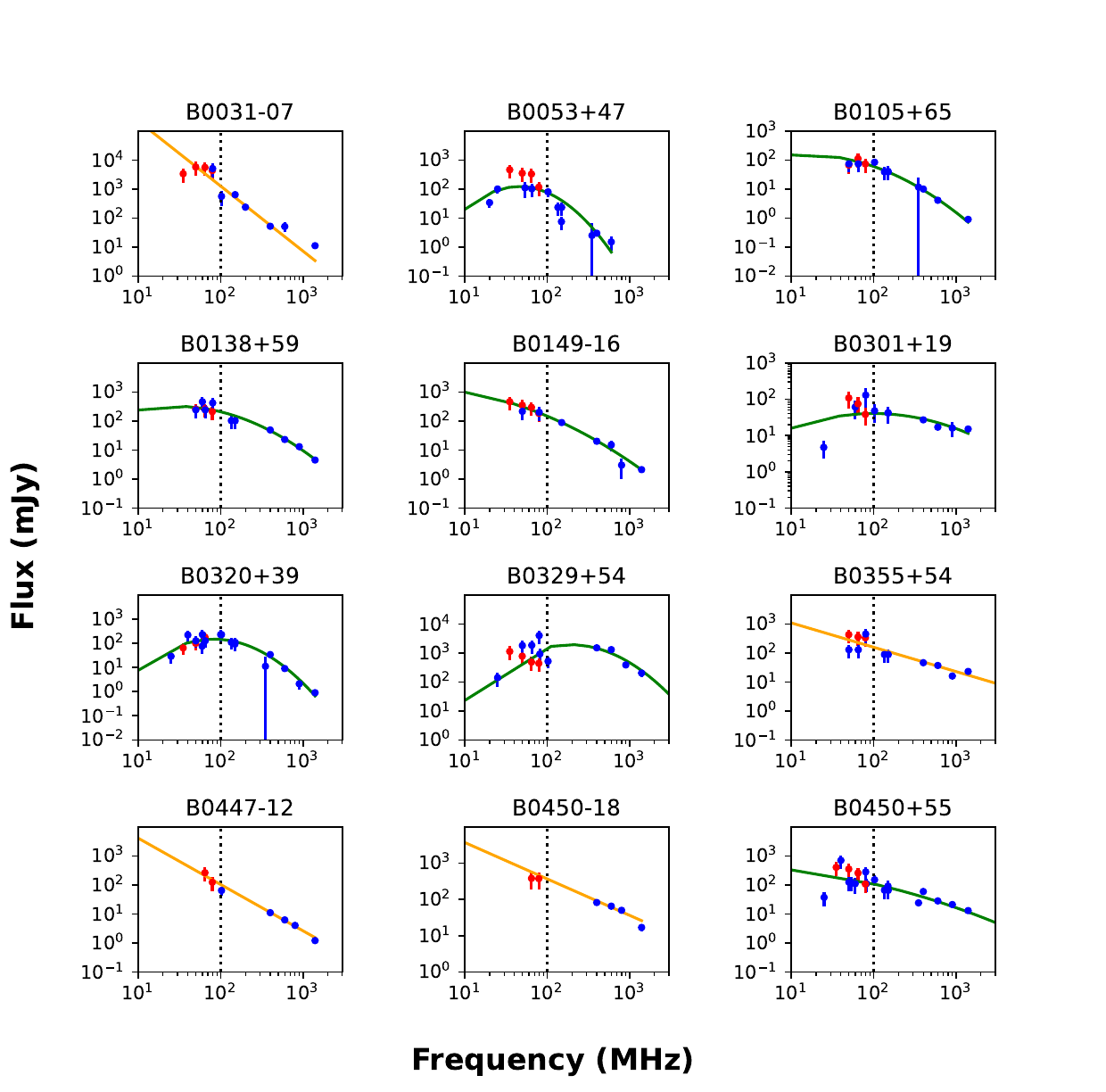}
\caption{LWA pulsar spectrum and flux density measurements at a variety of frequencies. Red and blue points indicate measurements from this study and from other catalogs respectively. Orange and green curves indicate a simple power law fit and power law with a low-frequency turnover applied to the data, respectively, and the dashed vertical line shows the spectral
fit reference frequency, with only the best fit shown in each case.}
\label{fig:psrspectrum}
\end{figure*}
\begin{figure*}[htbp!]
\centering
\includegraphics[width=\textwidth,height=22cm]{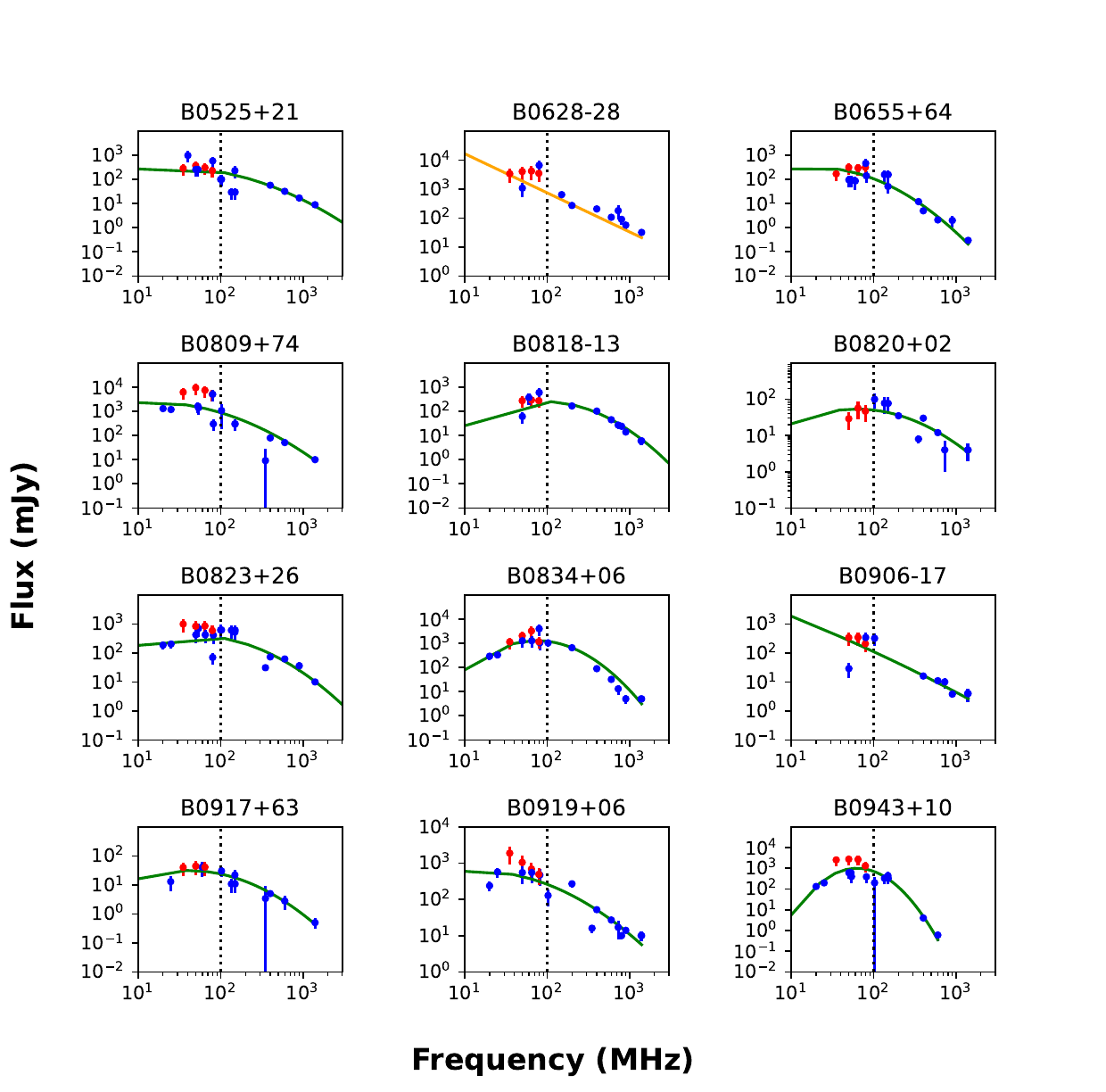}
\caption{Fig \ref{fig:psrspectrum} Continued}
\end{figure*}
\begin{figure*}[htbp!]
    \centering
    \includegraphics[width=\textwidth,height=22cm]{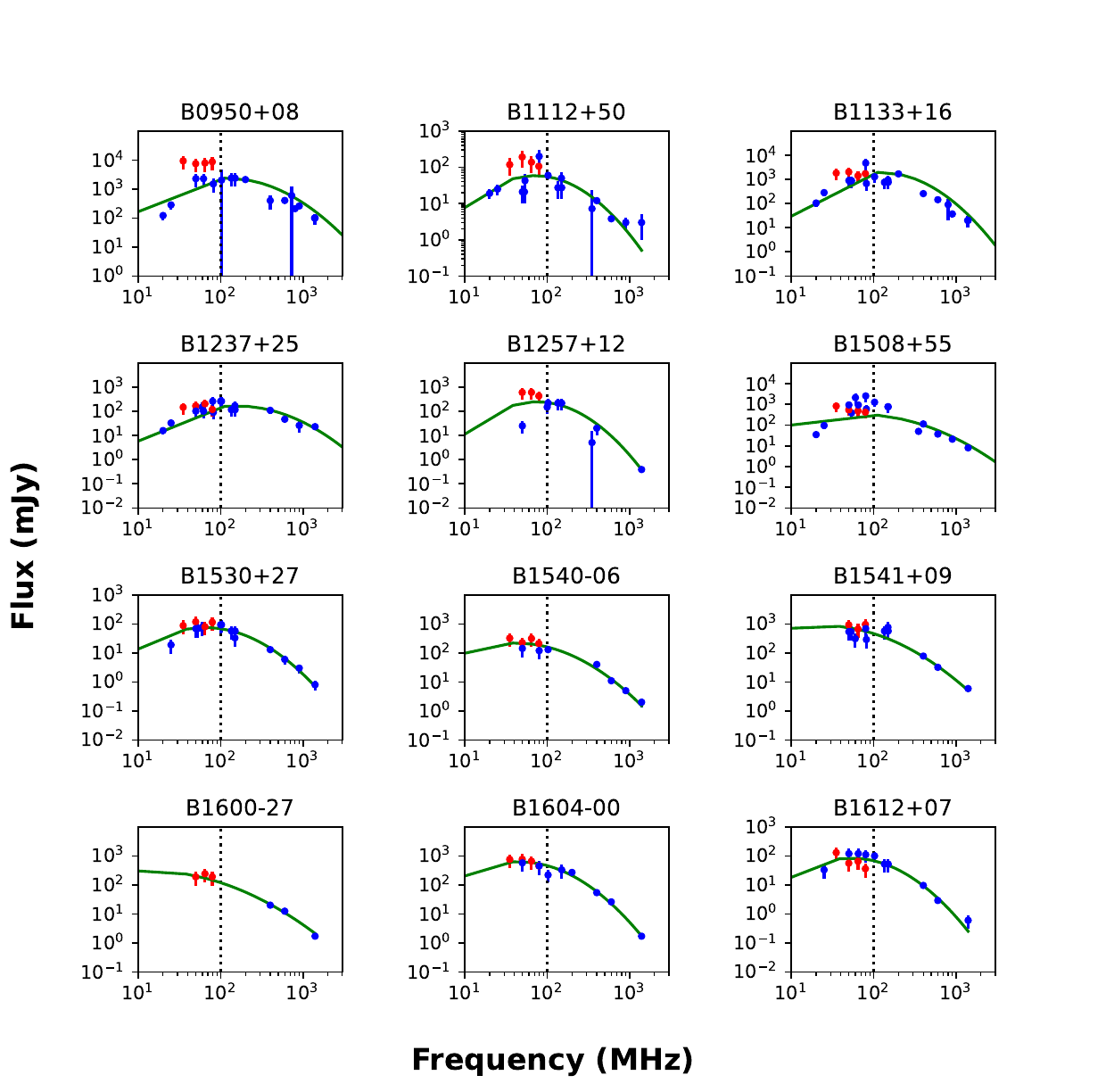}
    \caption{Fig \ref{fig:psrspectrum} Continued}
\end{figure*}
\begin{figure*}[htbp!]
    \centering
    \includegraphics[width=\textwidth,height=22cm]{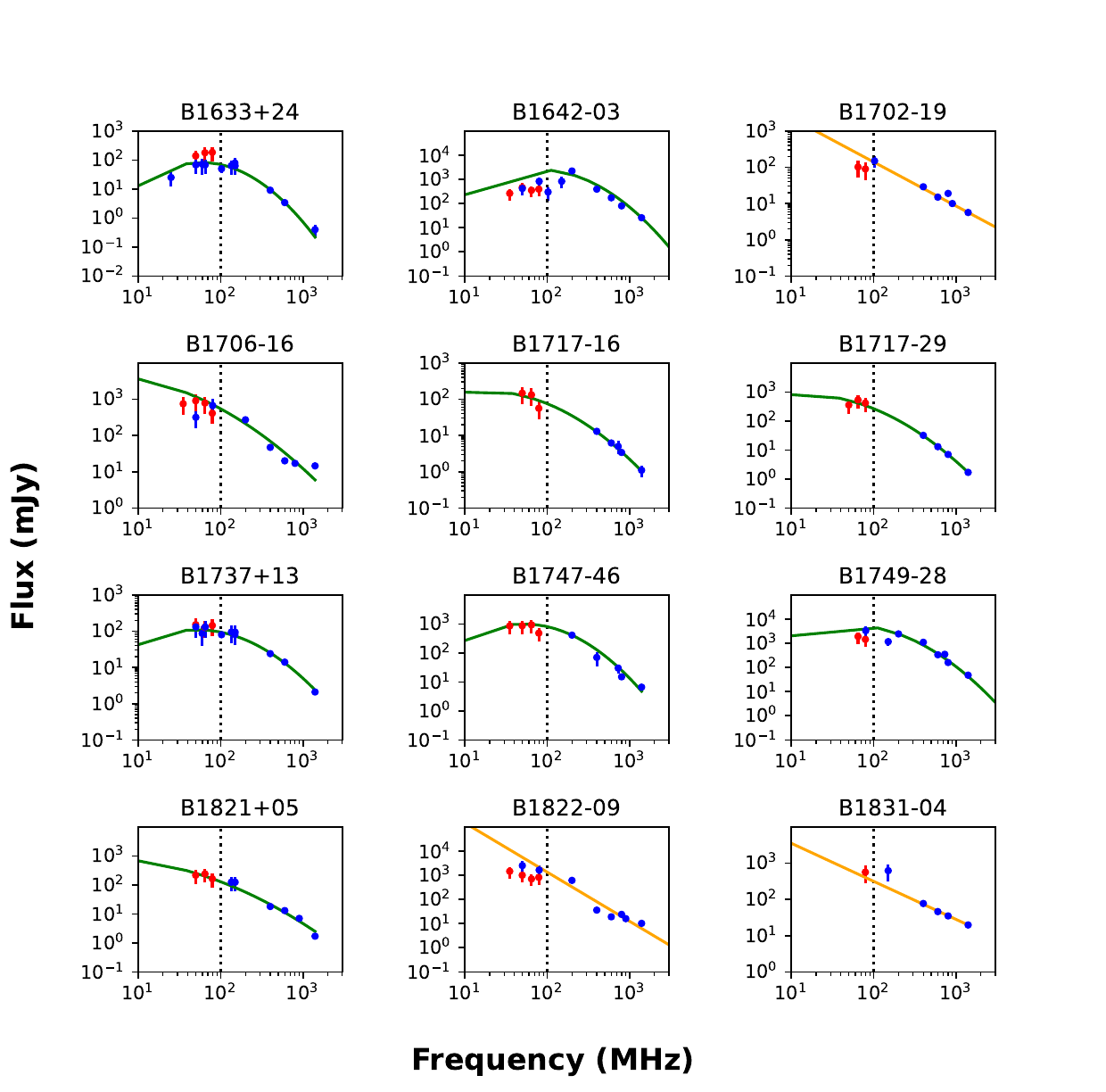}
    \caption{Fig \ref{fig:psrspectrum} Continued}
\end{figure*}
\begin{figure*}[htbp!]
    \centering
    \includegraphics[width=\textwidth,height=22cm]{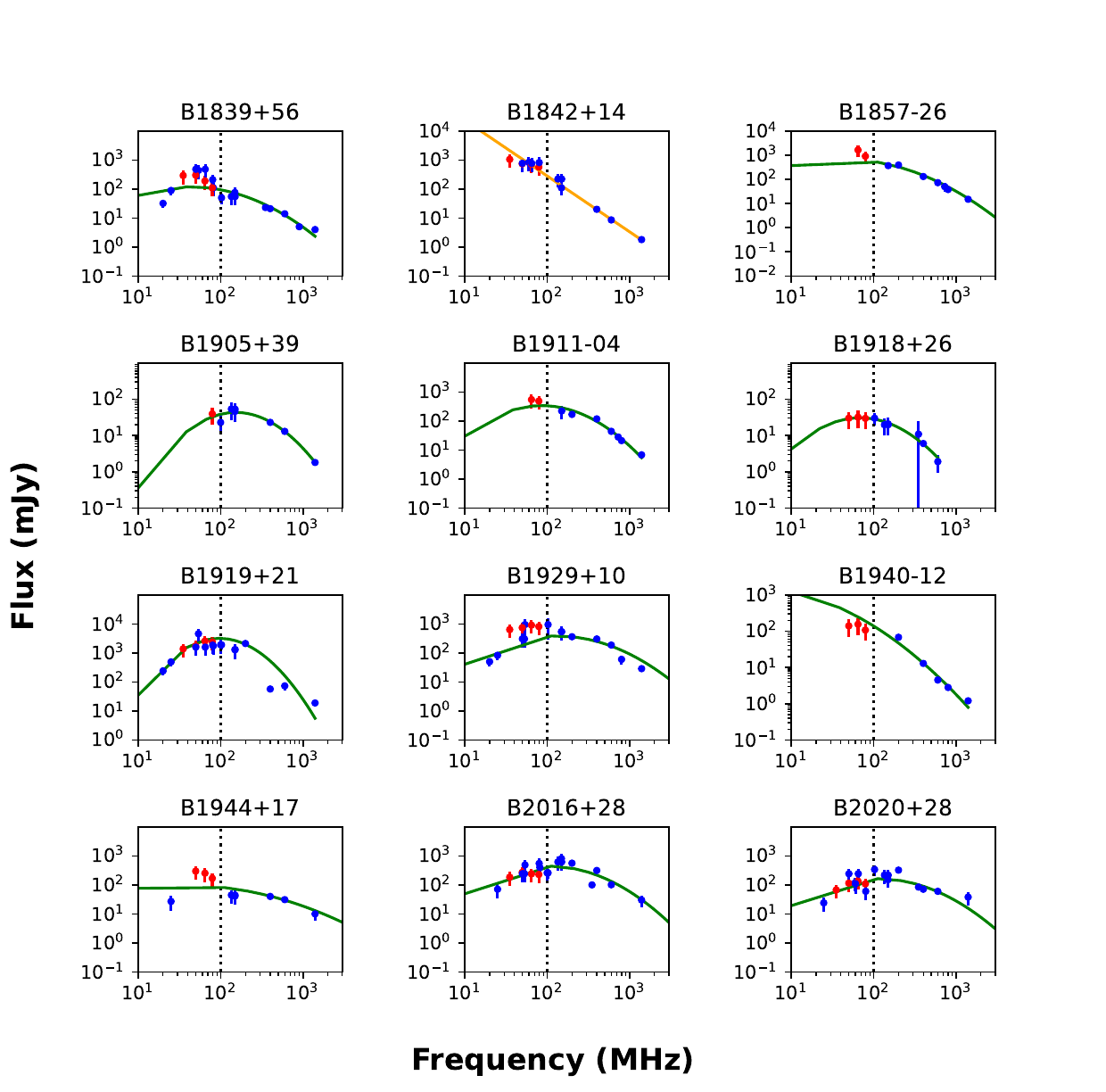}
    \caption{Fig \ref{fig:psrspectrum} Continued}
\end{figure*}
\begin{figure*}[htbp!]
    \centering
    \includegraphics[width=\textwidth,height=22cm]{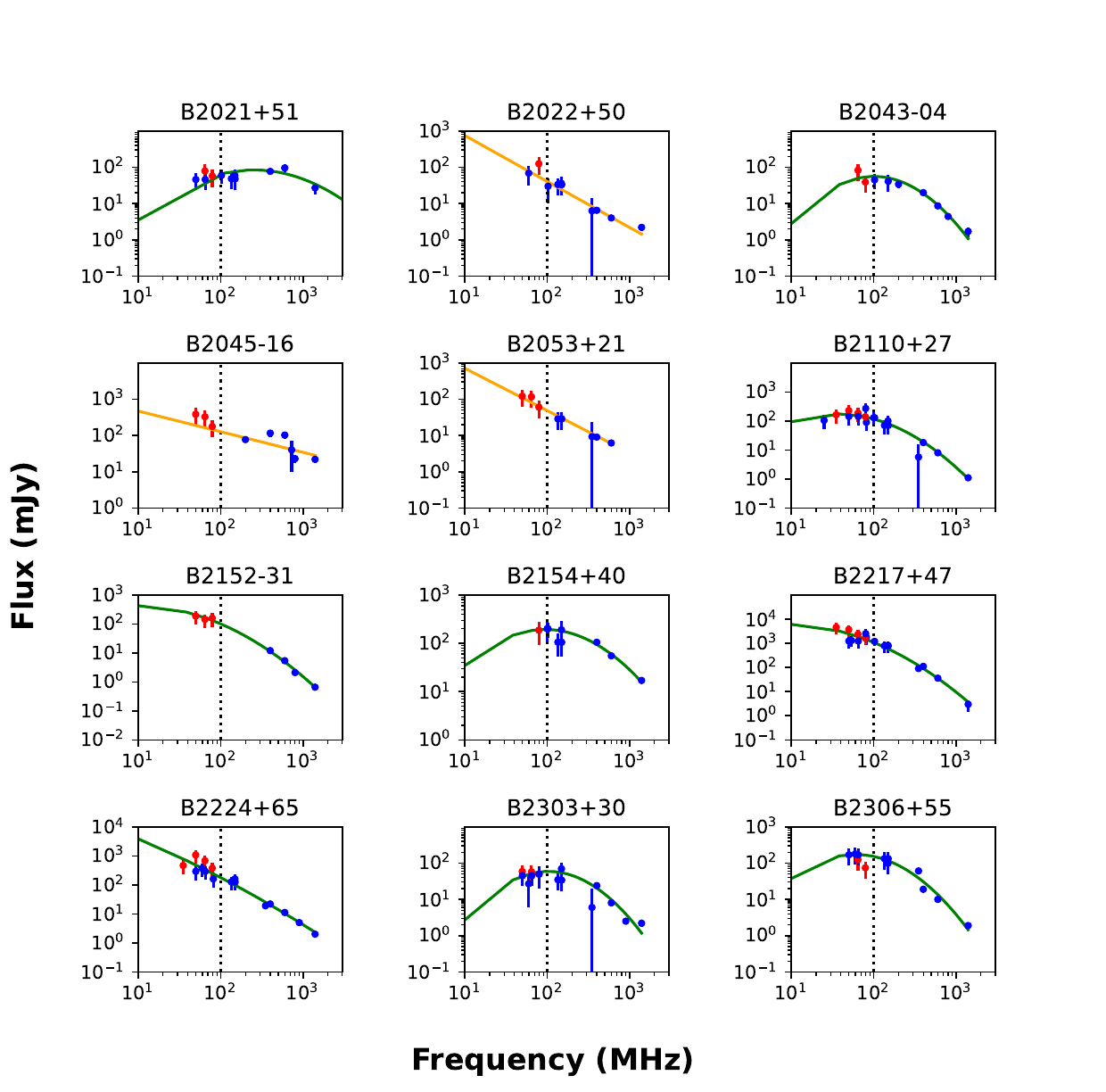}
    \caption{Fig \ref{fig:psrspectrum} Continued}
\end{figure*}
\begin{figure*}[htbp!]
    \centering
    \includegraphics[width=\textwidth,height=22cm]{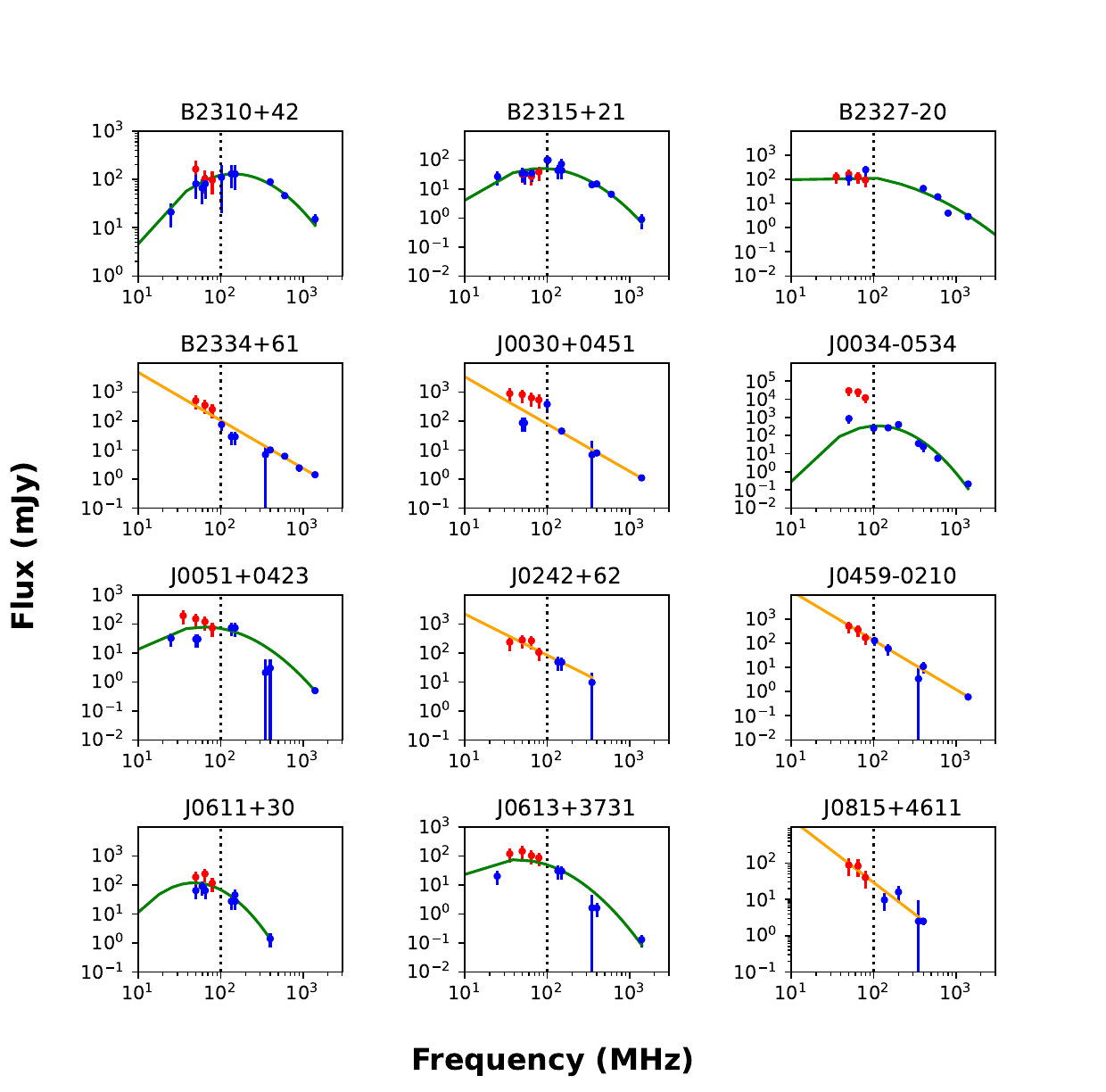}
    \caption{Fig \ref{fig:psrspectrum} Continued}
\end{figure*}
\begin{figure*}[htbp!]
    \centering
    \includegraphics[width=\textwidth,height=22cm]{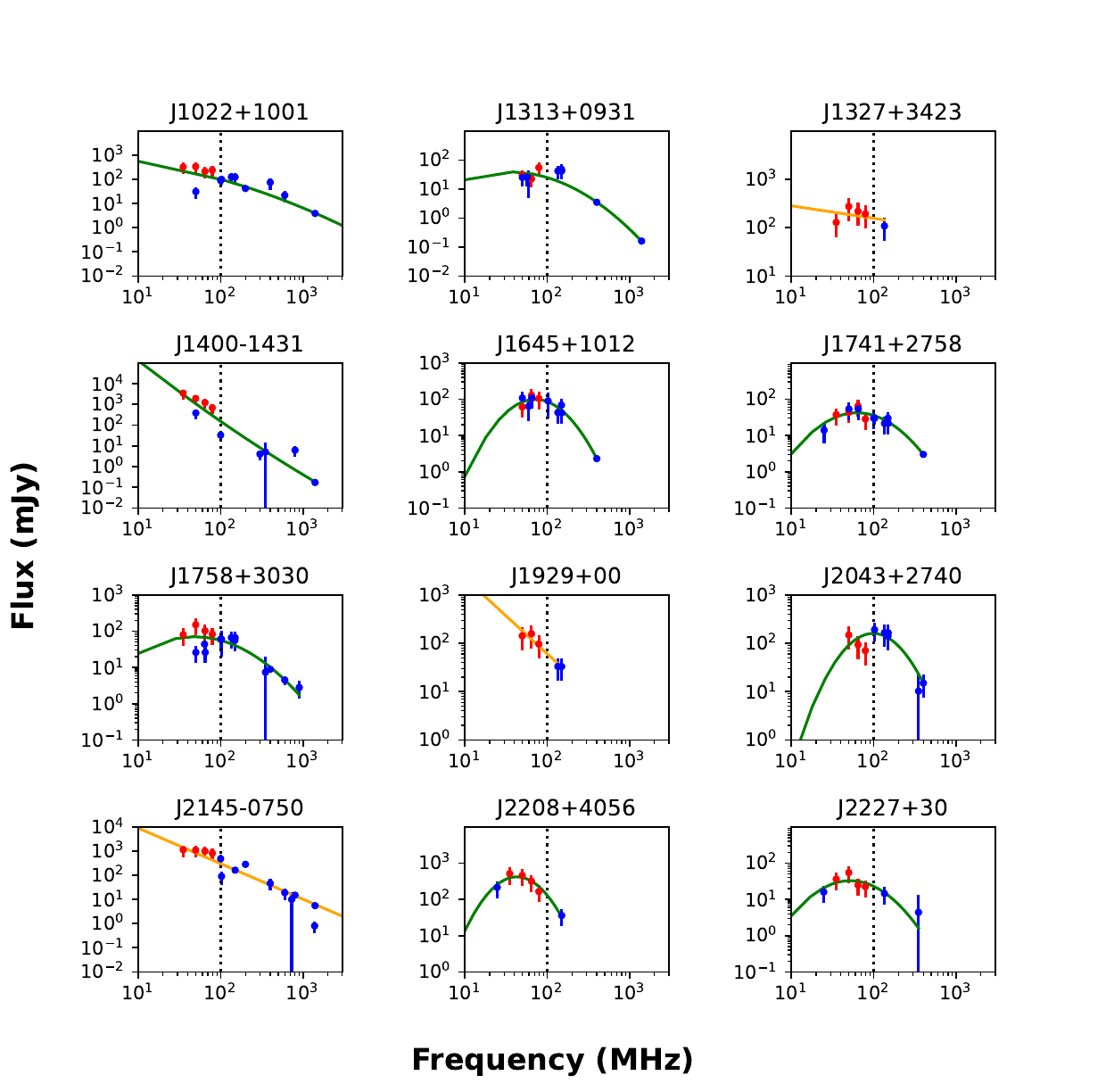}
    \caption{Fig \ref{fig:psrspectrum} Continued}
\end{figure*}
\begin{figure*}[htbp!]
    \centering
    \includegraphics[width=\textwidth,height=22cm]{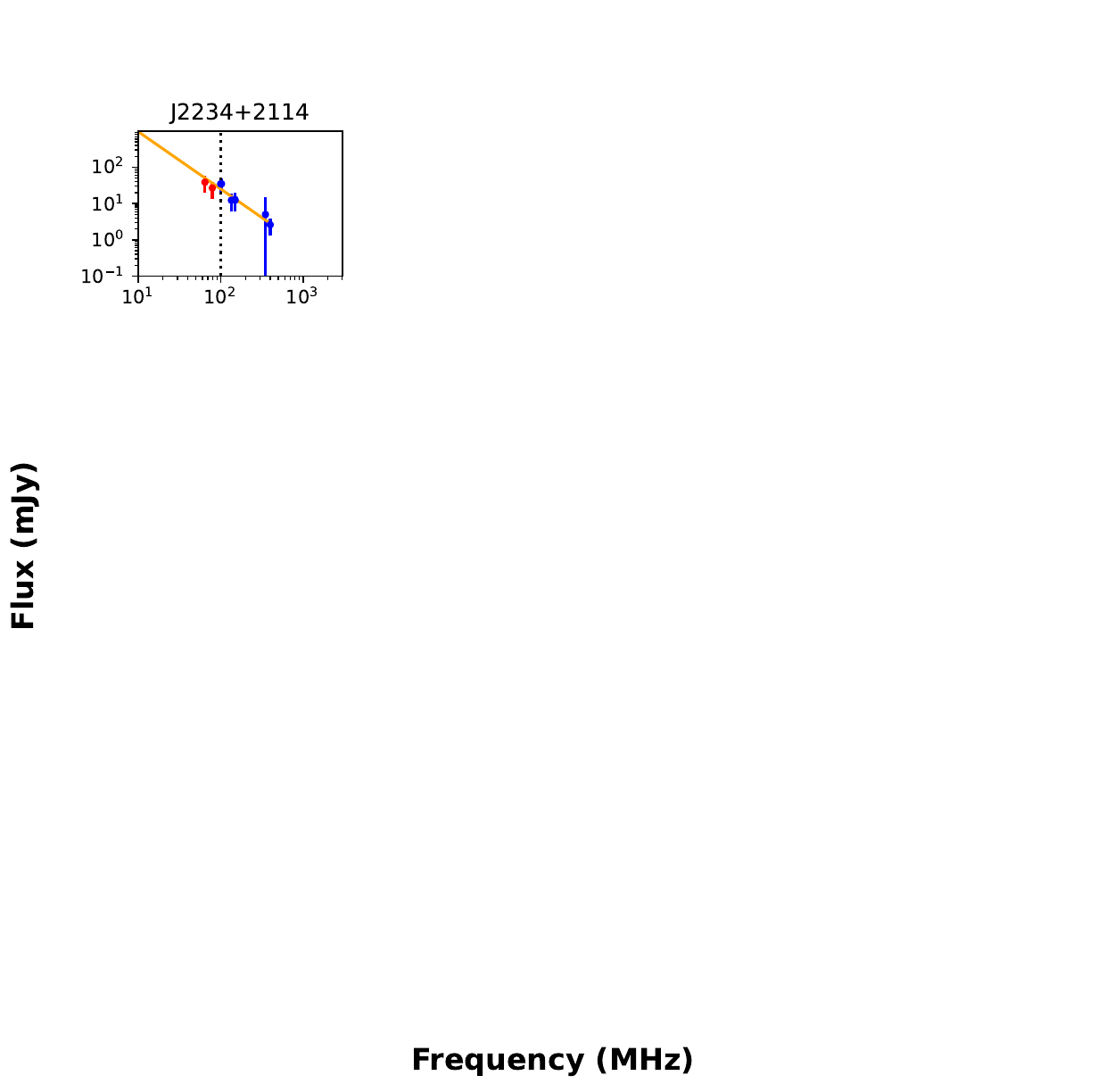}
    \caption{Fig \ref{fig:psrspectrum} Continued}
\end{figure*}
\clearpage

\section{Dispersion Measure Variation}
\begin{figure*}[htbp!]
    \centering
    \includegraphics[width=\textwidth,height=20.9cm]{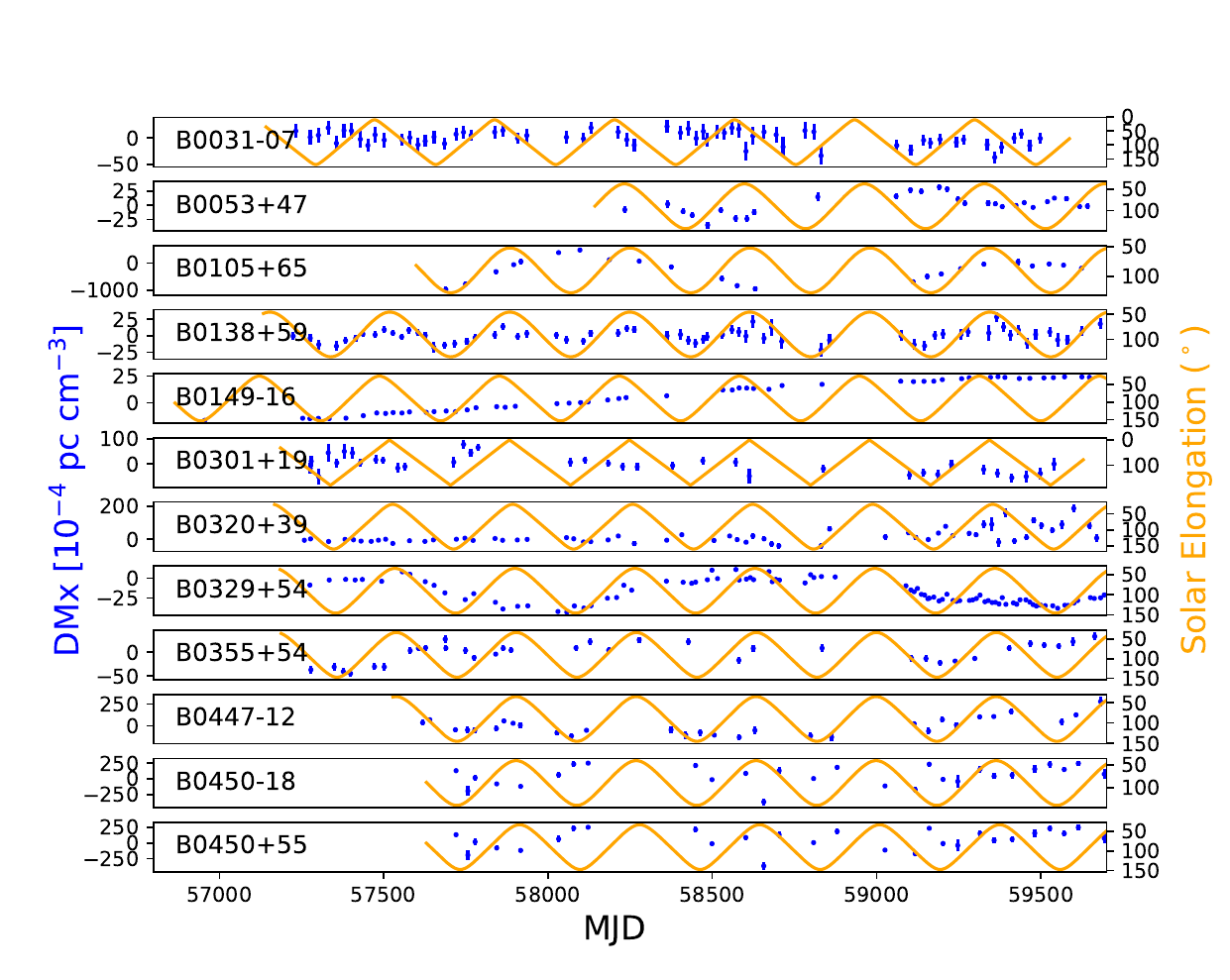}
    \caption{Dispersion measure (DM) variation for LWA pulsars. Blue points represent the excess DM with respect to the reported values in Table \ref{tab:dmperiod} and the orange curve represents the angular separation from the Sun.}
    \label{fig:dmtime}
\end{figure*}
\begin{figure*}
    \centering
    \includegraphics[width=\textwidth,height=22cm]{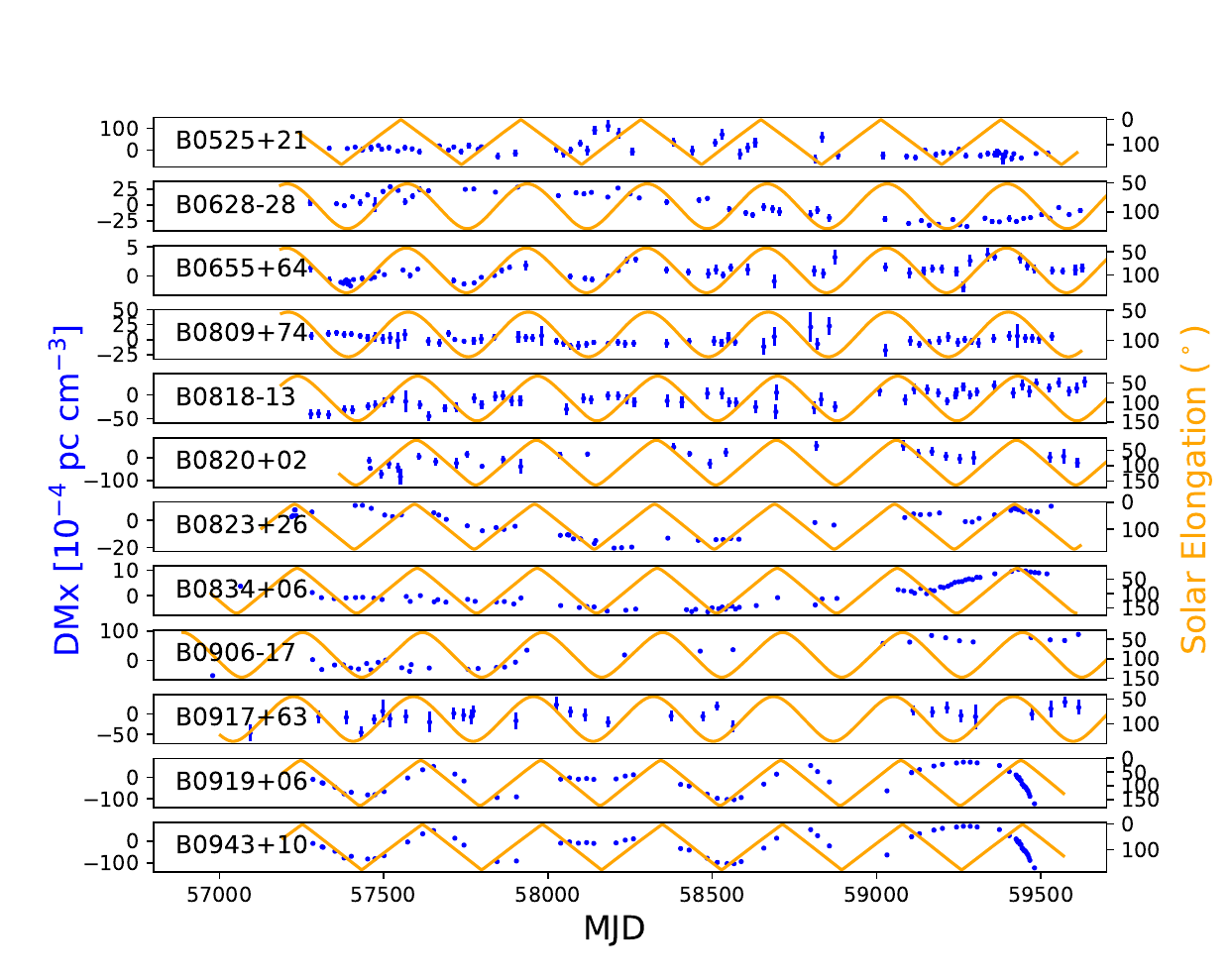}
    \caption{Fig \ref{fig:dmtime} Continued}
\end{figure*}
\begin{figure*}
    \centering
    \includegraphics[width=\textwidth,height=22cm]{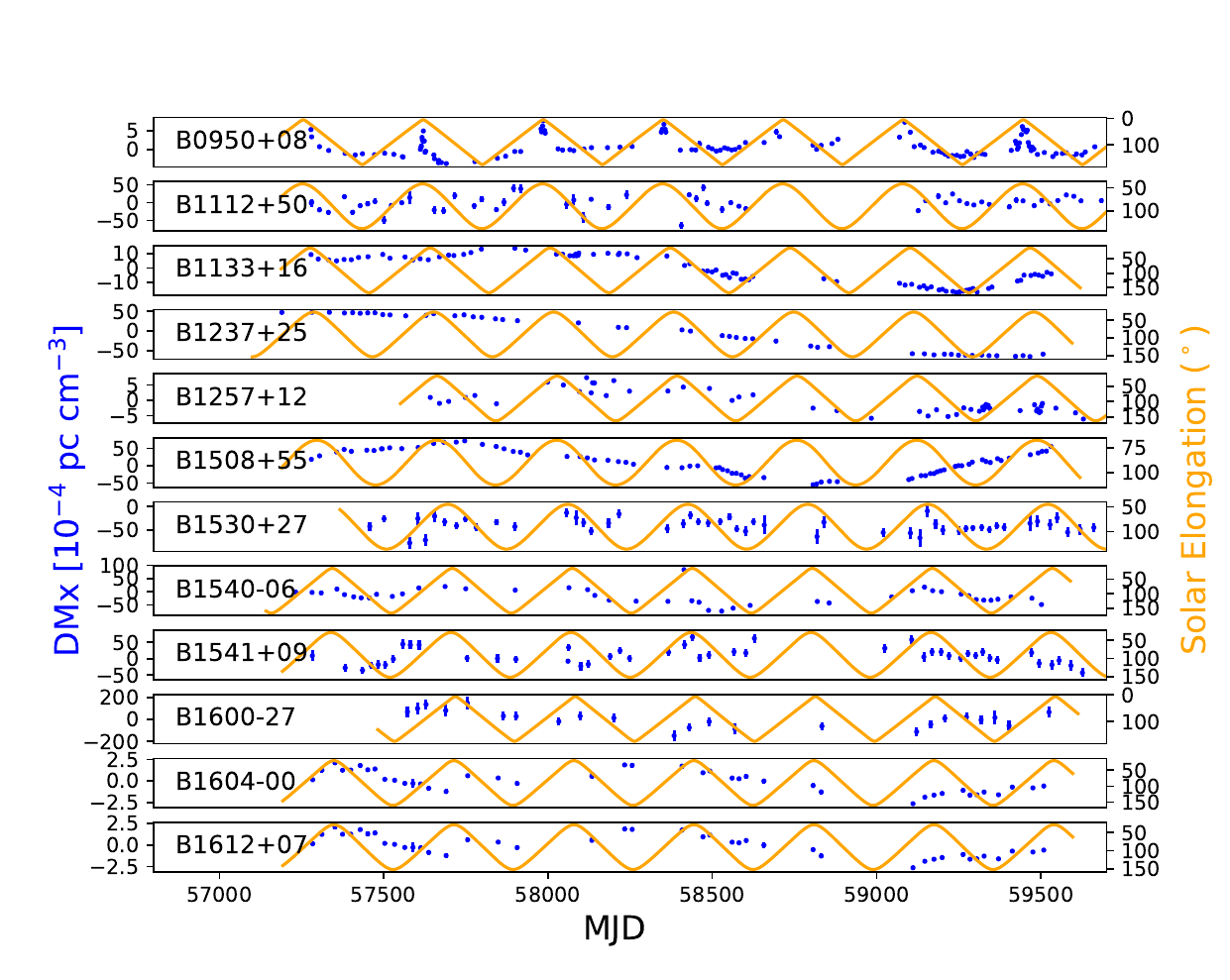}
    \caption{Fig \ref{fig:dmtime} Continued}
\end{figure*}
\begin{figure*}
    \centering
    \includegraphics[width=\textwidth,height=22cm]{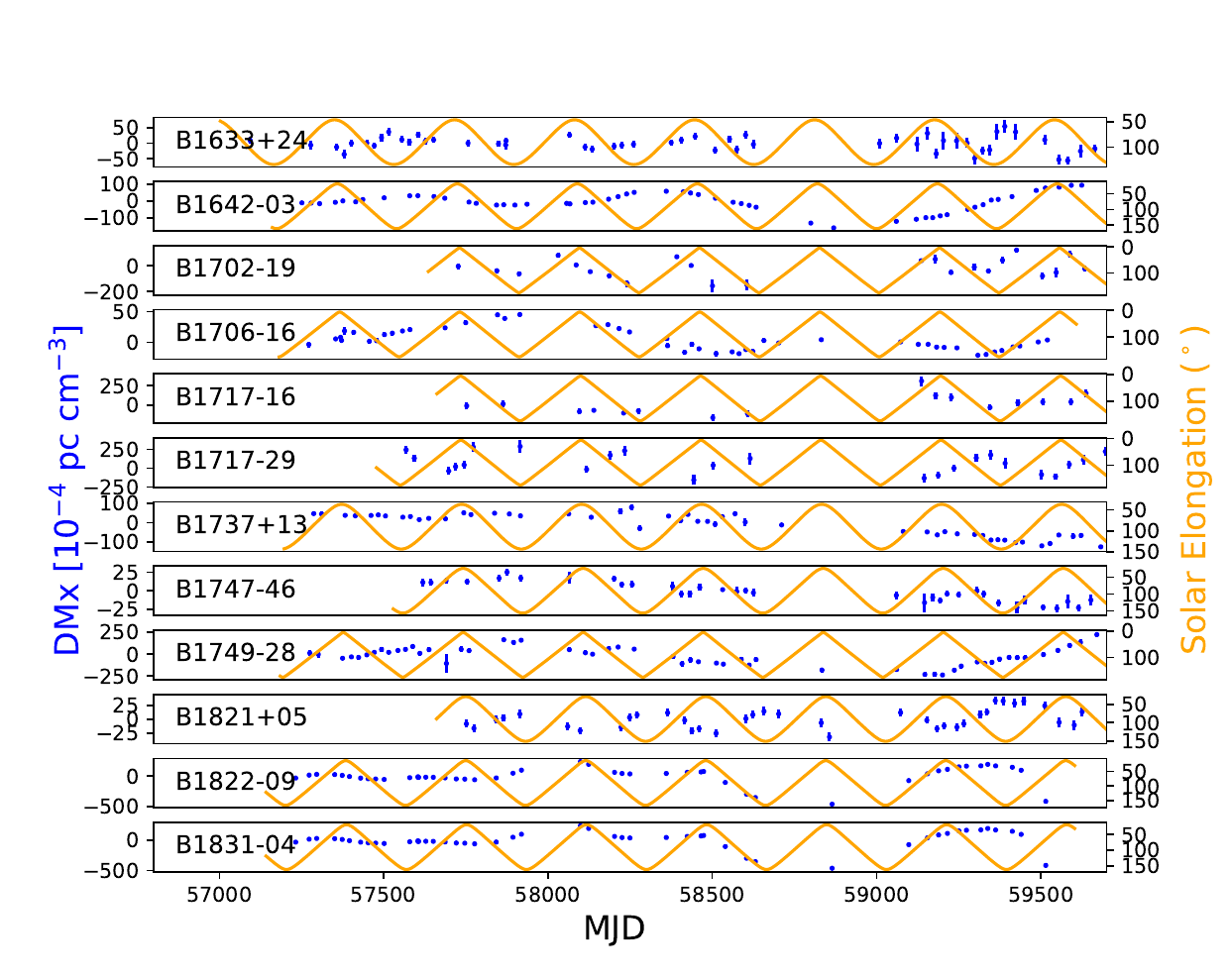}
    \caption{Fig \ref{fig:dmtime} Continued}
\end{figure*}
\begin{figure*}
    \centering
    \includegraphics[width=\textwidth,height=22cm]{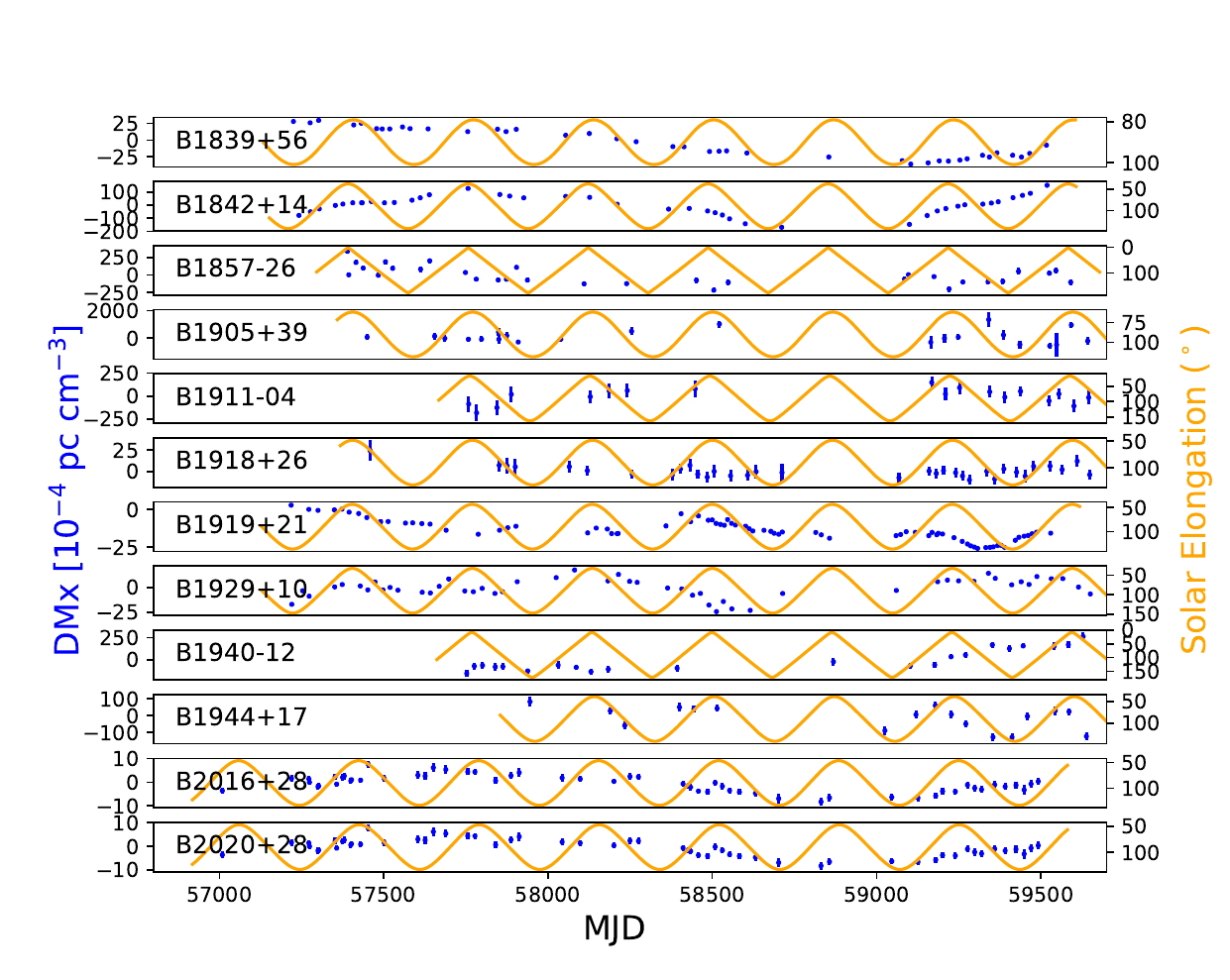}
    \caption{Fig \ref{fig:dmtime} Continued}
\end{figure*}
\begin{figure*}
    \centering
    \includegraphics[width=\textwidth,height=22cm]{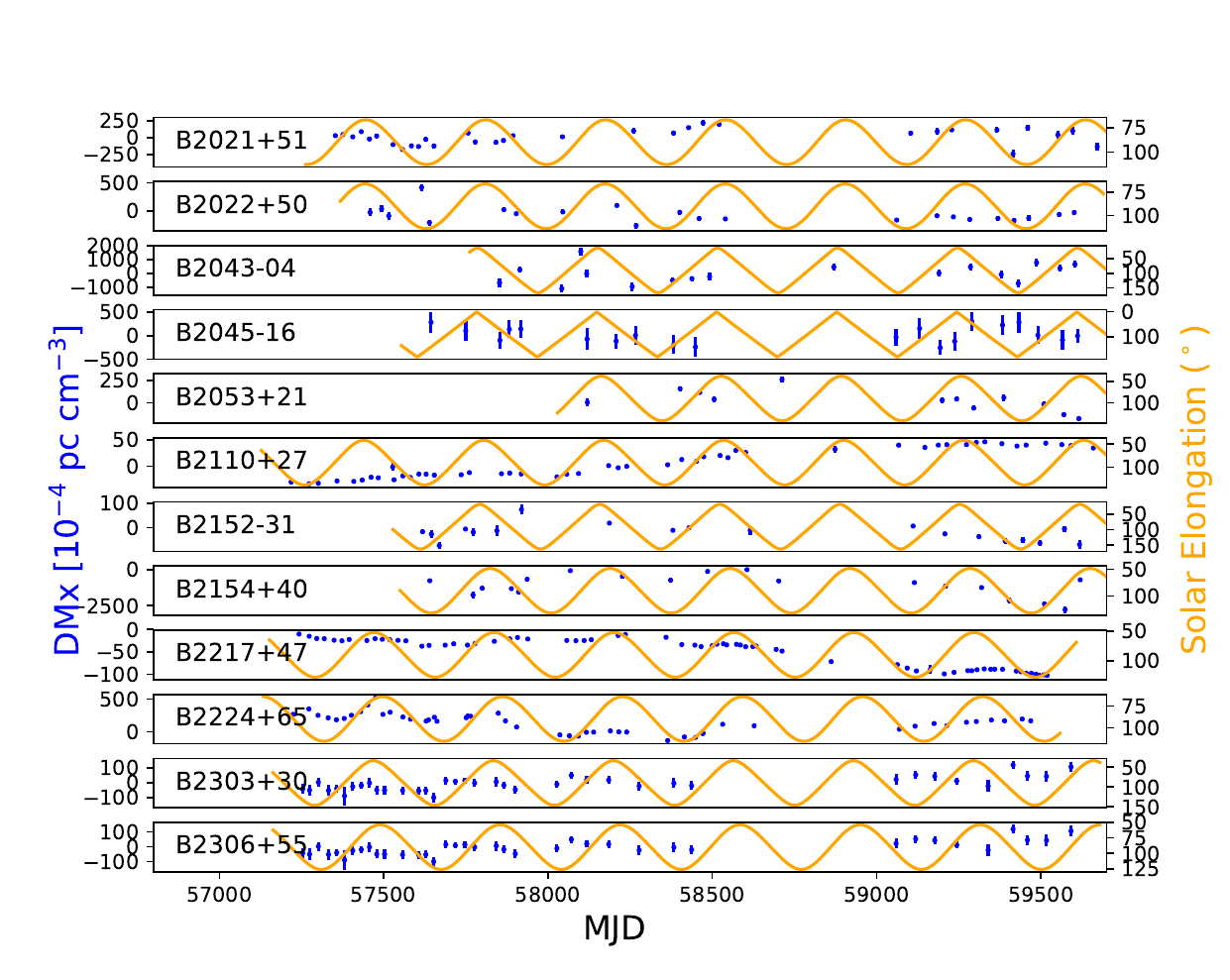}
    \caption{Fig \ref{fig:dmtime} Continued}
\end{figure*}
\begin{figure*}
    \centering
    \includegraphics[width=\textwidth,height=22cm]{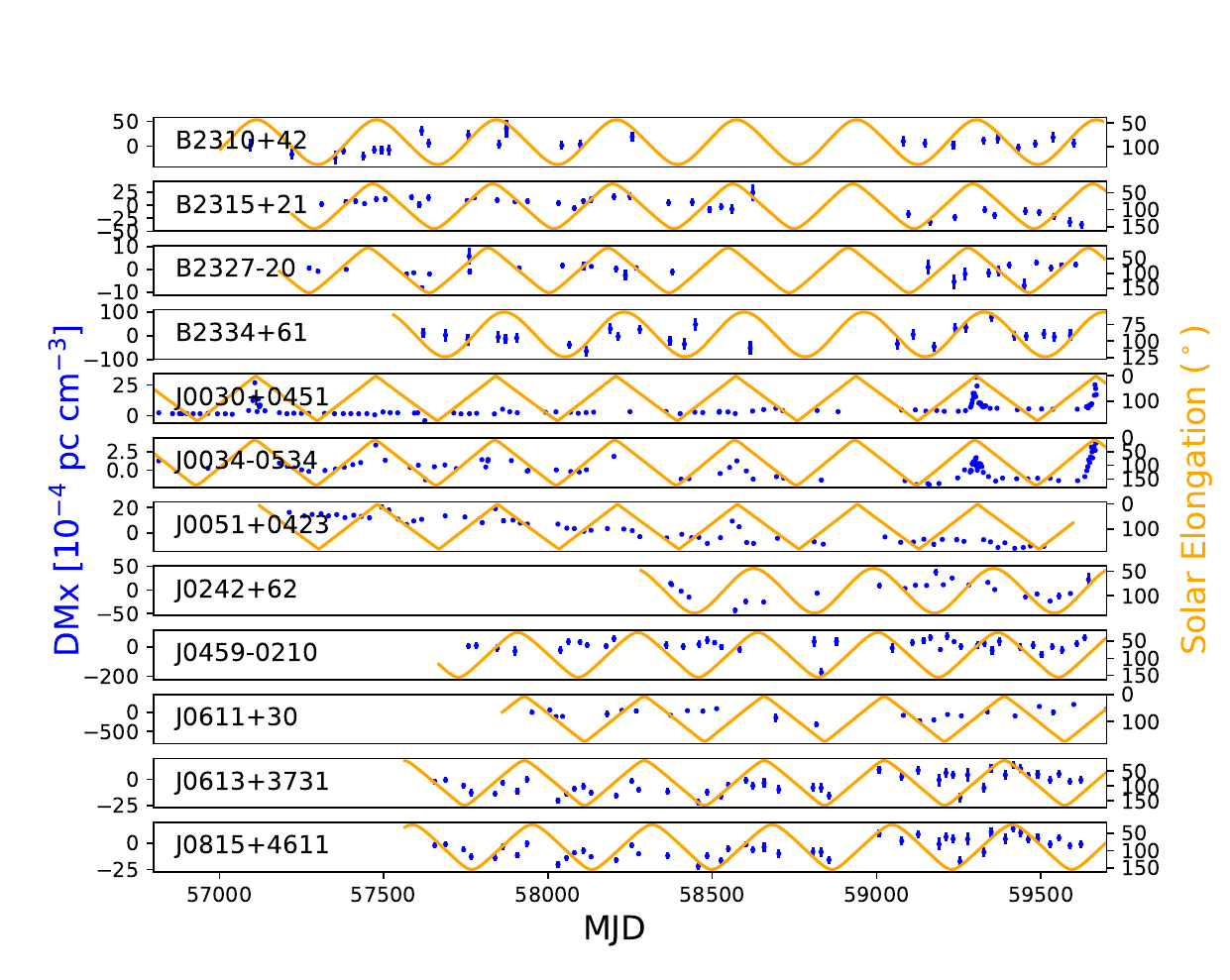}
    \caption{Fig \ref{fig:dmtime} Continued}
\end{figure*}
\begin{figure*}
    \centering
    \includegraphics[width=\textwidth,height=22cm]{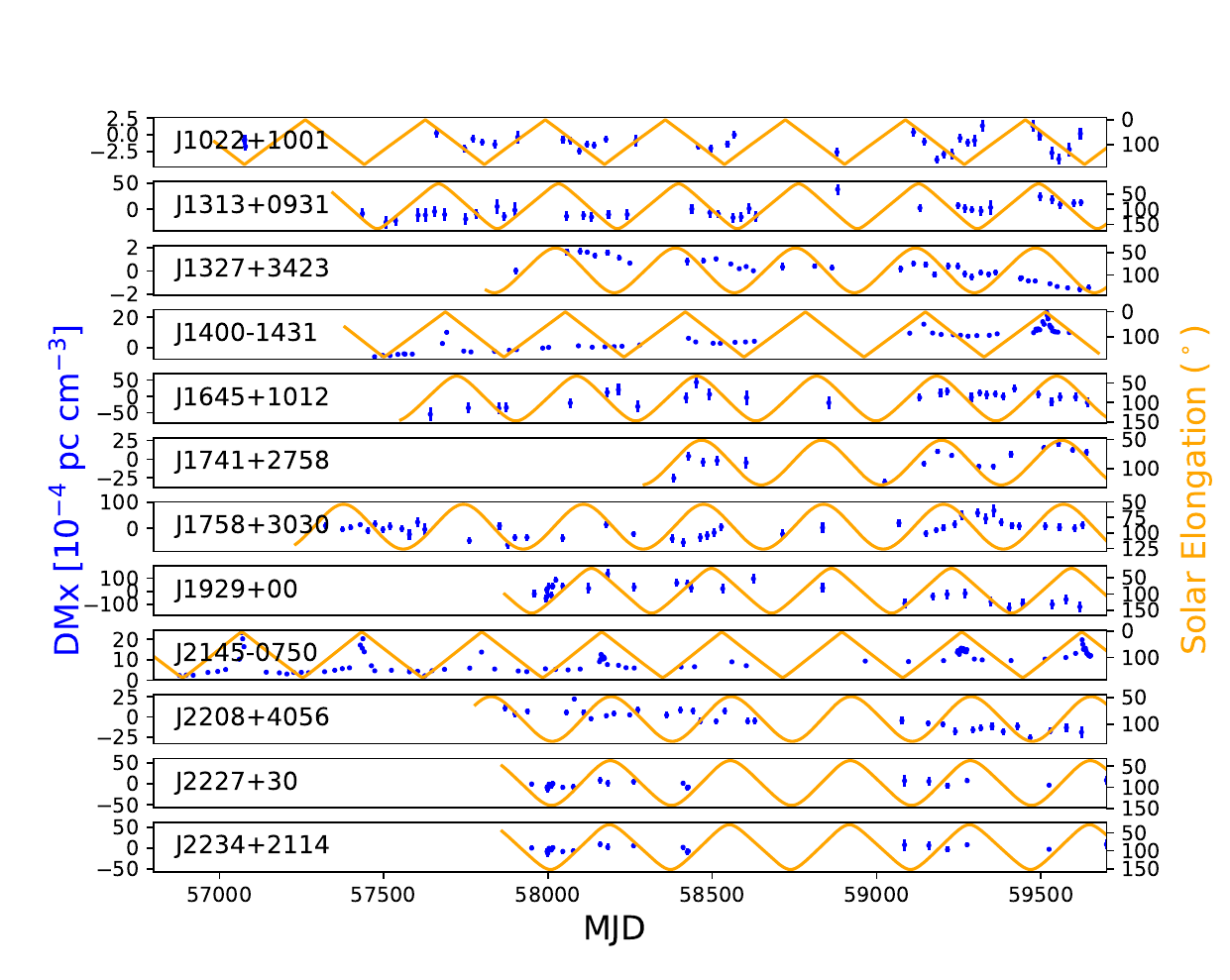}
    \caption{Fig \ref{fig:dmtime} Continued}
\end{figure*}
\clearpage

\section{LWA Pulsars Rotation Measure and Polarization Profiles}

\startlongtable
\centerwidetable
\begin{deluxetable*}{ccccccccccc}
\tablecolumns{11}
\tabletypesize{\small}
\tablewidth{0pt}
\tablecaption{Median Observed Rotation Measure for Pulsars where we detect polarization with the LWA, at Multiple Observing Frequencies ($\nu$) and the average Magnetic field strength of the ISM along the line of sight to the pulsar.\label{tab:rm}}
\tablehead{
    \colhead{Name} & \colhead{$\nu$} & \colhead{MJD} & \colhead{RM$_\mathrm{Obs}$} & \colhead{RM$_\mathrm{Ion}^\mathrm{GPS}$} & \colhead{RM$_\mathrm{Ion}^\mathrm{IGS}$} & \colhead{RM$_\mathrm{ISM}^\mathrm{GPS}$} & \colhead{RM$_\mathrm{ISM}^\mathrm{IGS}$} & \colhead{RM$_\mathrm{ISM}^\mathrm{ATNF}$} & \colhead{DM$_\mathrm{LWA}$} & \colhead{B$_\mathrm{||}^\mathrm{IGS}$}\\ 
    \colhead{ } & \colhead{(MHz)} & \colhead{ } &\colhead{(rad m$^{-2}$)} & \colhead{(rad m$^{-2}$)} & \colhead{(rad m$^{-2}$)} & \colhead{(rad m$^{-2}$)} & \colhead{(rad m$^{-2}$)} & \colhead{(rad m$^{-2}$)} & \colhead{(pc cm$^{-3}$)} & \colhead{($\mu$G)}
}
\startdata
B0031-07 & 49.8 & 57604 & 10.84(1) & ... & 0.84(12) & ... & 10.00(12) & 9.89(7) & 10.8833(14) & 1.132(14) \\ 
B0031-07 & 64.5 & 58108 & 10.80(3) & 0.55(7) & 1.03(8) & 10.25(8) & 9.77(9) &  & 10.8845(12) & 1.106(10) \\ 
B0031-07 & 79.2 & 59466 & 11.05(6) & ... & 1.37(11) & ... & 9.68(13) &  & 10.8832(11) & 1.096(15) \\ 
\hline
B0149-16 & 49.8 & 58099 & 7.75(1) & 0.72(9) & 1.51(15) & 7.03(9) & 6.24(15) & 6.60(500) & 11.9280(0) & 0.645(15) \\ 
B0149-16 & 64.5 & 58361 & 7.55(2) & 0.63(6) & 1.07(15) & 6.92(6) & 6.48(15) &  & 11.9286(1) & 0.669(15) \\ 
B0149-16 & 79.2 & 58361 & 7.55(4) & 0.63(6) & 1.07(15) & 6.92(7) & 6.48(16) &  & 11.9286(1) & 0.669(17) \\ 
\hline
B0329+54 & 64.5 & 58131 & -63.58(2) & 0.42(4) & 0.56(8) & -64.00(4) & -64.14(8) & -64.33(6) & 26.7617(1) & -2.953(4) \\ 
B0329+54 & 79.2 & 58230 & -63.21(1) & 1.09(3) & 1.11(8) & -64.30(3) & -64.32(8) &  & 26.7643(1) & -2.961(4) \\ 
\hline
B0450+55 & 49.8 & 58623 & 6.84(1) & ... & 0.92(11) & ... & 5.92(11) & 5.80(10) & 14.6031(3) & 0.499(9) \\ 
B0450+55 & 64.5 & 58036 & 6.49(1) & 0.33(4) & 0.61(5) & 6.16(4) & 5.88(5) &  & 14.5923(3) & 0.496(4) \\ 
B0450+55 & 79.2 & 58036 & 6.45(3) & 0.33(4) & 0.61(5) & 6.12(5) & 5.84(6) &  & 14.5923(3) & 0.493(5) \\ 
\hline
B0628-28 & 64.5 & 59160 & 47.12(3) & ... & 1.81(19) & ... & 45.31(19) & 46.53(12) & 34.4120(4) & 1.622(7) \\ 
B0628-28 & 79.2 & 58361 & 47.47(2) & 1.66(8) & 2.32(19) & 45.81(8) & 45.15(19) &  & 34.4157(4) & 1.616(7) \\ 
\hline
B0809+74 & 49.8 & 58260 & -13.05(1) & 1.28(8) & 1.06(8) & -14.33(8) & -14.11(8) & -14.00(7) & 5.76160(5) & -3.017(17) \\ 
B0809+74 & 64.5 & 58235 & -13.02(2) & 1.29(7) & 0.96(10) & -14.31(7) & -13.98(10) &  & 5.76150(6) & -2.989(21) \\ 
B0809+74 & 79.2 & 58235 & -13.01(1) & 1.29(7) & 0.96(10) & -14.3(7) & -13.97(10) &  & 5.76150(6) & -2.987(21) \\ 
\hline
B0818-13 & 64.5 & 57405 & -1.83(3) & ... & 1.49(9) & ... & -3.32(9) & -1.2(4) & 40.9806(10) & -0.100(3) \\ 
B0818-13 & 79.2 & 59240 & -1.39(5) & ... & 1.74(20) & ... & -3.13(21) &  & 40.9836(9) & -0.094(6) \\ 
\hline
B0823+26 & 49.8 & 58063 & 5.90(1) & 0.33(5) & 0.78(6) & 5.57(5) & 5.12(6) & 5.38(6) & 19.4781(1) & 0.324(4) \\ 
B0823+26 & 64.5 & 58254 & 6.78(2) & 1.17(7) & 1.73(10) & 5.61(7) & 5.05(10) &  & 19.4772(1) & 0.319(6) \\ 
B0823+26 & 79.2 & 58254 & 6.77(2) & 1.17(7) & 1.73(10) & 5.60(7) & 5.04(10) &  & 19.4772(1) & 0.319(6) \\ 
\hline
B0834+06 & 49.8 & 58266 & 26.86(0) & 1.33(10) & 2.06(13) & 25.53(10) & 24.80(13) & 25.32(7) & 12.8608(0) & 2.376(12) \\ 
B0834+06 & 64.5 & 58266 & 26.95(1) & 1.33(10) & 2.06(13) & 25.62(10) & 24.89(13) &  & 12.8608(0) & 2.384(12) \\ 
B0834+06 & 79.2 & 58266 & 26.97(1) & 1.33(10) & 2.06(13) & 25.64(10) & 24.91(13) &  & 12.8608(0) & 2.386(12) \\ 
\hline
B0919+06 & 35.1 & 58484 & 33.99(0) & ... & 1.05(11) & ... & 32.94(11) & 29.2(3) & 27.2920(7) & 1.487(5) \\ 
B0919+06 & 49.8 & 57401 & 33.62(1) & ... & 1.04(8) & ... & 32.58(8) &  & 27.2928(8) & 1.471(4) \\ 
B0919+06 & 64.5 & 58207 & 33.77(1) & 0.54(6) & 1.04(10) & 33.23(6) & 32.73(10) &  & 27.2990(7) & 1.477(5) \\ 
B0919+06 & 79.2 & 58799 & 34.30(3) & 1.08(9) & 1.38(13) & 33.22(9) & 32.92(13) &  & 27.3052(8) & 1.485(6) \\ 
\hline
B0943+10 & 49.8 & 58108 & 14.90(0) & 0.4(4) & 0.83(8) & 14.50(4) & 14.07(8) & 14.10(8) & 15.3249(9) & 1.131(6) \\ 
B0943+10 & 64.5 & 58108 & 14.89(0) & 0.39(4) & 0.83(8) & 14.50(4) & 14.06(8) &  & 15.3249(9) & 1.130(6) \\ 
B0943+10 & 79.2 & 58556 & 14.93(4) & 0.45(6) & 0.88(10) & 14.48(7) & 14.05(11) &  & 15.3333(14) & 1.129(9) \\ 
\hline
B0950+08 & 35.1 & 58256 & 2.82(0) & 0.95(7) & 1.61(11) & 1.87(7) & 1.21(11) & -0.66(4) & 2.97040(0) & 0.502(46) \\ 
B0950+08 & 49.8 & 58256 & 2.81(0) & 0.95(7) & 1.61(11) & 1.86(7) & 1.20(11) &  & 2.97040(0) & 0.498(46) \\ 
B0950+08 & 64.5 & 58256 & 2.81(0) & 0.95(7) & 1.61(11) & 1.86(7) & 1.20(11) &  & 2.97040(0) & 0.498(46) \\ 
B0950+08 & 79.2 & 58256 & 2.80(1) & 0.95(7) & 1.61(11) & 1.85(7) & 1.19(11) &  & 2.97040(0) & 0.494(46) \\ 
\hline
B1112+50 & 49.8 & 59231 & 3.1(1) & ... & 0.52(4) & ... & 2.58(4) & 2.63(5) & 9.18960(5) & 0.346(5) \\ 
B1112+50 & 64.5 & 59320 & 3.19(3) & ... & 0.73(4) & ... & 2.46(5) &  & 9.18740(5) & 0.330(7) \\ 
\hline
B1133+16 & 35.1 & 58044 & 5.17(0) & 1.11(7) & 1.40(11) & 4.06(7) & 3.77(11) & 3.97(7) & 4.84650(0) & 0.958(28) \\ 
B1133+16 & 49.8 & 58361 & 5.39(1) & 1.16(7) & 1.56(12) & 4.23(7) & 3.83(12) &  & 4.84630(0) & 0.974(31) \\ 
B1133+16 & 64.5 & 58361 & 5.38(1) & 1.16(7) & 1.56(12) & 4.22(7) & 3.82(12) &  & 4.84630(0) & 0.971(31) \\ 
B1133+16 & 79.2 & 58025 & 5.71(1) & 1.78(4) & 1.91(14) & 3.93(4) & 3.80(14) &  & 4.84650(0) & 0.966(36) \\ 
\hline
B1508+55 & 49.8 & 58407 & 2.41(1) & 0.78(4) & 1.04(8) & 1.63(4) & 1.37(8) & 1.43(5) & 19.6125(1) & 0.086(5) \\ 
B1508+55 & 64.5 & 58407 & 2.31(2) & 0.78(4) & 1.04(8) & 1.53(4) & 1.27(8) &  & 19.6125(1) & 0.080(5) \\ 
B1508+55 & 79.2 & 58057 & 2.73(5) & 1.04(4) & 1.30(9) & 1.69(6) & 1.43(10) &  & 19.6157(1) & 0.090(6) \\ 
\hline
B1540-06 & 49.8 & 58268 & -4.6(1) & 0.65(10) & 1.05(11) & -5.25(10) & -5.65(11) & -1.8(8) & 18.3760(6) & -0.379(7) \\ 
B1540-06 & 64.5 & 58268 & -4.59(1) & 0.65(10) & 1.05(11) & -5.24(10) & -5.64(11) &  & 18.3760(6) & -0.378(7) \\ 
B1540-06 & 79.2 & 58268 & -4.61(5) & 0.67(10) & 1.04(11) & -5.28(11) & -5.65(12) &  & 18.3760(6) & -0.379(8) \\ 
\hline
B1604-00 & 35.1 & 57848 & 6.93(0) & ... & 1.12(9) & ... & 5.81(9) & 6.5(10) & 10.6815(0) & 0.670(10) \\ 
B1604-00 & 49.8 & 58407 & 7.56(1) & 1.35(6) & 1.88(13) & 6.21(6) & 5.68(13) &  & 10.6817(0) & 0.655(15) \\ 
B1604-00 & 64.5 & 58407 & 7.48(2) & 1.35(6) & 1.88(13) & 6.13(6) & 5.60(13) &  & 10.6817(0) & 0.646(15) \\ 
B1604-00 & 79.2 & 58135 & 7.01(4) & 0.62(6) & 1.15(10) & 6.39(7) & 5.86(11) &  & 10.6816(0) & 0.676(13) \\ 
\hline
B1642-03 & 64.5 & 58056 & 17.43(3) & 1.50(8) & 1.97(13) & 15.93(9) & 15.46(13) & 15.8(3) & 35.7555(5) & 0.533(4) \\ 
B1642-03 & 79.2 & 58056 & 17.39(4) & 1.50(8) & 1.97(13) & 15.89(9) & 15.42(14) &  & 35.7555(5) & 0.531(5) \\ 
\hline
B1822-09 & 79.2 & 59188 & 69.93(6) & ... & 2.22(10) & ... & 67.71(12) &  & 19.4101(18) & 4.298(8) \\ 
\hline
B1839+56 & 49.8 & 57755 & -2.69(1) & ... & 1.24(5) & ... & -3.93(5) & -3.85(6) & 26.7732(1) & -0.181(2) \\ 
B1839+56 & 64.5 & 57755 & -2.74(3) & ... & 1.24(5) & ... & -3.98(6) &  & 26.7732(1) & -0.183(3) \\ 
B1839+56 & 79.2 & 57755 & -2.26(4) & ... & 1.24(5) & ... & -3.50(6) &  & 26.7732(1) & -0.161(3) \\ 
\hline
B1919+21 & 49.8 & 58210 & -16.07(1) & 0.68(3) & 1.08(7) & -16.75(3) & -17.15(7) & -16.99(5) & 12.4378(0) & -1.699(7) \\ 
B1919+21 & 64.5 & 58146 & -15.87(1) & 0.88(4) & 1.37(9) & -16.75(4) & -17.24(9) &  & 12.4381(0) & -1.708(9) \\ 
B1919+21 & 79.2 & 58146 & -15.89(1) & 0.88(4) & 1.37(9) & -16.77(4) & -17.26(9) &  & 12.4381(0) & -1.710(9) \\ 
\hline
B1929+10 & 49.8 & 57746 & -5.31(1) & ... & 1.83(14) & ... & -7.14(14) & -6.87(2) & 3.18470(2) & -2.762(54) \\ 
B1929+10 & 64.5 & 58271 & -6.40(1) & 0.41(4) & 0.76(11) & -6.81(4) & -7.16(11) &  & 3.18560(2) & -2.769(43) \\ 
B1929+10 & 79.2 & 58364 & -6.17(5) & 0.56(5) & 0.99(10) & -6.73(7) & -7.16(11) &  & 3.18500(1) & -2.770(43) \\ 
\hline
B2217+47 & 64.5 & 58543 & -34.69(2) & 0.90(3) & 1.06(9) & -35.59(4) & -35.75(9) & -35.93(6) & 43.4944(3) & -1.013(3) \\ 
B2217+47 & 79.2 & 58131 & -34.65(2) & 0.92(9) & 1.22(10) & -35.57(9) & -35.87(10) &  & 43.4956(4) & -1.016(3) \\ 
\hline
B2224+65 & 64.5 & 57629 & -22.19(3) & ... & 0.52(5) & ... & -22.71(6) & -22.99(7) & 36.4806(18) & -0.767(2) \\ 
B2224+65 & 79.2 & 58138 & -21.85(5) & 0.52(9) & 0.80(8) & -22.37(10) & -22.65(9) &  & 36.4637(16) & -0.765(3) \\ 
\hline
B2327-20 & 64.5 & 58378 & 10.17(2) & 0.88(9) & 1.38(16) & 9.29(9) & 8.79(16) & 9.20(80) & 8.45580(1) & 1.281(23) \\ 
B2327-20 & 79.2 & 57616 & 10.39(3) & ... & 1.78(1) & ... & 8.61(3) &  & 8.45510(1) & 1.255(4) \\ 
\hline
J0051+0423 & 64.5 & 58109 & -4.49(2) & ... & 0.94(8) & ... & -5.43(8) & ... & 13.9268(1) & -0.480(7) \\ 
\hline
J0613+3731 & 64.5 & 58208 & 17.16(1) & 0.82(5) & 1.06(7) & 16.34(5) & 16.10(7) & 16(2) & 18.9768(3) & 1.045(5) \\ 
J0613+3731 & 79.2 & 58457 & 16.84(5) & 0.57(5) & 0.73(8) & 16.27(7) & 16.11(9) &  & 18.9762(3) & 1.046(6) \\ 
\enddata
\tablenotetext{a}{RM$_\mathrm{Obs}$ and DM$_\mathrm{LWA}$ are the observed RM and DM of the pulsar on that MJD.}
\tablenotetext{b}{RM$_\mathrm{Ion}^\mathrm{GPS}$ and RM$_\mathrm{Ion}^\mathrm{IGS}$ are the ionospheric correction for RM using TEC measurement from GPS co-located with LWA and IGS model respectively.}
\tablenotetext{c}{RM$_\mathrm{ISM}^\mathrm{GPS}$, RM$_\mathrm{ISM}^\mathrm{IGS}$, and RM$_\mathrm{ISM}^\mathrm{ATNF}$ are the ionosphere corrected for RM using LWA GPS, IGS model, and ATNF catalog respectively.}
\tablenotetext{d}{B$_\mathrm{||}^\mathrm{IGS}$ is the average line-of-sight magnetic field of the ISM using TEC corrections from IGS model.}
\end{deluxetable*}
\clearpage

\begin{figure*}[h!]
    \centering
    \includegraphics[width=\textwidth,height=21cm]{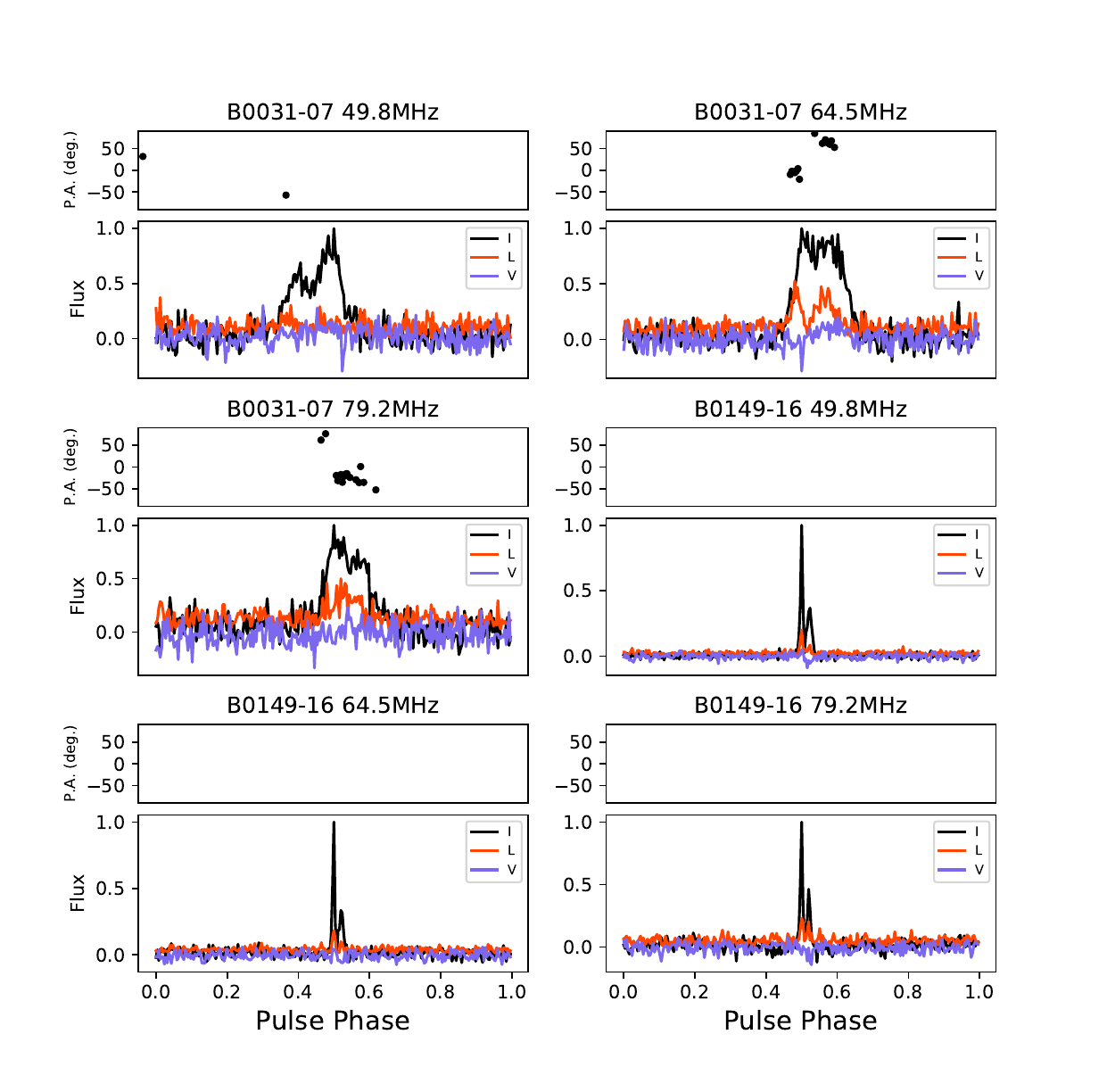}
    \caption{Profiles for pulsars at multiple frequencies where we detect polarization with the LWA. The bottom panel in each sub-figure shows the total intensity, linear, and circular polarization profiles in black, orange, and cyan respectively. The top panel shows the position angle of linear polarization in phase bins where fractional linear polarization is more than 30$\%$.}
    \label{fig:psrprofrm}
\end{figure*}
\begin{figure*}
    \centering
    \includegraphics[width=\textwidth,height=22.1cm]{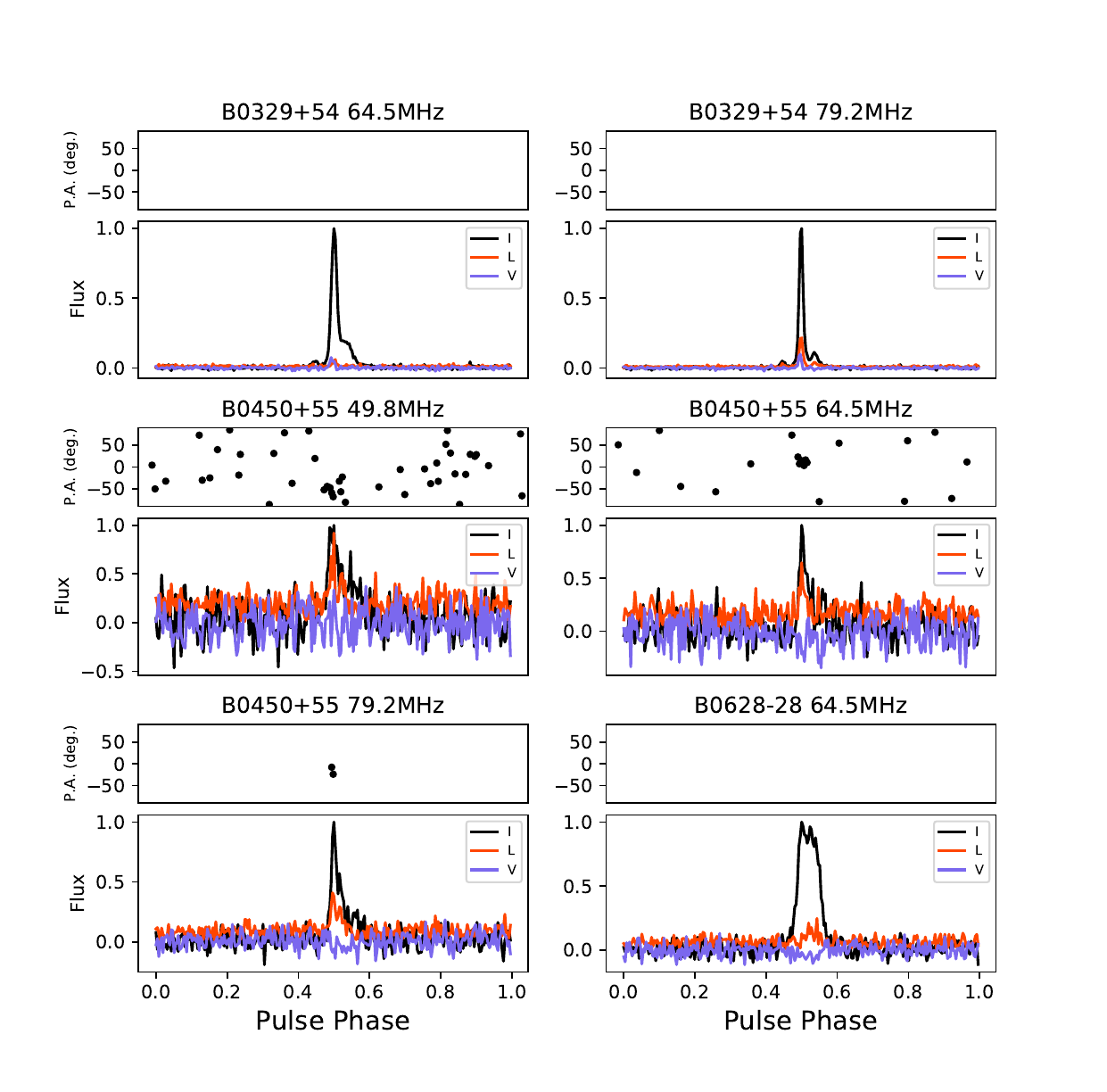}
    \caption{Fig \ref{fig:psrprofrm} Continued}
\end{figure*}
\begin{figure*}
    \centering
    \includegraphics[width=\textwidth,height=22.1cm]{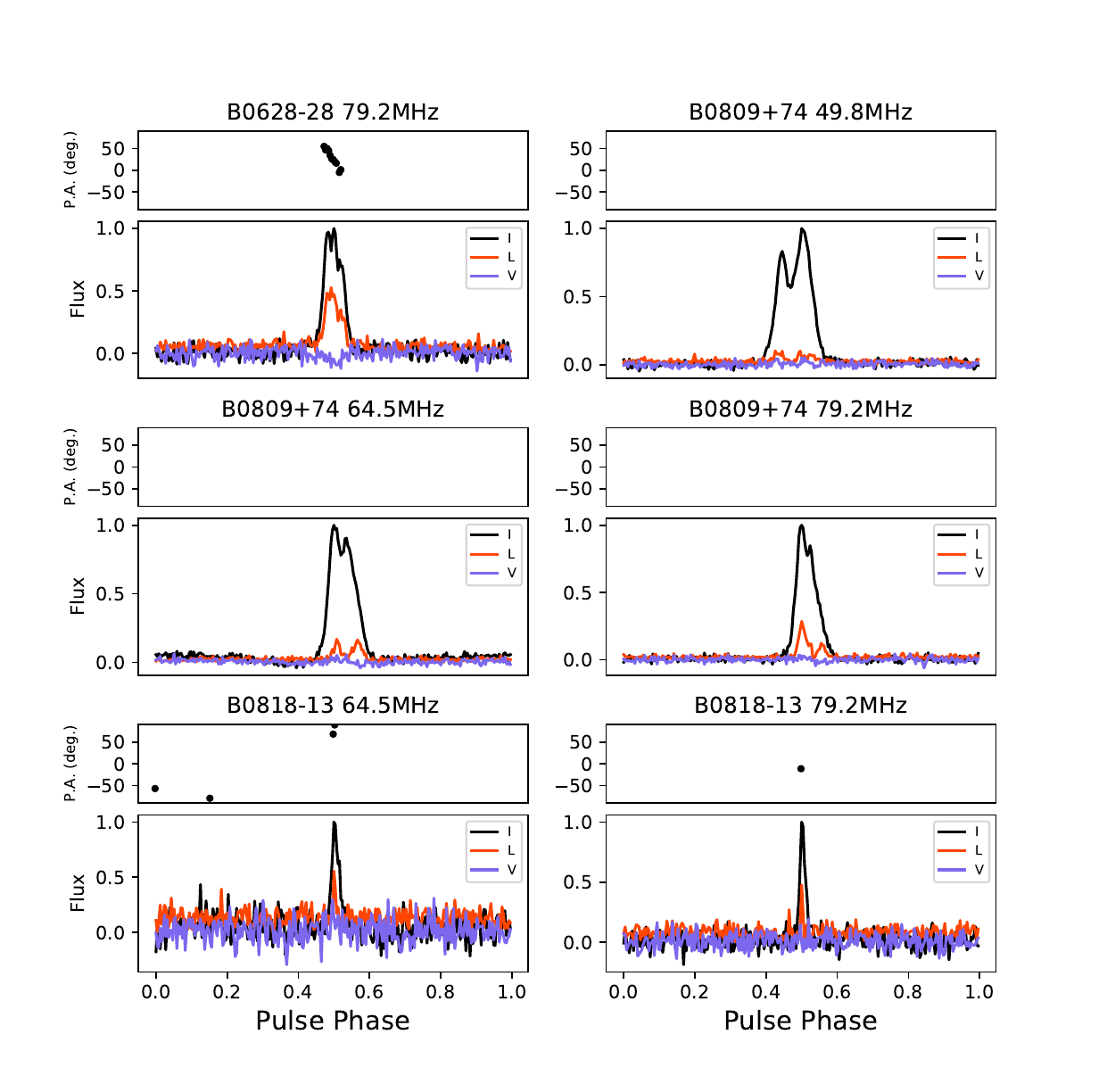}
    \caption{Fig \ref{fig:psrprofrm} Continued}
\end{figure*}
\begin{figure*}
    \centering
    \includegraphics[width=\textwidth,height=22.1cm]{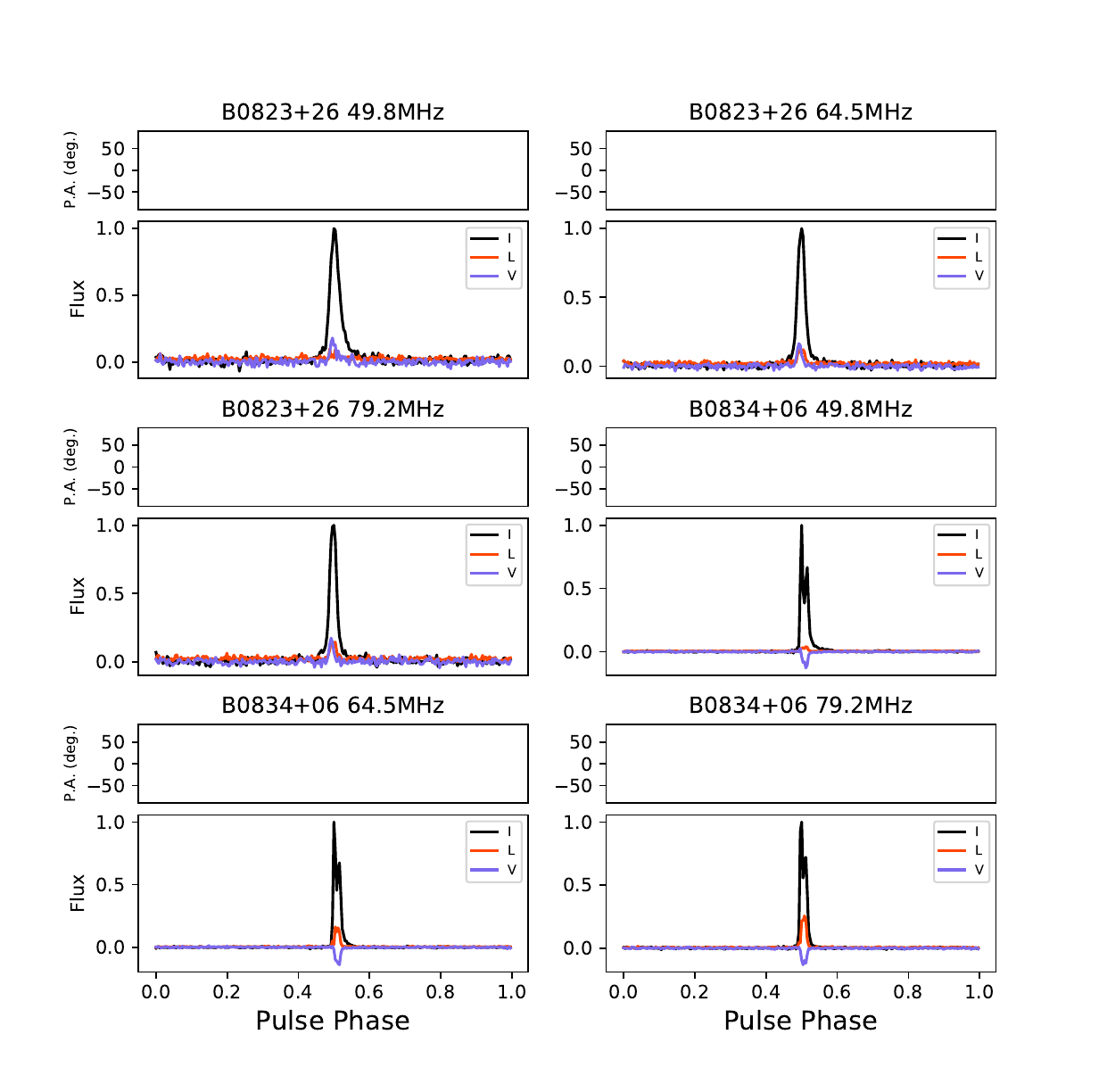}
    \caption{Fig \ref{fig:psrprofrm} Continued}
\end{figure*}
\begin{figure*}
    \centering
    \includegraphics[width=\textwidth,height=22.1cm]{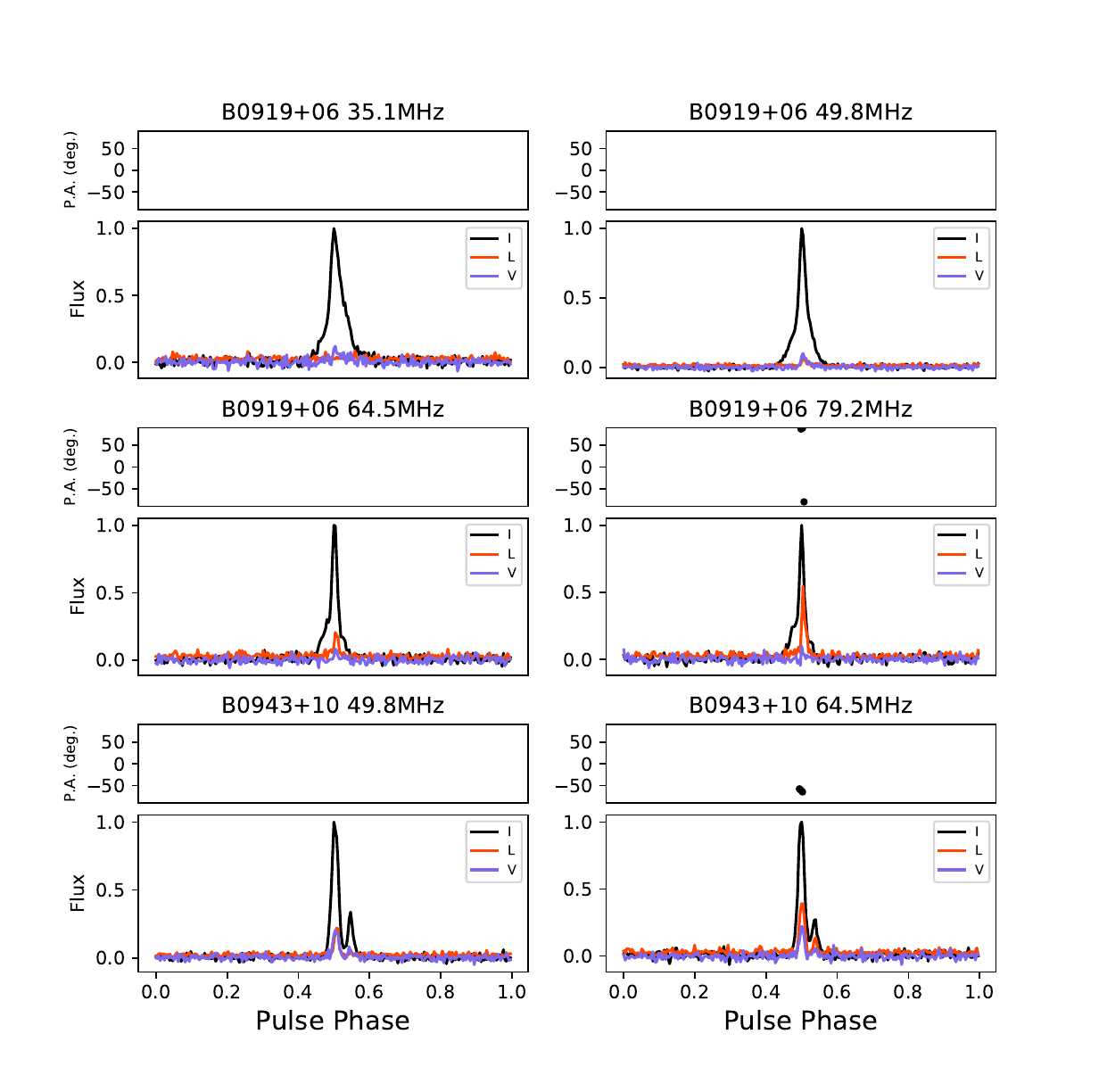}
    \caption{Fig \ref{fig:psrprofrm} Continued}
\end{figure*}
\begin{figure*}
    \centering
    \includegraphics[width=\textwidth,height=22.1cm]{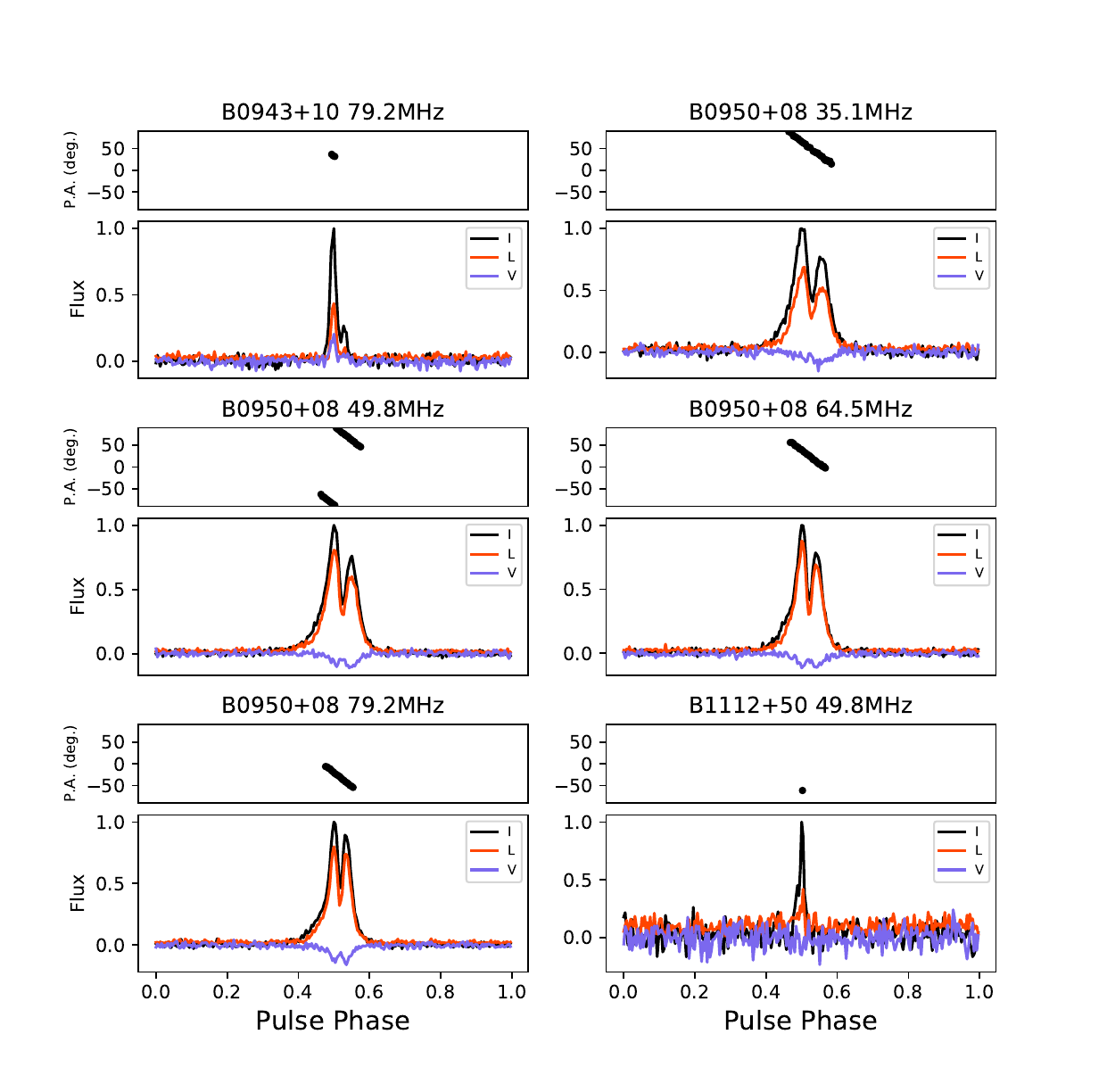}
    \caption{Fig \ref{fig:psrprofrm} Continued}
\end{figure*}
\begin{figure*}
    \centering
    \includegraphics[width=\textwidth,height=22.1cm]{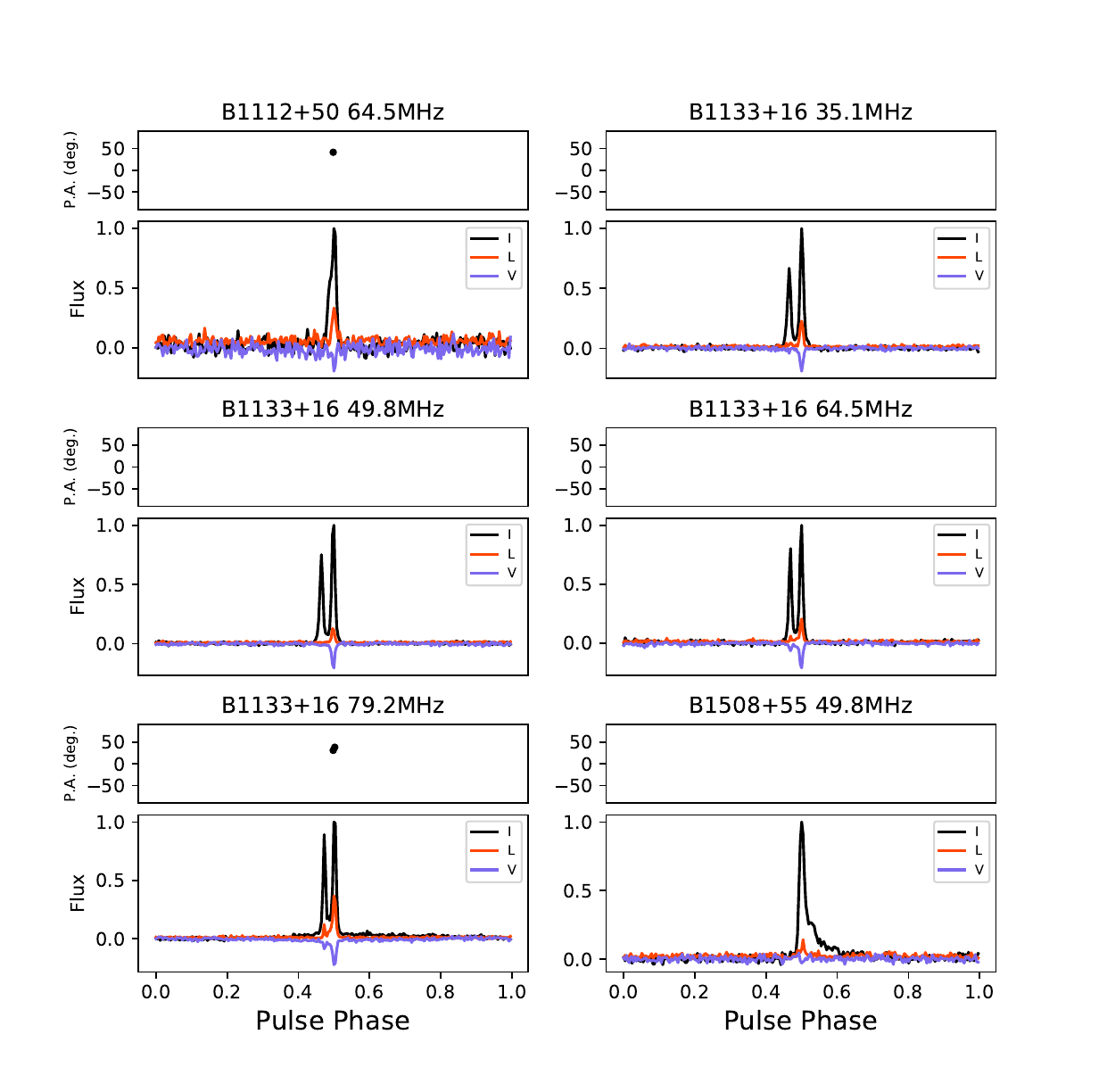}
    \caption{Fig \ref{fig:psrprofrm} Continued}
\end{figure*}
\begin{figure*}
    \centering
    \includegraphics[width=\textwidth,height=22.1cm]{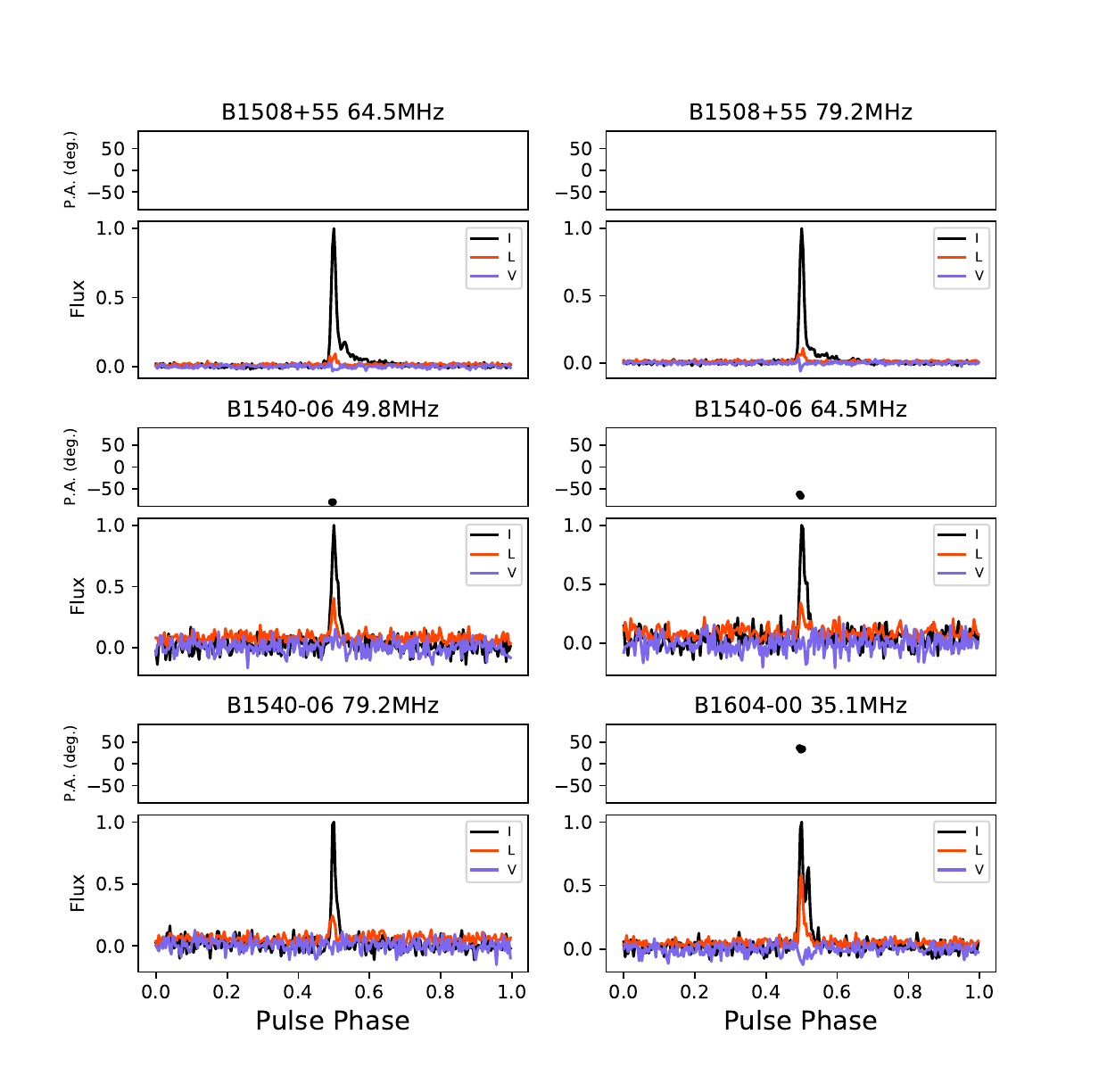}
    \caption{Fig \ref{fig:psrprofrm} Continued}
\end{figure*}
\begin{figure*}
    \centering
    \includegraphics[width=\textwidth,height=22.1cm]{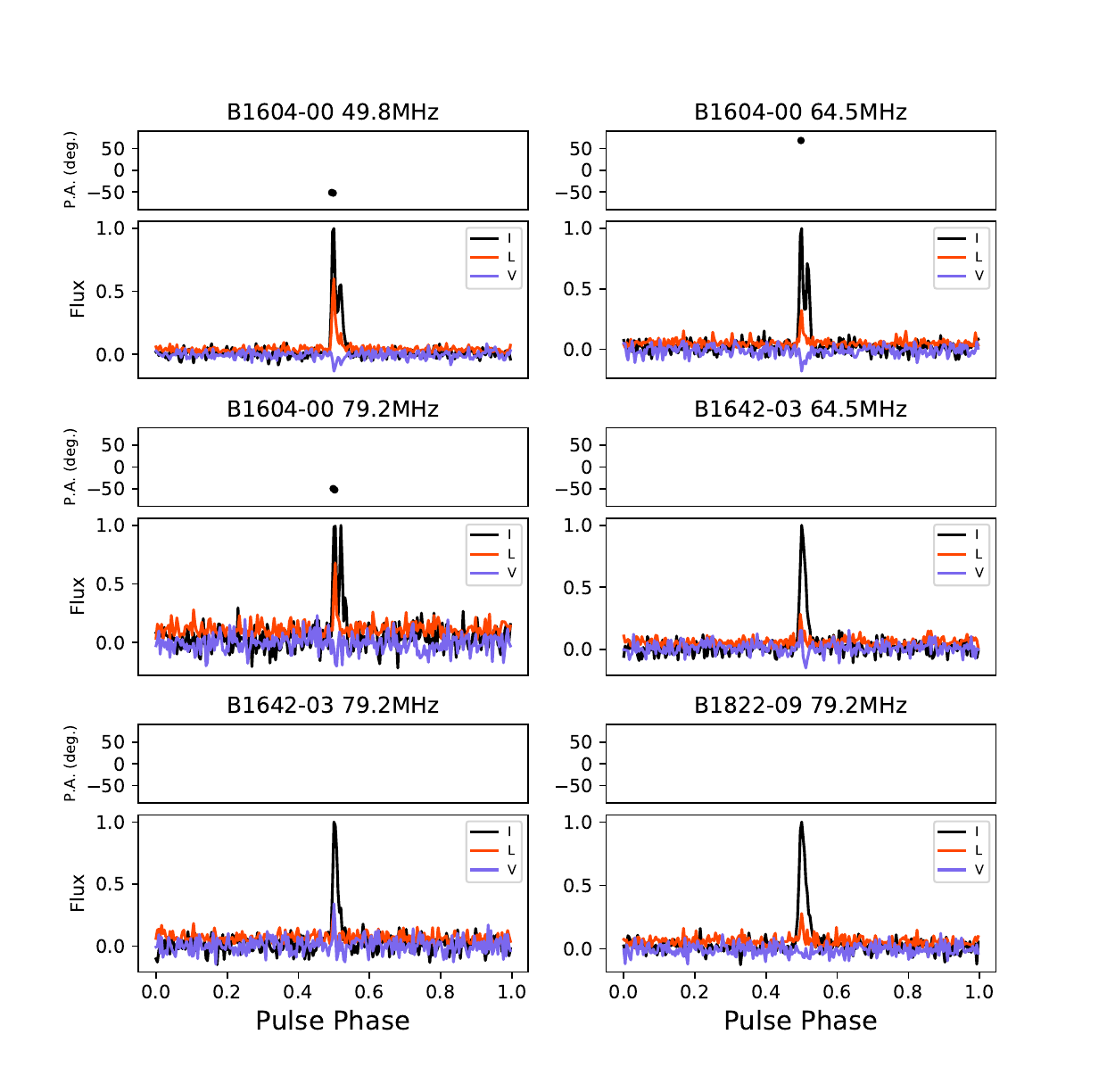}
    \caption{Fig \ref{fig:psrprofrm} Continued}
\end{figure*}
\begin{figure*}
    \centering
    \includegraphics[width=\textwidth,height=22.1cm]{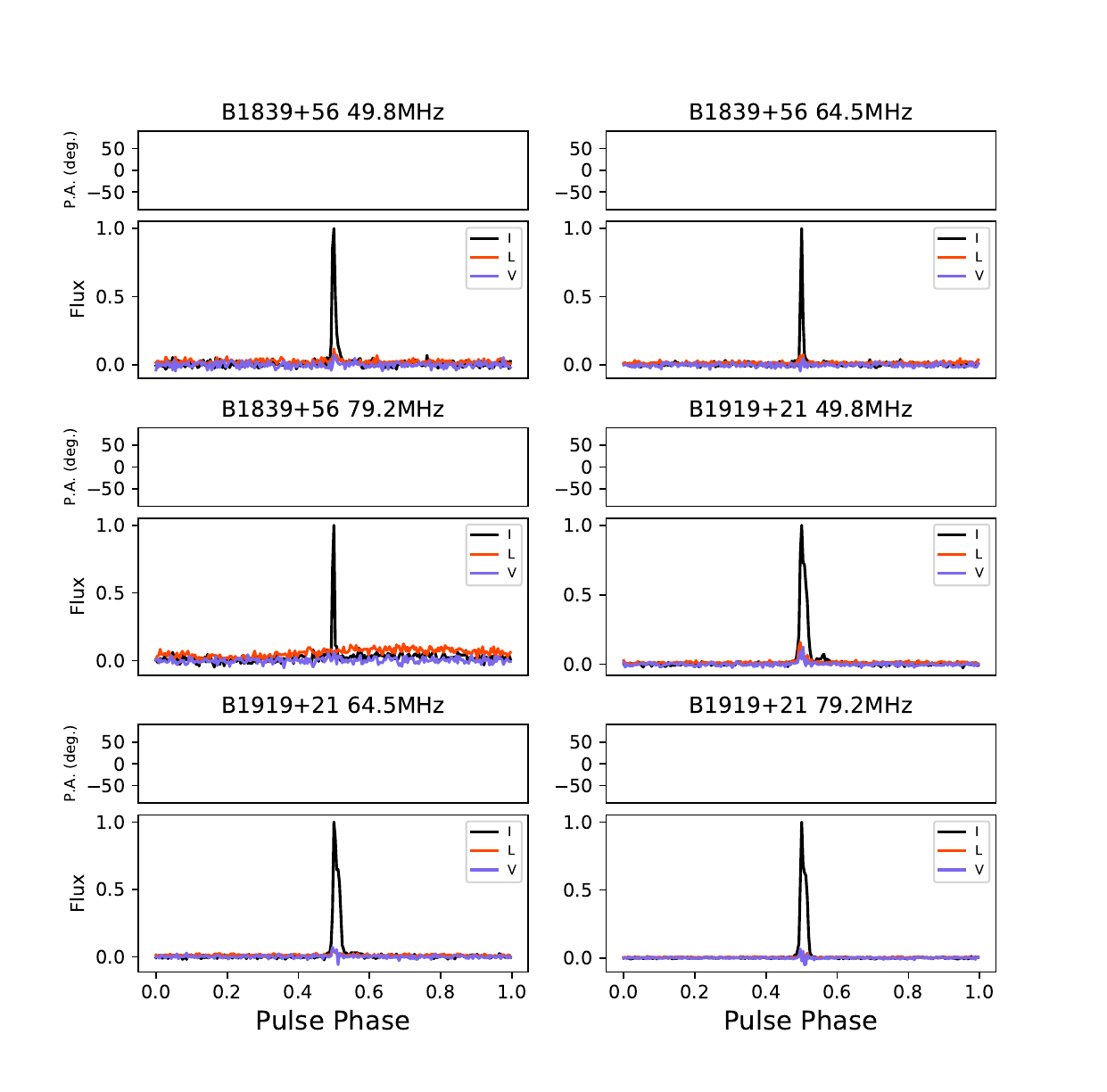}
    \caption{Fig \ref{fig:psrprofrm} Continued}
\end{figure*}
\begin{figure*}
    \centering
    \includegraphics[width=\textwidth,height=22.1cm]{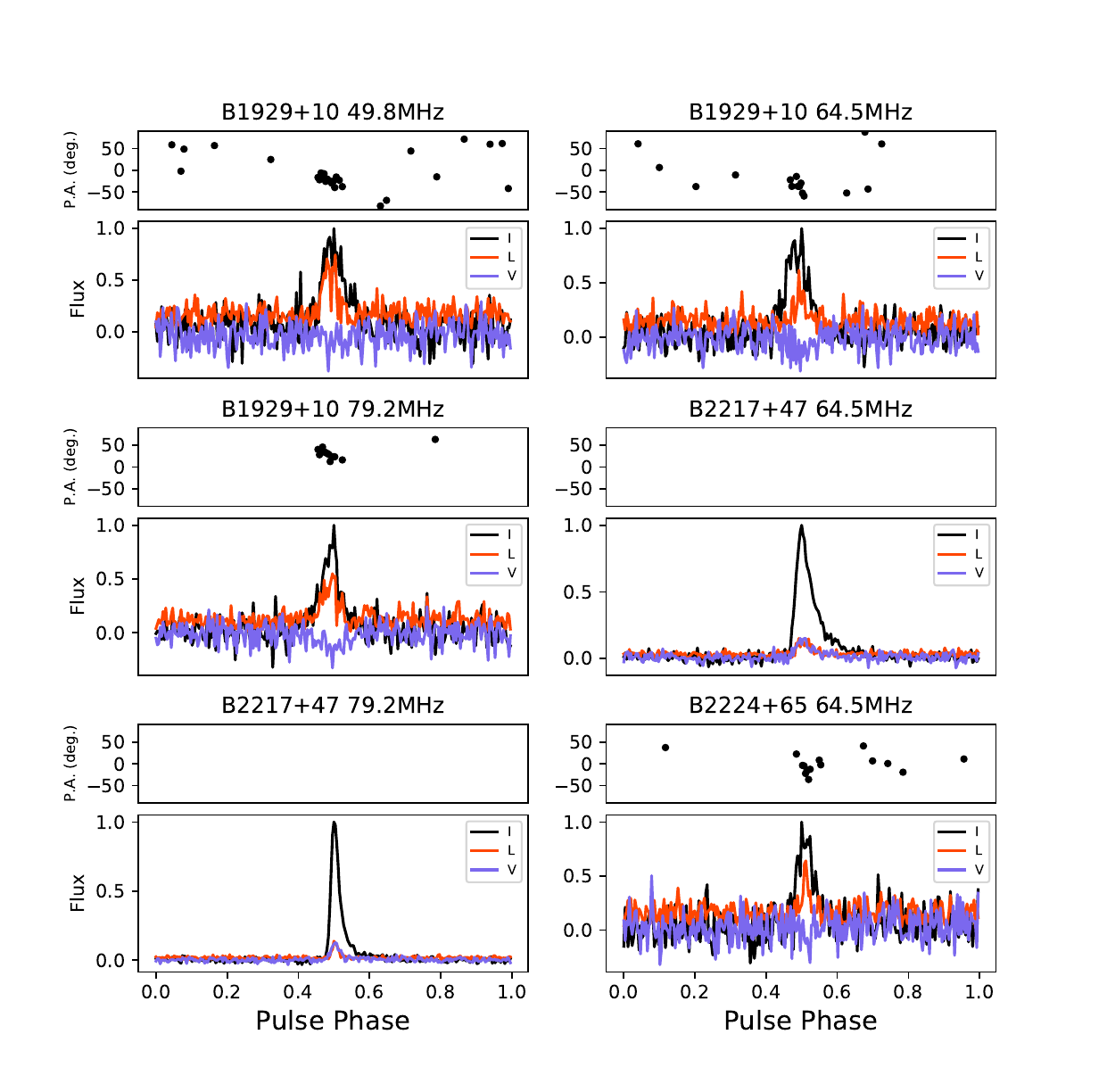}
    \caption{Fig \ref{fig:psrprofrm} Continued}
\end{figure*}
\begin{figure*}
    \centering
    \includegraphics[width=\textwidth,height=22.1cm]{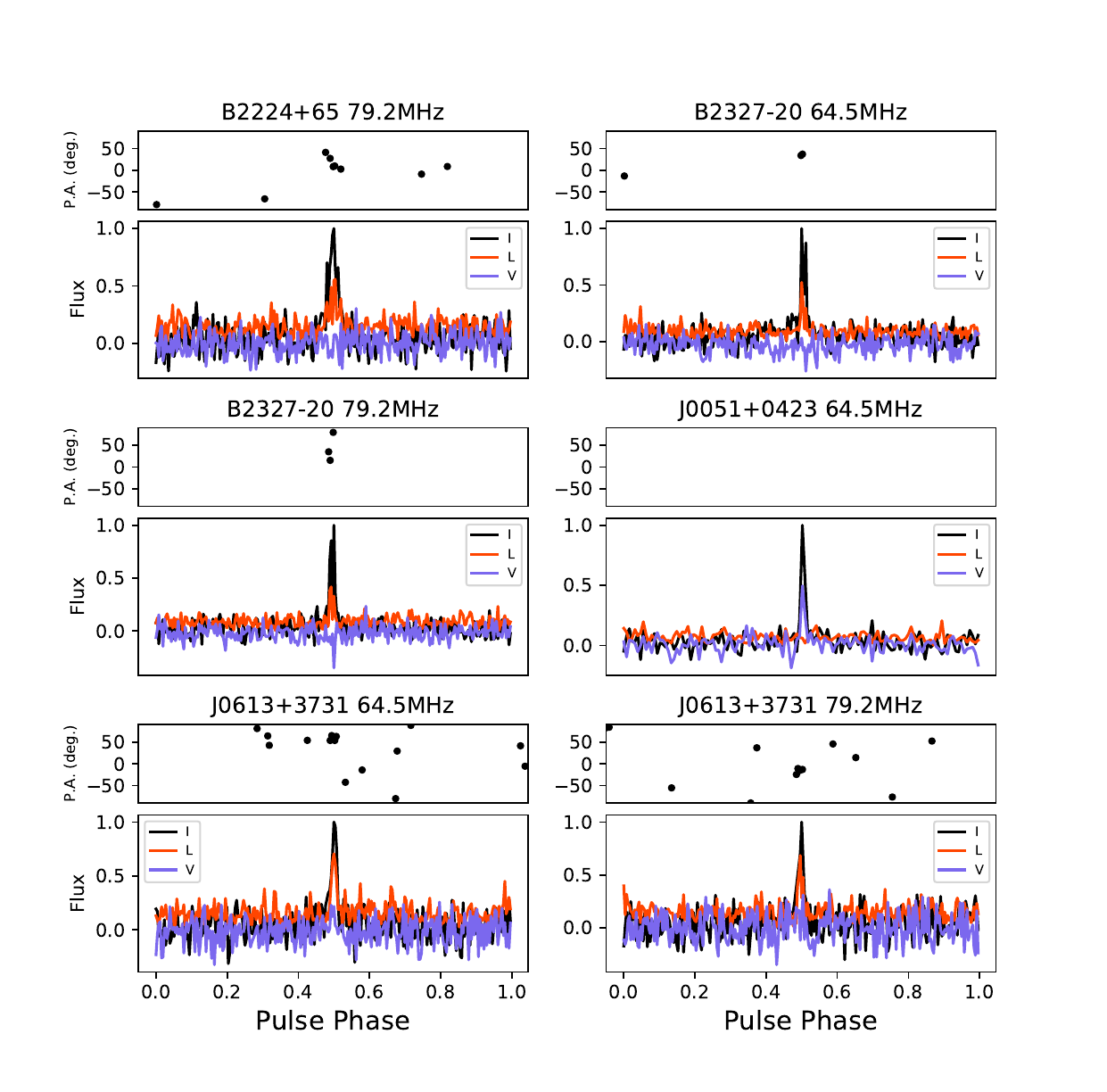}
    \caption{Fig \ref{fig:psrprofrm} Continued}
    \label{fig:J0051}
\end{figure*}
\clearpage

\section{Variation of Rotation Measure, and Average Magnetic Field}

\begin{figure*}[htbp!]
    \centering   
    \includegraphics[width=\textwidth,height=20.5cm]{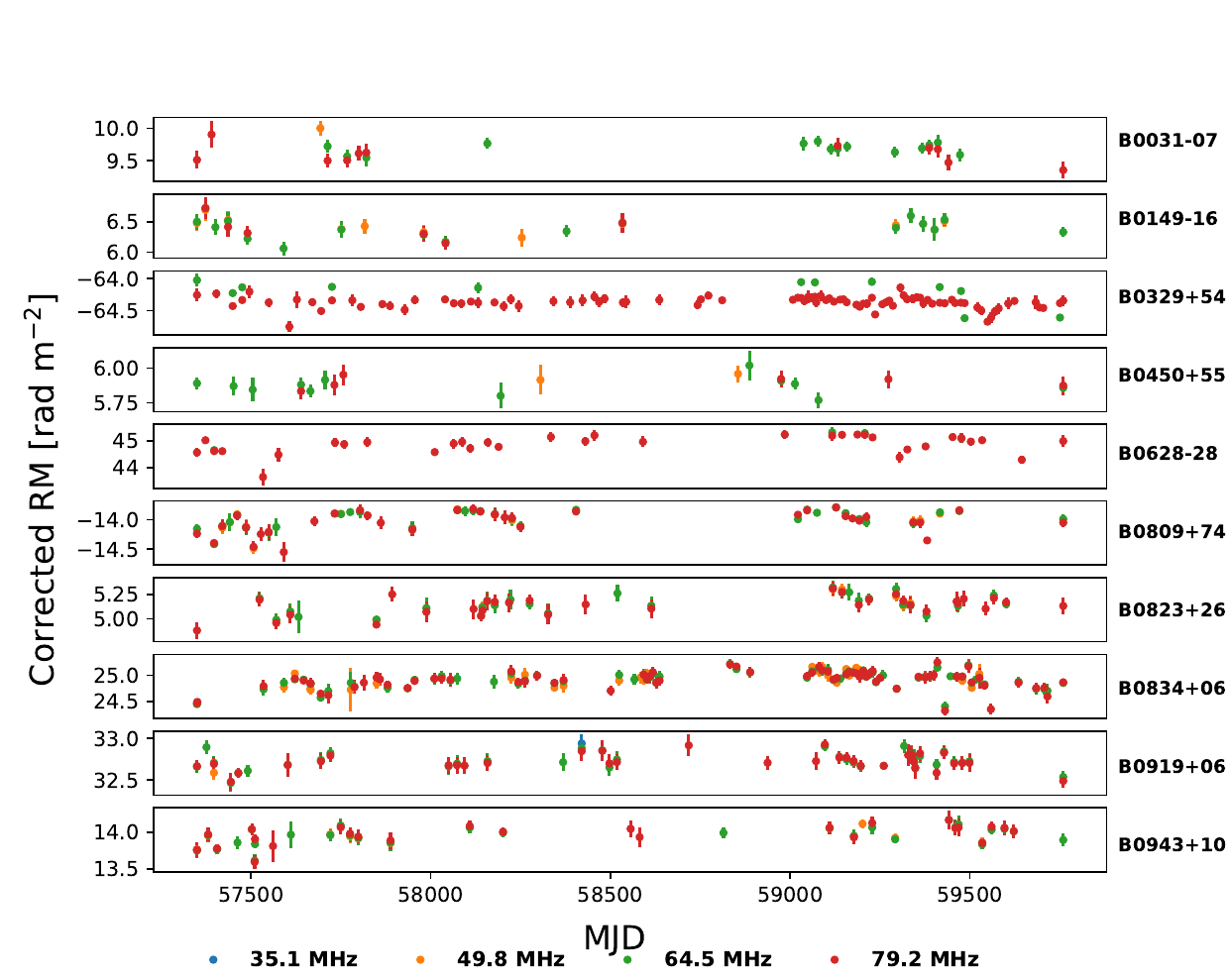}
    \caption{Variation of pulsar RM after correcting for ionospheric contribution using the TEC values from the IGS model. Colors in the bottom legend indicate the frequency at which the measurement was made.}
    \label{fig:rmtime}
\end{figure*}
\begin{figure*}[htbp!]
    \centering   
    \includegraphics[width=\textwidth,height=22.1cm]{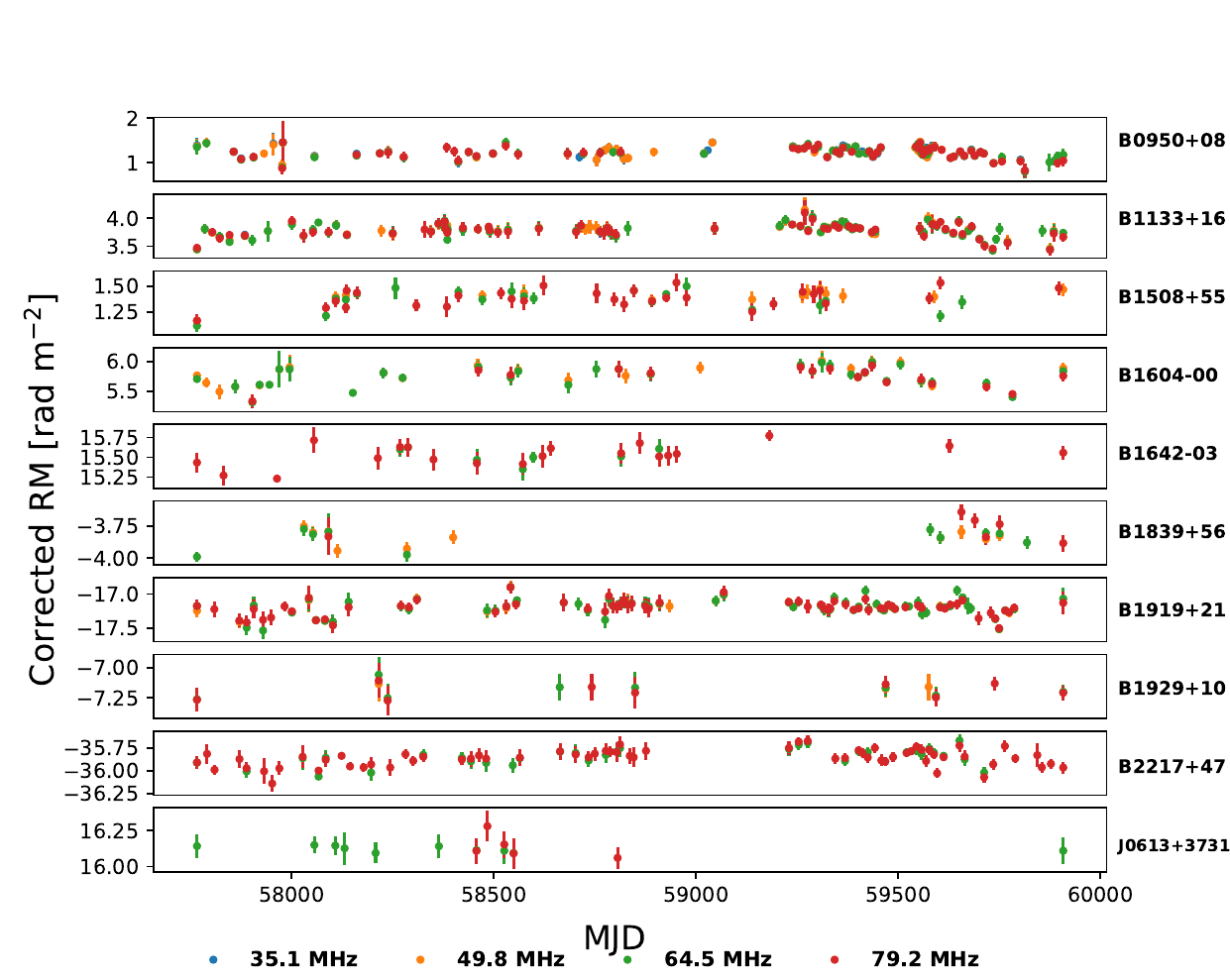}
    \caption{Fig \ref{fig:rmtime} Continued}
\end{figure*}
\clearpage

\begin{figure*}
    \centering   
    \includegraphics[width=\textwidth,height=20.5cm]{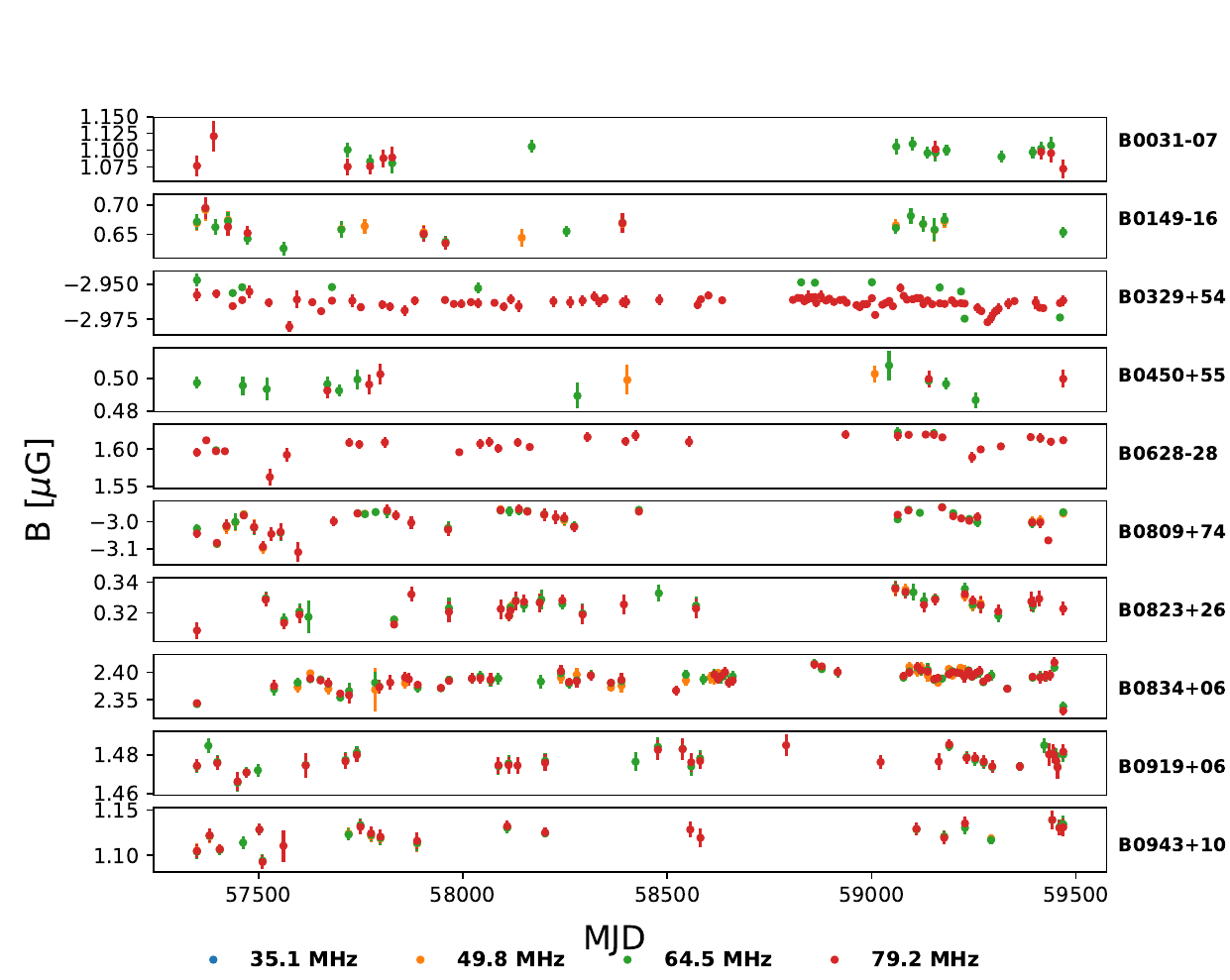}
    \caption{Variation of average magnetic field B$_{\parallel}$ calculated using RM and DM values per equation \ref{eq:5}.
    The error on B$_{\parallel}$ is the quadrature sum of errors in RM and DM. Colors in the bottom legend indicate the frequency at which the measurement was made.}
    \label{fig:Btime}
\end{figure*}
\begin{figure*}
    \centering   
    \includegraphics[width=\textwidth,height=22.0cm]{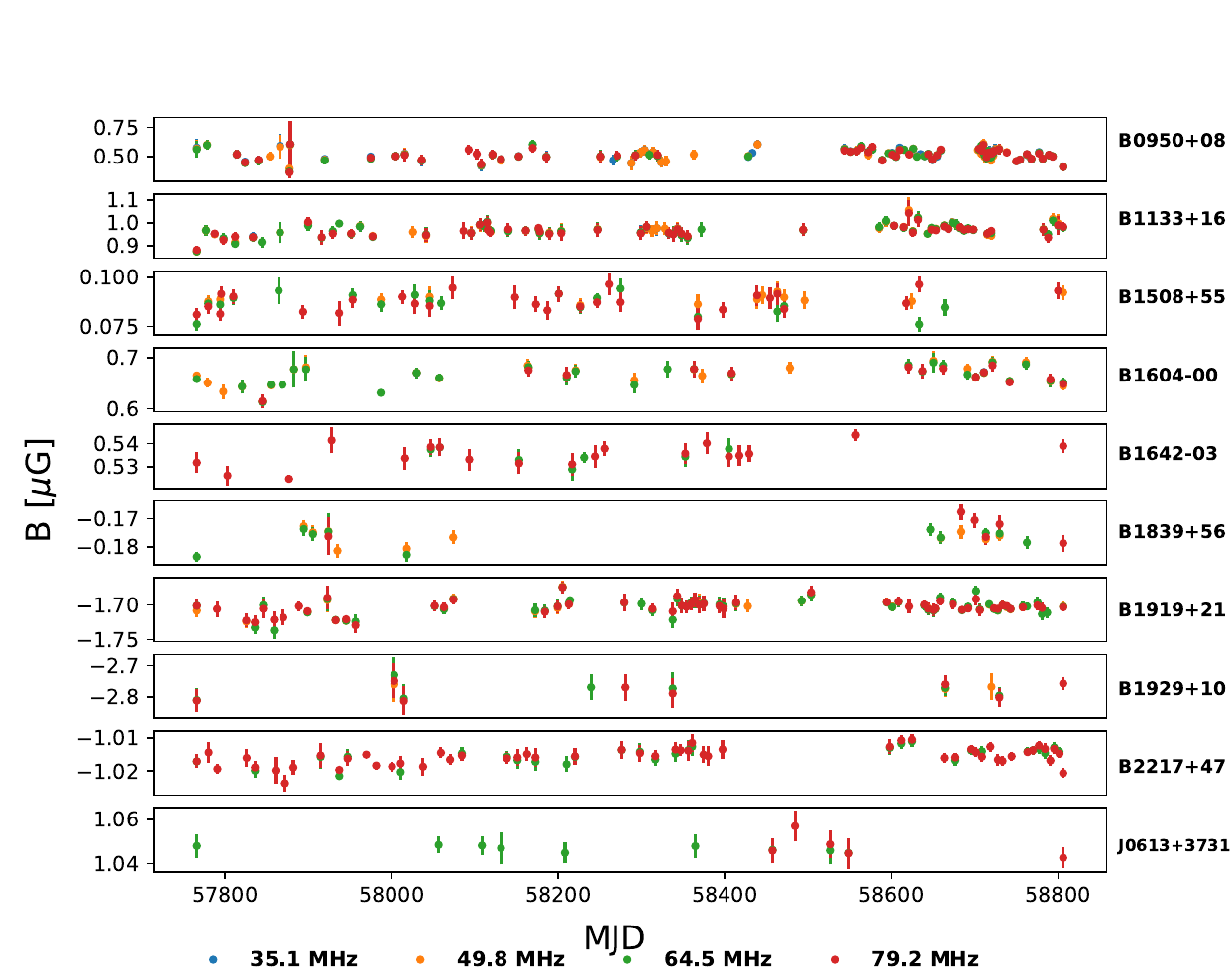}
    \caption{Fig \ref{fig:Btime} Continued}
\end{figure*}

\end{document}